# Optical Observations of Meteors Generating Infrasound – I: Acoustic Signal Identification and Phenomenology

Elizabeth A. Silber[1*], Peter G. Brown[1]

[1] Department of Physics and Astronomy, University of Western Ontario, London, Ontario, Canada, N6A 3K7, E-mail: esilber@uwo.ca

[*]Corresponding author:

E-mail: esilber@uwo.ca

Mailing Address:

Department of Physics and Astronomy
1151 Richmond Street
University of Western Ontario
London ON
N6A 3K7
CANADA
Ph: +1-519-661-2111 x86393
Fax: +1-519-661-2033

# Abstract

We analyze infrasound signals from 71 bright meteors/fireballs simultaneously detected by video to investigate the phenomenology and characteristics of meteor-generated near-field infrasound (< 300 km) and shock production. A taxonomy for meteor generated infrasound signal classification has been developed using the time-pressure signal of the infrasound arrivals. Based on the location along the meteor trail where the infrasound signal originates, we find most signals are associated with cylindrical shocks, with about a quarter of events evidencing spherical shocks associated with fragmentation events and optical flares. The video data indicate that all events with ray launch angles >117° from the trajectory heading are most likely generated by a spherical shock, while infrasound produced by the meteors with ray launch angles ≤117° can be attributed to both a cylindrical line source and a spherical shock. We find that meteors preferentially produce infrasound toward the end of their trails with a smaller number showing a preference for mid-trail production. Meteors producing multiple infrasound arrivals show a strong infrasound source height skewness to the end of trails and are much more likely to be associated with optical flares. We find that only about one percent of all our optically-detected meteors have associated detected infrasound and estimate that regional meteor infrasound events should occur of order once per week and dominate in numbers over infrasound associated with more energetic (but rarer) bolides. While a significant fraction of our meteors producing infrasound (~1/4 of single arrivals) are produced by fragmentation events, we find no instances where acoustic radiation is detectable more than about 60° beyond the ballistic regime at our meteoroid sizes (grams to tens of kilograms) emphasizing the strong anisotropy in acoustic radiation for meteors which are dominated by cylindrical line source geometry, even in the presence of fragmentation.

# 1 Introduction

Low frequency sound extending from below the human hearing range of 20 Hz and down to the natural oscillation frequency of the atmosphere (Brunt-Väisälä frequency) is known as infrasound (Beer, 1974; Jones, 1982). There are many sources of infrasound, both natural and anthropogenic. Infrasonic waves undergo little attenuation at ground level compared to audible sound because the attenuation is proportional to the square of frequency (Bass et al., 1972; Sutherland and Bass, 2004). This means that infrasound can be used for global monitoring of explosions. Since the mid-1990s the International Monitoring System (IMS) of the Comprehensive Test-Ban Treaty Organization (CTBTO) in Vienna, Austria, utilizes infrasound as one of its monitoring technologies. At present, the IMS has 45 certified and fully operational global infrasound stations (Christie and Campus, 2010; www.ctbto.org).

Meteors are one of the most elusive sources of infrasound. When small cosmic particles, also known as meteoroids, collide with the Earth's atmosphere at hypersonic velocities (11.2 – 72.8 km/s), they produce a wide range of phenomena, including heat, light, and ionization (collectively known as a meteor) as well as an atmospheric shock (Ceplecha et al., 1998). A typical visual meteor is produced by a particle larger than 1 mm; however, the size limit is a strong function of the entry velocity (Ceplecha et al., 1998).

The most notable historical example of meteor infrasound occurred on June 30, 1908, when a large meteoroid exploded over the Podkamennaya Tunguska River, generating an intense shockwave (Chyba et al., 1993). It was later discovered that infrasound generated during this massive explosion travelled twice around the globe and was recorded by microbarometers in Europe, primarily in England (Whipple, 1930). After the event, meteor infrasound observations became rare, only to be reinvigorated during the Cold War when infrasound was used to monitor nuclear explosions. It was realized however, that some explosive sources were not nuclear explosions, but in fact large meteoroid (1 – 10 m in size) airbursts (ReVelle, 1997; Silber et al., 2009). A theoretical treatment predicting the nature of infrasound generated by meteoroids was first developed in 1974 (ReVelle, 1974, 1976). However, the difficulty in unambiguously identifying infrasound produced by a particular meteor has left much of this theory unverified (e.g. Kraemer, 1977). Recently, Haynes and Millet (2013) have adapted the Whitham sonic boom theory (Whitham, 1974) to produce a theoretical model to predict the overpressure and period from meteor shocks.

In general, infrasound source characteristics (such as energy) are often estimated by purely empirical means (e.g. Mutschlecner and Whitaker, 2010); however, this process is of limited use for meteor infrasound where the source altitudes are very high and few empirical measurements exist. Consequently, a strong need exists for a large dataset of meteor events with independently known speed, trajectories and energies as a first step in validating theoretical frameworks.

During the late 1970s and early 1980s, there were attempts by several groups (McIntosh et al, 1976; Kraemer, 1977) to observe bright meteors simultaneously with optical, radar and infrasound instruments. In five years of observations only two events were positively detected (Kraemer, 1977). It was not until the inception of the IMS network that some well documented cases of infrasound from meteors were observed (ReVelle and Whitaker, 1999; Evers and Haak, 2003). More recently, several regional optical meteor networks emerged using modern technologies to monitor bright meteors (e.g. Oberst et al, 1998), resulting in an additional handful of meteor infrasound observations (Edwards et al, 2008).

In most cases, meteor infrasound signals have been associated with meteors whose flight characteristics were poorly known, limiting the ability to validate ReVelle's (1974, 1976) analytic meteor infrasound theory. In addition to validating existing models, the frequency of occurrence of meteor infrasound from any given location remains poorly known as does the diversity of the meteor infrasound source functions - either cylindrical (associated with the main ballistic wave) or spherical (associated with fragmentation event) and their relative importance. The relationship between the meteor energy deposition as a function of height and shock production as well as the effects of the varying atmospheric conditions on meteor infrasound propagation remain only partially explored.

To address these questions, we have measured a large dataset of meteors with the purpose of model testing and statistical studies. This has been accomplished by associating infrasound from meteors (also referred to as events) using optical measurements as a cue to search for meteor infrasound. We employed the Southern Ontario Meteor Network (SOMN) (Weryk et al., 2007; Brown et al., 2010) which uses integrated optical, and infrasound technologies to monitor, detect and measure the trajectory of bright regional meteor events. Between 2006 and 2011, a total of 6989 meteor events were recorded optically and of these 80 were also infrasonically detected. The advantage of studying short range (< 300 km) infrasonic events is that these direct signals are detected before they undergo substantial (and sometimes poorly defined) modifications during propagation due to atmospheric variability.

The specific goals of this coordinated optical-infrasound meteor study are to: (i) use astrometric optical measurements to positively identify infrasound from meteors; (ii) establish and constrain the point (and its uncertainty) along the meteor trail where the infrasound signal emanates; (iii) estimate the potential importance of atmospheric variability due to winds on meteor infrasound propagation; (iv) determine the type of shock production mechanism for meteor generated infrasound; and (v) classify meteor infrasound and correlate meteor infrasound classes using pressure-time waveforms. A major goal is to develop an observational foundation for future work to understand the underlying physical mechanisms which modify meteor infrasound signals and relate to sonic boom theory.

The second paper in this study will use the results from this work as the basis to critically evaluate the meteor weak shock theory of ReVelle (1974; 1976) as applied to meteors and use photometric measurements of infrasonically detected meteors to compare masses derived infrasonically from photometric/dynamic measurements.

Our paper builds upon an earlier study (Edwards et al., 2008) and extends it by using a large data base of optically detected meteor generated infrasound events (in the current study 71 vs. 12 simultaneously detected events in the earlier study). Our work also has an implementation of a new methodology for infrasonic signal association, verification and measurement and it uses an improved optical meteor astrometric measurement technique, hence providing better ground truth and constraints. With a larger ensemble of events we have also been able to develop a taxonomy of infrasound signal classification and define a new algorithm for determining the meteor shock source heights. Finally, this study takes into consideration atmospheric variations in meteor infrasound propagation and interpretation to constrain the uncertainty in source height.

Our global goals in this and the forthcoming paper are to: (i) critically examine the weak shock theory developed for meteors (ReVelle, 1976) experimentally, (ii) use weak shock theory to provide a bottom up estimate (using the infrasound signal alone) of the meteor blast radius and compare this with the equivalent blast radius from entry modelling as determined photometrically, and (iii) develop a homogenous dataset of meteor infrasound detections with known source characteristics (trajectory, energy, speed) for statistical examination of shock characteristics. Point (iii) will also allow others to test and compare infrasound shock models, both analytic and numerical.

In the following sections, we first present an optical and infrasound measurement methodology (with further details described in the supplemental material), and a proposed meteor infrasound classification system. Next we discuss the identification of the source height for our meteor infrasound signals using a ray trace model with a Monte Carlo implementation of gravity waves to simulate wind variability. In the final section we present our overall results, together with a discussion and our conclusions. In Appendix A (supplementary material) we present detailed optical and infrasound measurement methodology, our criteria for meteor infrasound detection and association, our new algorithm for calculating shock source heights, and the hardware aspects of the infrasound array and the all-sky camera systems used in this study.

## 2 Methodology

**2.1 Astrometry, Meteor Infrasound Signal Identification and Measurements**

The All-Sky and Guided Automatic and Realtime Detection (ASGARD) camera network is currently comprised of 10 stations throughout Southwestern Ontario. During the study period (2006-2011) however, up to 7 stations were in place. The precise position in the sky of each optically observed meteor in our data set was directly measured following the procedure outlined in Appendix A (Supplemental material). Briefly, the camera calibration plates for each event were generated using Meteor Analysis (*METAL*) software (Weryk et al., 2007; Weryk and Brown, 2012) and applied to camera images containing the meteor. The astrometric solution and precise position of the meteor in the sky were then generated using the software *MILIG* (Borovička, 1990).

The Elginfield Infrasound Array (ELFO), a four sensor tripartite array, is located 20 km from London, Ontario, Canada (43º.1907N, 81º.3152W, 322 m). The array spacing was optimized for detecting events with a peak frequency of approximately 1 Hz. The technical and operational details about the camera network and the infrasound array are given in Appendix A.

Two software packages are used to identify possible infrasonic signals, *MatSeis1.7* (Harris and Young, 1997; Young et al., 2002) and the Progressive Multi-Channel Correlation (*PMCC*) algorithm (Cansi, 1995; Cansi and Klinger, 1997; Garcès et al., 2003). Relative advantages of the two software packages, as well as the detection parameters we employed to identify and measure impulsive infrasound signals from meteors are discussed in Appendix A.

When a positive infrasound signal detection is found (here defined as a signal-to-noise ratio (SNR) of at least 3 dB) in the correlation indicating a coherent infrasonic wave crossing the array, additional information is generally required before the infrasonic signal can be confidently associated with a specific source. An infrasonic signal at range of less than 300 km from a typical local meteor is usually short in duration (1–10 s is typical), and in the majority of cases appears as a single N-wave (DuMond et al., 1946) with a duration on the order of seconds. Thus, it is imperative to have some sort of discriminative methodology which allows for the convincing association of meteor events with an infrasound signal.

In our study, we use the automated optical trajectory solutions from each detected meteor for each night, with the date, time, begin and end coordinates and altitudes as inputs, to calculate the following values for both the meteor begin and end point as seen from ELFO: the expected back azimuth, range (great circle path between the source and receiver) and expected travel and arrival time for tropospheric (0.340 km/s average speed or celerity), stratospheric (0.285 km/s) and thermospheric (0.220 km/s) signals. Using these values as guides, we then perform a targeted search for possible infrasonic signals at ELFO associated with the meteor. A typical ground-projected distance for most of the camera-detected meteors is on average 120 km (and up to 300 km) from ELFO, and since the meteor trail has a significant horizontal length (on order of tens of kilometers), the expected back azimuth, range and travel time windows may vary significantly between the begin point and the end point. Note that we check all meteors that were optically detected independent of brightness, where the optical network limiting meteor magnitude is near -2 (corresponding to gram-sized meteoroids at a speed of 40 km/s). The arrival time search 'window' thresholds are bounded by the expected signal travel times given the fastest and the slowest infrasonic celerity (tropospheric and thermospheric, respectively) at the begin point and end point, as well as the closest range from the meteor begin/end points to the array with an added five seconds of buffer. The search 'window' thresholds for back azimuth are given by the azimuthal fan sweep from the begin point up to the end point (with a 5° buffer at each end to account for other possible deviations due to the measurement uncertainty, atmospheric effects and array response).

For each optical meteor, a search for possible infrasonic detections is performed using both MatSeis and PMCC within the expected arrival time and back azimuth windows as described in Appendix A. These two quantities are expected to be much more constrained than trace velocity (or signal elevation angle) (McIntosh and ReVelle, 1984).

Once a positive detection is declared correlated with an optical meteor following the procedure described in Appendix A, a manual optical astrometric solution is used to re-run the *MATLAB*® program in order to refine the timing, distance and back azimuth predictions. These new values are then used to further check the observations against the predicted quantities according to our criteria. If necessary, the entire process of signal detection is repeated. This happens in cases where there is significant difference between the automated and manual optical astrometric trajectory solutions, which in turn affects the expected back azimuth and propagation time. In only five cases did this secondary check produce a rejection after an initial acceptance on the basis of the automated solutions (of ~ 80 initial events), suggesting a <10% loss rate due to poor initial automated optical solutions. Following positive infrasound signal correlations with an optical meteor, the meteor infrasonic signal parameters were measured, using the measurement technique described in Ens et al. (2012) and in Appendix A.

## 2.2 Raytracing and Atmospheric Variability Modelling

Having linked infrasonic signals with optical meteor events and performed signal measurements, the next step in event characterization is establishing where along the optical meteor trail the infrasonic signal originates. This was done by raytracing using the *SUPRACENTER* program (Edwards and Hildebrand, 2004) and a range independent real atmospheric profile for the day of each event. The raytracing results provide expected model timing and model arrival direction of the infrasound signal at ELFO, which are then compared with the observed signal. The total uncertainty in the signal arrival time consists of the event start time uncertainty (generally <1s except in two cases; see Appendix A) and the uncertainty produced by signal processing (1s). A non-isothermal and vertically inhomogeneous realistic atmospheric profile for each event was generated using wind data from the United Kingdom Meteorological Office (UKMO) assimilated dataset (Swinbank and O'Neill, 1994). The UKMO data extends to only 50-70 km in altitude; thus to reconstruct the atmospheric profile (atmospheric pressure, temperature and horizontal winds) from that point to 200 km altitude, the HMW95 (Horizontal Wind Model; Hedin et al., 1996) and the NRL-MSIS00 (Naval Research Laboratories – Mass Spectrometer and Incoherent Scatter Radar; Picone et al., 2002) models were combined by a smooth spline interpolation following the same procedure as adopted by Edwards et al. (2008).

Typically, optical meteor events last for 30 video frames or less; each frame corresponds to a different portion along the meteor's path. For raytracing, we used 100 discrete heights along the meteor trajectory (latitude, longitude and altitude) as source 'points' for raytracing. A version of *SUPRACENTER* (Edwards and Hildebrand, 2004) was used to follow rays from each point along the meteor trajectory launched toward the infrasound station and find probable arrivals, which we define as any rays emanating from any source point which travel to within 1 km from the central element, horizontally or vertically. Details of each ray that our propagation code indicates should reach the receiver are then saved, including the travel time, back azimuth and elevation angle at the receiver (station) and ray takeoff angle $\beta$. The latter is the angle between the meteor velocity and the ray wave vectors as well as the vertical and infrasound ray elevation angle at the source. Note that we only follow direct arrivals; ducted arrivals are not computed. The modelled infrasound travel times are adjusted for the finite meteor flight time.

Using our average atmosphere we found that many events showed no model propagation path or only a model propagation path from the part of the trajectory when winds were applied, though a signal was clearly recorded at the infrasound array. This emphasizes how the atmospheric variability and scattering can play a significant role in infrasound signal propagation (Balachandran et al., 1971; Brown and Hall, 1978; Green et al, 2011). Hourly, daily and seasonal variations in atmospheric infrasound propagation have been observed and well documented (e.g. Le Pichon et al., 2005; 2009). Many factors affect the propagation and detection of infrasound including attenuation (e.g. molecular absorption of sound in the air), non-linear effects, atmospheric turbulence, the effect of the ground surface on acoustic reflection, temperature and wind stratification, barriers, scattering, and atmospheric tidal effects (Brown and Hall, 1978; Ostashev, 2002; Sutherland and Bass, 2004; Kulichkov, 2010; Hedlin et al., 2012). Among these, gravity waves in particular (e.g. Hines and Reddy, 1967; Fritts and Alexander, 2003; Nappo, 2012) have been previously identified as having a significant impact on infrasound signal propagation as they perturb the local wind field on short timescales (Chunchuzov, 2004;

Ostashev et al., 2005; Green et al, 2011). Furthermore, gravity waves are especially important in perturbing the average windfield in the stratosphere and lower thermosphere, since this is where the gravity wave amplitudes and scale heights are the largest (e.g. Walterscheid and Hocking, 1991; Gardner et al., 1993; Fritts and Alexander, 2003; Bhattacharyya et al., 2003; Mutschlecner and Whitaker, 2006).

In more general terms, the scattering of infrasonic energy in the middle and upper atmosphere has also been recognized (Chunchuzov, 2004; Kulichkov, 2004; Ostashev et al., 2005; Millett et al., 2007), albeit still not fully understood. Case studies involving well documented large explosive events, such as the Buncefield oil depot explosion in the UK (Ceranna et al., 2009) and the Misty Picture experiment at White Sands Missile Range in New Mexico, USA (Gainville et al., 2010) have explicitly demonstrated noticeable effects on propagation produced by gravity wave perturbations.

However, the implications of gravity-wave induced wind perturbations on regional (near-field, <300 km) infrasound propagation from high altitude explosive sources have not been comprehensively investigated due to a lack of sufficient data. Because we expect wind perturbations to affect the propagation of meteor infrasound *a priori*, we have implemented a Monte Carlo - type approach to estimate the sensitivity of our raytrace solution to uncertainties in the wind field produced by gravity waves. We note that this is only one of several sources of uncertainty in infrasound propagation at regional distances; scattering, diffraction and local reflections may also play a role, but we do not explore these further in this work.

Many physical models of gravity waves exist (e.g. Mengel et al., 1995; Fritts and Alexander, 2003); however for the purpose of this study we used the gravity wave wind perturbation scheme implemented in the *InfraMap* raytracing software (Norris and Gibson, 2001; Gibson and Norris, 2000; 2003). Note that the perturbation to the temperature field (and hence indirectly to the sound speed) due to gravity waves is ignored, as this is much smaller than the direct perturbations to the effective sound speed through gravity wave-induced wind variability at our heights of interest.

In this approach the vertical mean wind profile is perturbed, using the Gardner et al. (1993) gravity wave model to simulate the spectral characteristics of gravity waves with varying height and a random-phase technique (Peitgen and Saupe, 1998). This model has been shown to explain propagation in shadow zones and counter-wind returns for other infrasound sources (Green et al, 2011). In our case, the gravity wave wind perturbation model provides an estimate of the variation in modelled travel-time and the back azimuth due to gravity wave induced variability in the atmosphere, resulting in a lower bound to our source height uncertainty.

To estimate the maximum expected deviation in the model travel time and back azimuth due to gravity waves, perturbations along the great circle propagation path (tailwind or headwind) and transverse to the great circle propagation path (crosswind) were applied to each atmospheric profile (Gibson and Norris, 2000). In total, 1400 perturbation realizations in the vertical wind profile per event were generated and then applied to each UKMO-HWM95 profile, forming a sample of 1400 individual atmospheric profiles, upon which raytracing was performed once again for each event. With the exception of a very small number of events (~10), nearly all meteors showed accessible propagation paths along the entire meteor trail. An example of some gravity wave perturbation realizations are shown in Figure S1.

## 2.3 Analysis of Raytracing Results

From the raytracing for each point on the meteor trajectory, we have the expected travel time, back azimuth, arrival elevation angle, the horizontal range (along the great circle path), total range, signal celerity and ray deviation from the trajectory (angle between the meteor velocity vector and the infrasound ray launch direction) at ELFO (Figure 1). When combined with the infrasound meteor measurements this produces residuals in the time, the back azimuth and the elevation angle between the model rays and the observations at the ELFO array. Our final dataset consists of 71 meteor infrasound events having a common optical record. To determine where along the meteor trail the infrasound signal originates, the raytracing model results were compared with the observed travel time, back azimuth and elevation angle measured at ELFO.

The ray deviation from the trajectory is important in understanding the nature of the shock production. On theoretical grounds, we expect that there are two types of shock production mechanisms – ablational, due to fragmentation; and ballistic, due to the production of cylindrical line source shock along the entire trail (ReVelle, 1976; Bronshten, 1983). A 90 degree ray deviation angle (±25°) is indicative of the ballistic shock (ReVelle, 1976; Brown et al., 2007) and a cylindrical line source geometry, while the ablational shock is expected to be more omnidirectional. However, it remains unknown which of these shock modes is dominant at small meteoroid sizes and what the range of allowable deviation from 90° for true ballistic shocks is. Quantifying these unknowns is one of the goals of this study.

Uncertainties in the signal arrival elevation at ELFO were computed assuming a possible variation of up to 7°C in the local temperature, which was measured either by a weather station located at ELFO or by UKMO model estimates of the surface temperature. Our raytracing model using Monte Carlo perturbation realizations produces a 'cloud' of possible airwave arrivals from each point along the trajectory. To decide from this ensemble the most likely shock source heights, we developed an algorithm to find the best estimate for the source shock height and its associated uncertainty.

The algorithm works as follows: from our model runs, four modelled arrival quantities were extracted to compare with the observations – travel time, back azimuth, elevation angle and ray deviation from the trajectory. The travel time was used as the primary source height discriminant while the back azimuth was used as a secondary discriminant. We note that the ray arrival elevation angle can vary significantly and is the least reliable measure of all observed quantities (McIntosh and ReVelle, 1984), as atmospheric turbulence, for example, may cause local temperature changes of 5°C in a few seconds (Embleton, 1996). As a result, it was not used to determine the final source heights, only as a tertiary check with the first two quantities. The best estimate of source height then produces an estimate for the ray launch angle, and its deviation from 90 degrees gives an indication of the shock type (ballistic = 90 ± 25 degrees, ablational = any angle). Further operational details pertaining to the algorithm and source height solution types are given in Appendix A.

While the majority of the solutions appeared to have a height solution (i.e. fall within the range of the observed quantity and its uncertainty), some solutions were degenerate (could have two possible height solutions), or showed no solution (i.e. did not fall within the observed quantity and our range of adopted uncertainty). In the cases where two source height solutions were possible for the modelled travel times, a subjective determination of the best fit was made, using the best fit height from the back azimuth to isolate the most likely 'true' source height. We note that the high fidelity astrometric solutions (±0.2 km for begin and end points in horizontal and vertical directions) in this study allow for accurate trajectory measurements and event timing, thus the final source heights matched to the model travel times will have atmospheric variability as the main source of uncertainty. Furthermore, direct arrival, short range (<300 km as in this study) infrasound signals do not undergo significant ducting and thus do not suffer from additional modification as seen in far-field (long distance) propagation.

## 3 Results and Discussion

### 3.1 Infrasound Signal Taxonomy and Phenomenology

The initial 80 optical meteors simultaneously detected by infrasound were reduced to 71 because some events did not have useable optical astrometric solutions while for others there were no direct arrivals. Out of 71 events, 55 had associated single infrasound arrivals and 16 produced multiple infrasound arrivals (three events were triple arrivals, while the remaining 13 were double arrivals). The summary of infrasonic signal characteristics as measured for meteor events producing single infrasound arrivals and multi infrasound arrivals are shown in Table S1 and Table S2, respectively. All analyzed signal celerities are consistent with direct path propagation from the source to the receiver. Our final dataset of events and their astrometric measurements are shown in Table S3. The detection efficiency of the array is summarized in Appendix A.

Using our dataset of 71 meteor infrasound events, a new taxonomic classification was developed based on the pressure-time record of the signal. While the majority of the signals in our dataset showed N-wave (DuMond et al., 1946) type arrivals typical of sonic boom signatures (Whitham, 1952; Seebass, 1967; Pierce, 1968; Crow, 1969; Gottlieb and Ritzel, 1988), some signals have double or even triple N-waves, and others exhibit more complicated features, such as variability in amplitude, wave appearance, ripples, and humps. Additionally, there were some signals which appeared 'diffuse', meaning that they have four or more pressure cycle maxima, all with comparable amplitudes where it is difficult to discern specific N-wave features. The amplitude and duration of the N-wave signature are linked to the body's shape and speed following Whitham's F-function theory and the sonic boom area rule (Whitham, 1952; Seebass, 1967). However, we do not specifically investigate these here, as the main focus of our study is the signal phenomenology, our intent being to lay the observational foundations for future theoretical studies.

Our infrasound taxonomic signal classification is described in Table 1 and shown in Figure 2. Note that we only use the apparent shape of pressure vs. time for classification. The theoretical interpretation and treatment of such profiles are left for future studies. Class I signals clearly dominate the data set. In total, 57% of all arrivals were Class I signals, while 22% were Class II, 7% Class III and 14% Class IV signals. All signal and source height properties associated with each meteor infrasound taxonomic class are summarized in Table S4.

Meteor generated infrasound signals with slowly decaying wavetrains (in this study referred to as subclass 'a'), as well as double and triple brief cycle pulses with nearly identical amplitudes were also observed and reported by Edwards (2010). These peculiar types of multiple N-wave signals are defined in our classification table as Class II and Class III. Edwards (2010) suggested that these repeating N-wave like pulses could be due to the relaxation of the atmosphere as it restores itself to the ambient levels. Alternatively, local turbulence in the lowest layers of the atmosphere could cause some of these distortions as has been found for sonic booms (e.g. Pierce, 1968; Hayes et al., 1971).

These waveform signatures, however, are not unique to meteor generated infrasound. Gainville et al. (2010) modelled the infrasonic waveforms from the Misty Picture experiment and showed the measured waveforms as recorded at the Alpine, White River and Roosevelt sites, all of which were within a few hundred kilometers from the source (248 km, 324 km and 431 km, respectively). The waveform from Alpine bears a close resemblance to the Class II signals typically observed in our study (irrespective of the frequency content and signal amplitude). The signals in our study originated at high altitudes and had a direct propagation path to the detection location, while the signal from Alpine was a thermospherically refracted arrival. We note that the typical signal amplitudes in our study are at least two orders of magnitude smaller than those of the Misty Picture experiment. Gainville et al. (2010) showed that both scattering and nonlinear effects influence the shock front evolution and propagation and local effects near the receiver array may affect our meteor infrasound signals in a similar manner.

Brown et al. (2007) suggested that a very high temperature region near the edge of the blast cavity induced by rapid deposition of energy as the meteoroid propagates through the atmosphere may enable, due to non-linear refractive effects, outward shocks to have ray deviations up to 25° from the ballistic regime. Successive shocks, potentially originating at different points along the hypersonic path (cylindrical line source) may then undergo interference (e.g. converging shocks) and produce a complex flow pattern, similar to that around an axisymmetric slender body (e.g. Whitham, 1952). For example, a ballistic shock (which reaches an observer on the ground) from a certain height may get 'mixed' with a refracted ray emanating from an adjacent portion of the trail, thus forming complex shock features at the source. Meteoroids which fragment and have two or more distinct fragments with independent shocks might also produce multiple N-wave signatures, provided the transverse spread is on order of 100m or more, values consistent with transverse spread for some larger meteorite producing fireballs (Borovička and Kalenda, 2003). Moreover, events with ballistic arrivals having fragmentation episodes within a few kilometers of the specular point would have the cylindrical and spherical N-waves arrive close in time. Further studies and numerical modelling are needed to investigate the connection between the non-linear and refractive effects, as well as the shock pattern at the source and their subsequent manifestations in the waveform as received by the observer.

Infrasound observations from the large sample of events in this study suggest that Subclass 'a' can be generated anywhere in the meteor region (middle and upper atmosphere) and it does not show any particular association to source altitudes. Signal class also shows no correlation with source heights, suggesting that there is no association between these two parameters.

The correlation between the signal class and ray launch deviation angles is shown in Figure S2. While meteor infrasound taxonomic classes I, II and IV are evenly spread over nearly all ray launch deviation angles, all class III events occur only when ray deviations are strictly around the ballistic regime of 90°. The significance of this correlation is hard to gauge as there are a very small number of class III signals compared to the other classes, and the triple N-wave pressure signal is not obviously linked to any purely ballistic process.

The correlation between the signal peak-to-peak amplitude and the signal class is shown in Figure 3. There is a general pattern in the peak-to-peak signal amplitude, which shows an apparent decrease as the taxonomic class increases. This may hint at a change in the acoustic radiation efficiency with an angle from the ballistic direction or that longer duration shocks showing more N wave cycles have a larger wavetrain to "spread" similar energy and hence the peak amplitudes appear to be smaller. Indeed, classical Whitham's theory indicates that the quantity *pressure x duration* is conserved and hence the longer durations of the higher classes would expect to be correlated with attenuated amplitudes. However, the underlying cause of this correlation for meteor infrasound remains unclear. 'Outliers', which appear to be far beyond the peak-to-peak amplitude limits of any of the other meteors, are associated with higher energy bright fireballs which exhibit gross fragmentation events. One such outlier is the Grimsby meteorite-producing fireball (Brown et al., 2011). These high-energy events more strongly reinforce the correlation between the signal class and peak-to-peak amplitude. The dominant signal frequency shows a clear relationship with range, as expected (Figure S3). When searching for direct meteor infrasound arrivals, it is useful to define a zeroth order empirical discriminant via the upper envelope (Figure S3), which shows the estimated maximum likely detection range (*R*) for typical regional meteor events as a function of the dominant frequency:

$$Frequency = 2.05 + 290 \exp(-R/31) \qquad (1)$$

where the dominant signal frequency in Hz and *R* is in km.

### 3.2 Shock Source Heights and Entry Velocity Distributions

Using our new algorithm to determine the shock source heights from raytracing using a sequence of 1400 *InfraMap* gravity wave perturbation realizations, we found best estimates for the height (and its uncertainty) along the meteor path where the infrasound was produced and subsequently detected at ELFO for single arrivals and multi arrival events. These results are summarized in Table S5 and Table S6, respectively. For the purpose of quantifying the shock source parameters, each infrasound arrival was treated as a separate event. An overview map with the ground trajectories showing the point along the meteor trail where the infrasound is produced as detected at ELFO is shown in Figure S4. Overall, we find that the travel time is the most robust estimator for the true shock source heights; secondary estimates from other parameters usually agreed with the height found from travel time residuals. Taking into account gravity-wave induced variability through our model we were able to match all observed arrivals, demonstrating in a model sense that wind perturbations can be a major factor in permitting infrasound propagation paths which do not otherwise exist using an average atmosphere only. Typically, the model predicted spread in the travel time due to atmospheric variability was between 2s and 7 s, and the spread in the back azimuth was $1° - 4°$. Most events had source height agreement between both travel time and back azimuths, while others showed poorer agreement.

We compared our results with a previous study which explored 12 common optical-infrasound events recorded during the period between 2006 and the early 2007 (Edwards et al., 2008). These were also analysed as part of our study. In terms of raytracing, the main difference between the two studies is that in the current study there is a much smaller overall spread in the source height uncertainties per event. The shock source heights derived in this study differ from those found previously by 7 km on average. We attribute this difference to three major factors: (i) better astrometric solutions (e.g. improvements in the processing software have been made since 2007); (ii) the incorporation of atmospheric variability in raytracing solutions; and (iii) the method of finding the source heights using our new algorithm which differs from that used in the Edwards et al. (2008) study.

The luminous trails for all meteoroids in this study start at altitudes between 75 – 140 km, and end as low as 35 km for single arrivals and 20 km for multi-arrivals. It is evident from Figure 4 that there are two distinct meteoroid populations in terms of entry velocity; one population has entry velocities < 40 km/s (31 single arrival events or 56%), while the other population has entry velocities > 40 km/s (24 single arrival events or 44%). The same trend can be seen in the multi arrival events; nine out of 16 events are in the slow entry velocity population, while the remaining seven events are in the fast entry velocity population. These local peaks in the apparent speed of meteoroids at the Earth have been recorded by other systems, both optical and radar (e.g. McKinley, 1961). They represent the asteroidal/Jupiter family comet material (low speed peaks) and the near-isotropic comet or the Halley-type comet material (high speed peak) following the classification convention of Levison (1996). In this study, the slow meteoroid populations (<40 km/s) often produce fireballs with long lasting luminous trails (>2 s), which penetrate substantially deeper into the atmosphere, consistent with their stronger structure (Ceplecha and McCrosky, 1976). In contrast, the fast meteoroid population (>40 km/s) usually produce shorter lasting luminous trails (~1 s), and have end heights above 63 km, due to their cometary origins (Borovička, 2006). Furthermore, high velocity meteors are more likely to produce flares, which are associated with fragmentation processes (Ceplecha et al., 1998). There is a strong correlation between meteoroid entry velocity, ablation heights (luminous trail) and the duration of luminous flight in the atmosphere (Ceplecha et al., 1998) (Figure 4), demonstrating the correlation between entry velocity, total luminous path length and meteoroid population types (Ceplecha and McCrosky, 1976). A summary of average properties and standard deviations as well as extreme values for the two velocity populations in the single arrival category are given in Table S7. The infrasound signal characteristics reflect the velocity-duration-altitude interdependence, namely higher speed meteoroids ablate higher and have shorter durations because they are also lower in mass for a similar brightness threshold.

The shock wave, as it propagates away from the meteor source height, undergoes attenuation proportional to the altitude, with the high frequency content preferentially removed due to absorptive losses, turbulence, heat conduction and molecular relaxation (ReVelle, 1976). As higher velocity meteoroids deposit more energy at high altitudes and have correspondingly lower frequency signals reaching the ground, there is a direct relationship between meteor entry velocity and dominant period and dominant frequency of the infrasonic signal received at the array for these small regional events. In general, we find that the dominant signal period is significantly smaller and confined to below one second in the slow entry velocity meteor population. Conversely, the high velocity population shows much more scatter; however, we observe an apparent upward trend in limiting signal period as a function of entry velocity. Only low velocity meteors (< 40 km/s) produced infrasonic signals with dominant frequencies > 4 Hz (Figure S5). This characteristic might be exploitable for stand-alone meteor infrasound measurement as a means to roughly constrain entry speed. Additionally, there is a strong relationship between the shock source height and the dominant signal frequency (Figure S5).

Compared to the slow velocity, deep penetrating meteoroids, the high velocity, very high altitude (>100 km) meteoroid population is less likely to produce infrasound that propagates to the ground, though other studies have reported rare instances of infrasound from high altitude meteors, especially those associated with meteor showers (McIntosh et al., 1976; ReVelle and Whitaker, 1999; Brown et al., 2007). Out of the total population of 71 single arrival events, 7 high altitude (>100 km) events generated infrasound detectable at the ground, suggesting that high-altitude ablation from meteoroids, even though not in the continuum flow, may still be capable of producing infrasound detectable at the surface, an effect also noted in studies of high altitude rocket-produced infrasound (Cotten and Donn, 1971).

Finally, we note that meteoroids with shallow entry angles are more likely to produce detectable infrasound due to the line source geometry having better propagation paths to the ground. The entry angle distributions for all events in this data set, as well as the ASGARD data, are shown in Figure S5.

## 3.3 Definition of $S_h$ Parameter and Estimation of Shock Type

To examine the possible shock source types (spherical vs. cylindrical) we first define a source height parameter ($S_h$) as:

$S_h = L_{Bh-S}/ L$      (2)

where $L_{Bh-S}$ is the path length along the trajectory from the begin point of the meteor to the shock source point, and L is the total path length of the entire visible meteor trajectory. Thus if $S_h \approx 1$, the shock originates at the meteor's luminous end point and as $S_h \to 0$, the shock originates closer to the luminous begin point of the meteor trail. The distribution of $S_h$ for all arrivals exhibits two peaks, one around the middle of the trail, and another one closer to the end point (Figure S7). As expected, there is a weak inverse relationship between the shock source height and the $S_h$ parameter with the shock source heights for the $S_h$ parameter < 0.5 constrained to the altitude region above 60 km (Figure S8).

We expect the statistical behaviour of the $S_h$ parameter to be diagnostic of the shock production mechanism, and thus examine this supposition further. For ballistic (cylindrical line) sources any portion of the trail is geometrically likely, though we expect a bias toward the end of the trail due to lower attenuation from lower source heights. The point of fragmentation, associated with spherical source geometry, depends on many factors, such as the meteoroid velocity, composition, tensile strength, etc. (Ceplecha, 1998, Ceplecha and ReVelle, 2004), but such fragmentation points tend to occur lower in the trail, particularly for asteroidal-type meteoroids. Several well documented meteor events have provided valuable information shedding more light on the fragmentation and breakup characteristics of larger meteoroids (e.g. Borovička and Kalenda, 2003) during meteoroid flight through the atmosphere (e.g. Brown et al., 2003; ReVelle et al., 2004), which may be similar to several deep penetrating events in our multi arrival population (e.g. Brown et al., 2011), but not representative of the entire population.

In our study, five multi arrival events and eight single arrival events have ray deviation angles (β) greater than 115°, which may indicate spherical shock production based on earlier interpretations (e.g. Edwards et al., 2008). If a fragmentation episode (which we expect to act as a quasi omni-directional acoustic emission source) can be linked directly to the shock production point along the trail, then the ray deviation angle is not expected to be confined to the ballistic regime (90°±25°) (see Brown et al. (2007) for a discussion of the theoretical variance in the ballistic angle from considerations of the expected gradient introduced in the local sound speed by the cylindrical shock).

Edwards et al. (2008) suggested the presence of a quasi-ballistic regime, defined as the region with ray deviation, $\beta$, of 110° – 125°, where the waveform exhibits ballistic shock features (interpreted as a typical N-wave appearance), but does not fall within the true ballistic regime (in Edwards et al. (2008) referred to as 90°±20°). Following Edwards et al. (2008) and Edwards (2010), any signals beyond 125° would therefore suggest spherical shock production at the source.

In this study we find an absence of a clear boundary in the waveform features that would phenomenologically distinguish ballistic shock from non-ballistic shock (identification based solely on the basis of N-wave appearance), or place it in the transition region. Spherical shocks, depending on the overall geometry and propagation effects, are also expected to produce N-wave signals and have ray deviations within the ballistic region (ray deviation angle 90°±25° as defined by Brown et al. (2007)), as the decay of any shock at large distances tends to exhibit N-wave behaviour due to the cumulative effects of non-linearity in the waveform, independent of the source characteristics (e.g. Whitham, 1972). This implies that the infrasound pressure-time signal alone may not be sufficient enough to clearly identify the shock source type (point vs. line source) and that some information about the geometry of the source has to be known. In our dataset, a number of events with ray deviations larger than 125° (beyond the quasi-ballistic as defined by Edwards et al. (2008)) still show Class I signals (i.e. N-wave signature), interpreted earlier as being typical of the ballistic regime. Conversely, it is not unusual to find a Class IV signal within the ballistic region (though this might still be due to a fragmentation event).

To investigate this question, along with the behaviour of the $S_h$ parameter as a function of shock type, we tried to determine if flares (produced by fragmentation) in a meteor lightcurve were a source of infrasound at the ground, as such fragmentation episodes are expected to produce quasi-spherical shock sources. The photometric lightcurve for each event was produced following standard photometric procedures (e.g. Brown et al., 2010). Since the visual magnitude was not calibrated to an absolute visual magnitude at this stage, it was normalized to -1 for each event and the differential lighturve used to identify flares. An example of a brightness vs. meteoroid height lightcurve for two representative events is shown in Figure S9.

If the height of the optical flare falls within the uncertainty bounds of the infrasonically estimated shock source height derived from raytracing, then this suggests that the shock type is spherical in nature. We assume that infrasound shock source heights showing no association with flares are more likely to be cylindrical line sources. In all but one case (event 20071004) it was possible to determine the most probable nature of shock production based on correlation with the lightcurve. This one ambiguous event, which was excluded from the lightcurve shock-source type analysis, shows continuous fragmentation in the video records which is indicative of a spherical shock generated by a rapidly moving point source; however, there is an absence of a clearly defined flare in the light curve. The distributions of shock type in the single arrival and multi arrival event categories, as well as the distribution of the $S_h$ parameter as a function of shock type, are shown in Figure 5.

The photometric lightcurves were also used to establish the height of the peak brightness; this is the region associated with the maximum energy deposition and ablation/mass loss along the trail (Zinn et al., 2004). The location of the peak brightness along the trail was used to define an $M_p$ parameter, which is defined (in analogy with the $S_h$ parameter) as:

$M_p = L_{Bh-Mp}/ L$ (3)

where $L_{Bh-Mp}$ is the path length along the trajectory from the begin point of the meteor to the point of peak luminosity, and L is the total path length of the entire visible meteor trajectory. This information was then used to examine the correlation between the $S_h$ parameter and the $M_p$ parameter (Figure S10). We note that if the shock source height occurs at the point of peak luminosity $S_h=M_p$.

Our examination of the correlation between optical meteor flares and infrasound production suggests:

(i) There are events with ray deviation angles, β, in the ballistic regime (90°±25°) which are most likely generated by a spherical shock (rapidly moving point source or terminal burst) based on their association with a visible flare or fragmentation - the fraction of these events differs between the single and multiple meteor infrasound arrival populations as shown in Figure 5;

(ii) From (i), it follows that the infrasound produced in the single arrival population is predominantly generated by a meteor's hypersonic passage through the atmosphere (cylindrical line source), while a significant portion of multi arrivals tend to be produced by fragmentation. This suggests that most meteor infrasound direct signals showing multiple arrivals are not due to atmospheric multi-pathing, but fragmentation based on our observed optical correlation to flares;

(iii) All events with large ray launch deviation angles ($\beta>117°$) were found to be associated with optical fragmentation points (flares) and hence likely originate from a spherical type of shock. It should be remarked that it may not be possible to determine this based on the infrasound signal alone;

(iv) A small number of events (12% of cases, or 7 arrivals out of 60) which were not associated with flares and hence we interpret to be generated as a cylindrical line source have their shock source height closely associated with the height of maximum brightness along the luminous trail (Figure S10). Taking into account the uncertainty in the $M_p$ parameter, this percentage may actually be closer to ~25%.

(v) The $S_h$ parameter for events showing flares is skewed towards larger values (i.e. closer to $S_h=1$), while events showing no flares (which are most likely ballistic shocks) are predominantly generated in the region around the middle of the trail (i.e. $S_h \sim 0.5$) (Figure 5).

As shown by Zinn et al. (2004), the region of maximum luminosity is also where energy deposition peaks and hence is the point where the blast radius is the largest. This would also be the location where the refractive effects (i.e. gradient in local sound speed produced by the shock) are the largest. Nevertheless, point source shocks are still present in the ballistic region, based on the occurrence of flares in the trail along the ballistic launch zone, indicating that it is not only necessary to investigate the type of signal, but also understand the geometry of each event to uniquely associate it with a meteor infrasound signal and a particular source shock type.

### 3.4 Multi Arrival Event Population

Meteor infrasound signals showing multiple arrivals display source height and shock type characteristics different than those events with single meteor infrasound arrivals as shown in Table S8. In the majority of cases, the main arrival (i.e. arrival with the maximum pressure amplitude) is associated with the lowermost part of the meteor trajectory (higher value of $S_h$). There are two distinct categories of multi arrival events. One category (9 out of 16 events) shows shock source heights which are nearly identical within uncertainty for at least two arrivals (Group M1), while the other group (7 out of 16 events) is associated with clearly different shock source heights for each infrasound arrival (Group M2). Three events in the M1 group were most likely generated by a cylindrical line source, while the remaining six we associate with spherical shocks based on a common association with optical flares.

The M2 group may be interpreted as either a ballistic shock and/or a spherical shock occurring at different locations along the meteor trail at different heights, while the M1 group may be associated with different ray paths from a single source, similar to the cause of secondary sonic booms (Rickley and Pierce, 1980). The former was observed in several well documented meteorite producing fireballs (e.g. Brown et al., 2011; Brown et al., 2003). Fragmentation processes may lead to complex shock configurations (Artemieva and Shuvalov, 2001; Borovička and Kalenda, 2003). Multiple fragments, if sufficiently large, may each generate individual shocks (Artemieva and Shuvalov, 2001) that could therefore produce distinct infrasonic arrivals possibly appearing to originate from the same altitude, though only the Grimsby event in our dataset is large enough for this mechanism to be plausible. Among our multi arrival events, only one was found to have a shock source height at an altitude greater than 90 km (20100530); all others occur at lower altitudes with a mean height of 72±15 km. This suggests that meteoroids capable of producing multiple infrasonic signals detectable at the ground are both typically slower (and hence more likely to be of asteroidal origin) and/or may be intrinsically more energetic than a typical single arrival producing fireball. This is further supported by the observation that the maximum dominant frequency for the multi arrival population does not exceed ~4 Hz.

# 4 Conclusions

In this first paper in a two part series, we present the experimental investigation, phenomenology and analysis of a data set of meteors simultaneously detected optically and infrasonically. The specific goals of this coordinated optical-infrasound meteor study were to: (i) use astrometric optical measurements to positively identify infrasound from meteors; (ii) establish and constrain the point (and its uncertainty) along the meteor trail where the infrasound signal emanates; (iii) examine the role of atmospheric variability due to winds on the infrasound solution; (iv) determine the type of shock production mechanism for meteor generated infrasound; and (v) classify meteor infrasound and correlate meteor infrasound classes with structures seen in meteor infrasound signals to establish the foundation for future work to understand the underlying physical mechanisms and possibly relate to sonic boom theory. Our main conclusions related to these initial goals are:

(i) Seventy one optically recorded meteor events were positively identified with infrasound signals using the observed travel time and arrival azimuths in comparison to the expected values from the measured trajectories. Among these data, some 16 events also produced more than one distinct infrasound arrival. We have found that there are two entry velocity meteor populations which produced infrasound detectable at the ground, slow (< 40 km/s) and fast (> 40 km/s), which we associate with predominantly asteroidal and cometary populations, respectively. These exhibit different astrometric, source height, and signal characteristics. The findings in this study also suggest that infrasound from high altitude meteors may be more common than previously thought. The slow and fast velocity meteoroid populations also exhibit an apparent upper limit on the dominant infrasound signal period and frequency. The slow velocity population appears to be capped at a dominant signal frequency of ~14 Hz, while the fast velocity population is limited to the upper dominant signal frequency bound of <4 Hz. There is a strong inverse relationship between the shock source height and the dominant signal frequency.

(ii) By comparing travel time and arrival azimuths with raytracing model values and incorporating the perturbations in the wind field, we determined the altitude and uncertainty of the infrasound source height production and its location along the meteor trail. We find that meteors preferentially produce infrasound toward the end of their trails with a smaller number showing a preference for mid-trail production.

(iii) We show that the atmospheric variability may play a nontrivial role even for short range (< 300 km) infrasound propagation from meteors, with our model variations due to gravity wave induced winds producing spreads in the travel time between 2 s and 7 s, and spreads in the back azimuth from 1° – 4° for our meteor dataset which averaged ~150 km in range and 70-80 km in source altitude. Some of the differences observed between our raytracing predictions and observations may also be due to scattering, diffraction and range dependent changes in the atmosphere which we do not explicitly examine.

(iv) We linked the type of shock production (cylindrical and spherical) at the source based on whether or not the infrasound source height corresponds with an optical flare. We find that large deviation angles ($\beta$>117°) are generally associated with spherical sources (point source or a moving point source). Approximately one quarter of single arrival meteor infrasound events are associated with fragmentation episodes (spherical shocks); while almost half of all multi arrival events are correlated with fragmentation events. Notably, the multi arrival population shows a strong source height skewness to the end of trails; this reflects the larger masses involved as well as the predominance of asteroidal meteoroids in the multi-arrival category. We find that the ray deviation angle cannot be used as a sole discriminant of the type of shock at the source, as both spherical and hypersonic line sources, depending on the geometry and orientation with respect to the observer, may produce signals which apparently have ray deviation angles confined to the ballistic regime.

(v) We have developed a new signal taxonomic scheme based on the appearance and qualitative characteristics of the waveforms. This taxonomic scheme may be extended to infrasound signals from other explosive sources located at a relatively short distance from the source (within ~300 km, to remain consistent with our dataset). For example, N-waves are common features of all shocks at some distance from their sources (not only in meteors). Further studies, theoretical and experimental, are required to better explain certain shock features, such as double and triple N-waves (Table 1), and link them to the source and propagation effects. We also found an association between the signal peak-to-peak amplitude and the signal class, which may relate to the sonic boom equal area rule. The signal amplitude tends to decrease as a function of signal class; this is clearly observable when all our data is plotted. All four classes of signals can be found in any source (spherical or line source) category.

We find that about 1% of all our optically detected meteors have associated infrasound. For a typical infrasound station such as ELFO we estimate that regional meteor infrasound events should occur on the order of once per week and dominate in numbers over infrasound associated with more energetic (but rarer) bolides detectable at large (>500 km) propagation distances. While a significant fraction of our meteors generating infrasound (~1/4 of single arrivals) are produced by fragmentation events, we find no instances where acoustic radiation is detectable more than 60° beyond the ballistic regime at our meteoroid sizes (grams - tens of kilograms). Indeed, the average deviation angles among our population are within 10°-15° of purely ballistic. This emphasizes the strong anisotropy in acoustic radiation for meteors which are dominated by cylindrical line source geometry, even in the presence of fragmentation.

## Acknowledgements

Funding in support for part of this project was provided by the CTBTO Young Scientist Award funded through the European Union Council Decision 2010/461/CFSP IV. EAS thanks Dr. W. K. Hocking for discussions about gravity waves, Zbyszek Krzeminski for his help with astrometric reductions and Jason Gill for his help with setting up a parallel processing machine to run raytracing. Funding in support of this work from NASA Co-operative agreement NNX11AB76A, from the Natural Sciences and Engineering Research Council and the Canada Research Chairs is gratefully acknowledged.

## Tables and Table Captions:

Table 1: Proposed meteor infrasound signal classification taxonomy.

| Class I | **Single N-wave**<br>- Clean, no noticeable features, a single most prominent and complete cycle with the P2P amplitude significantly above any other cycles within the signal portion of the waveform (any other peak which is less than one ½ of the P2P amplitude would not be counted) |
|---|---|
| **Class II** | **Double N-wave**<br>- Clean, no noticeable features, two prominent and complete cycles similar in size, with the main (dominant) cycle significantly above any others within the signal segment and with the second peak at least 1/2 the P2P amplitude (any other peak which is less than one ½ of the P2P amplitude would not be counted) |
| **Class III** | **Triple N-wave**<br>- Clean, no noticeable features, similar in size, classification same as above |
| **Class IV** | **Diffuse**<br>- 4+ complete cycles all within 1/2 the P2P amplitude of the dominant cycle<br>- If the signal is barely above the noise (low SNR), then it is classified as the class IV |
| **Subclasses** | **a** - with a slowly decaying wavetrain and/or trailing reverberations (noticeable for additional 1+ cycles)<br>**b** – subtle to noticeable additional features within the main wave/cycle (humps, ripples, spikes, dips, widening (but not a typical U-wave), etc) confined to the main cycle<br>**c** – complex. Presence of 'sub-cycles' and other complex features in the main cycle and beyond. 'c' can also be considered a combination of 'a' + 'b'. Presence of leading reverberations.<br>**d** – mixed U- and N- waves (instead of a typical sharp turn as seen in N-waves, a U-wave displays significant widening, after which it gets its destination) |
| **Signal phase**<br>(first motion above RMS) | (+) Positive phase<br>(-) Negative phase |
|  | Positive phase (+)<br>Negative phase (-)<br>Indistinguishable (indeterminate) phase (*) |

# Figures and Figure Captions:

**Figure 1:** An example composite plot showing the travel time (upper left), back azimuth (upper right) and arrival elevation angle (lower left) for an event on May 11, 2007 occurring at 07:41 UT. The lower right plot is the ray launch deviation angle as a function of height along the meteor trail. The blue points represent modelled arrivals for 1400 gravity wave realizations for this event. The yellow model points in all plots are the simulation means along the meteor trail for each height (averaging along the x-axis). The vertical solid black line corresponds to the observed quantity with its uncertainty (dotted line), except in the case of ray deviation angle, which is simply a reference to the expected ballistic angle (90°). This particular event is an example which shows a unique solution as defined by our height selection algorithm. The composite plot also shows that the back azimuth determined model source height differs from the model source height determined by the travel time residuals.

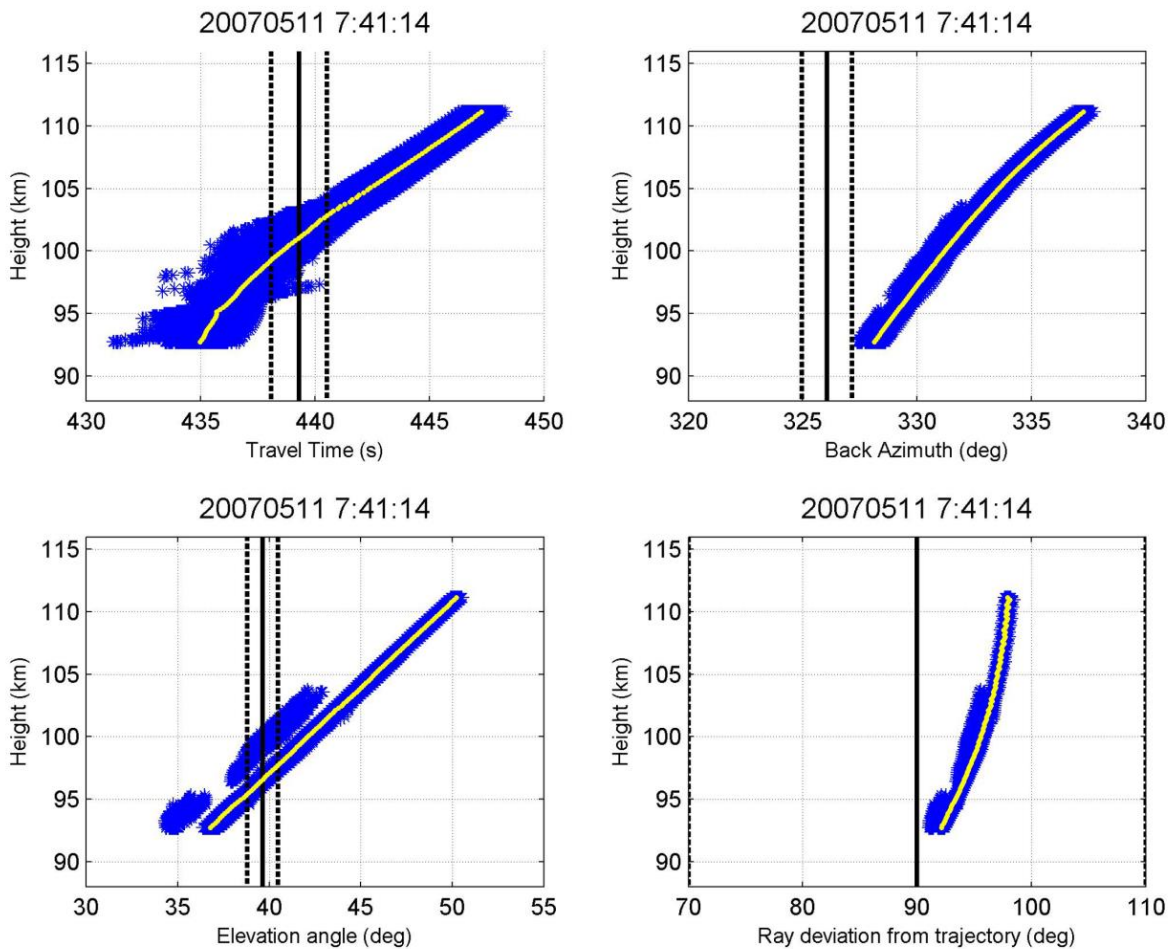

**Figure 2:** Examples of each taxonomic class.

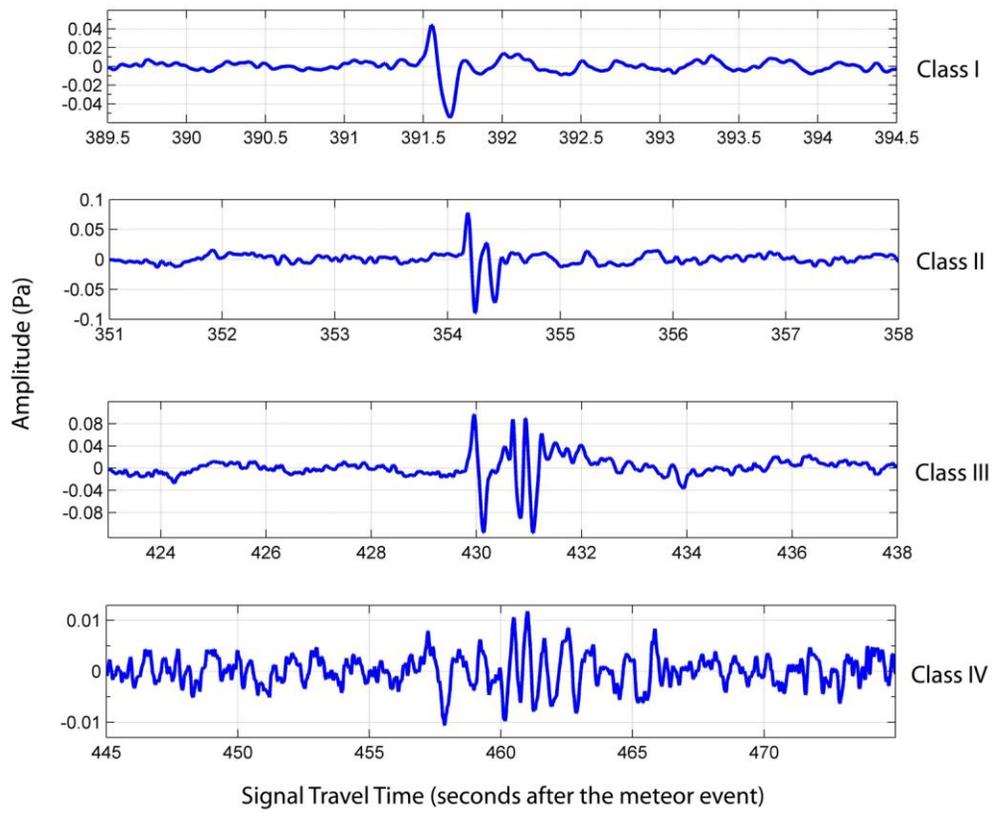

**Figure 3:** The relationship between the peak-to-peak amplitude of the meteor infrasound signal as a function of signal class. There is an evident fall-off in amplitude as the signal class increases.

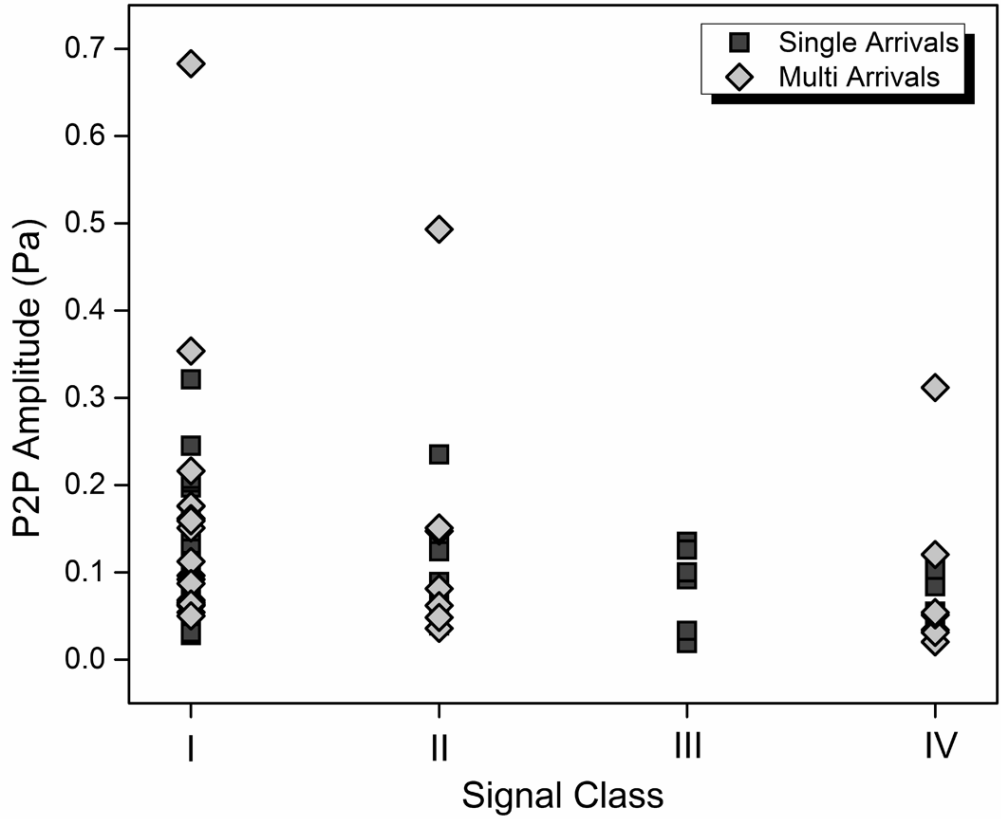

**Figure 4:** Meteor entry velocity populations. There appears to be a dip between 36 – 43 km/s, a feature seen in most raw meteor speed distributions (e.g. Brown et al., 2004). The slow meteor population has their velocities < 40 km/s, while the fast population is defined as having the entry velocity > 40 km/s.

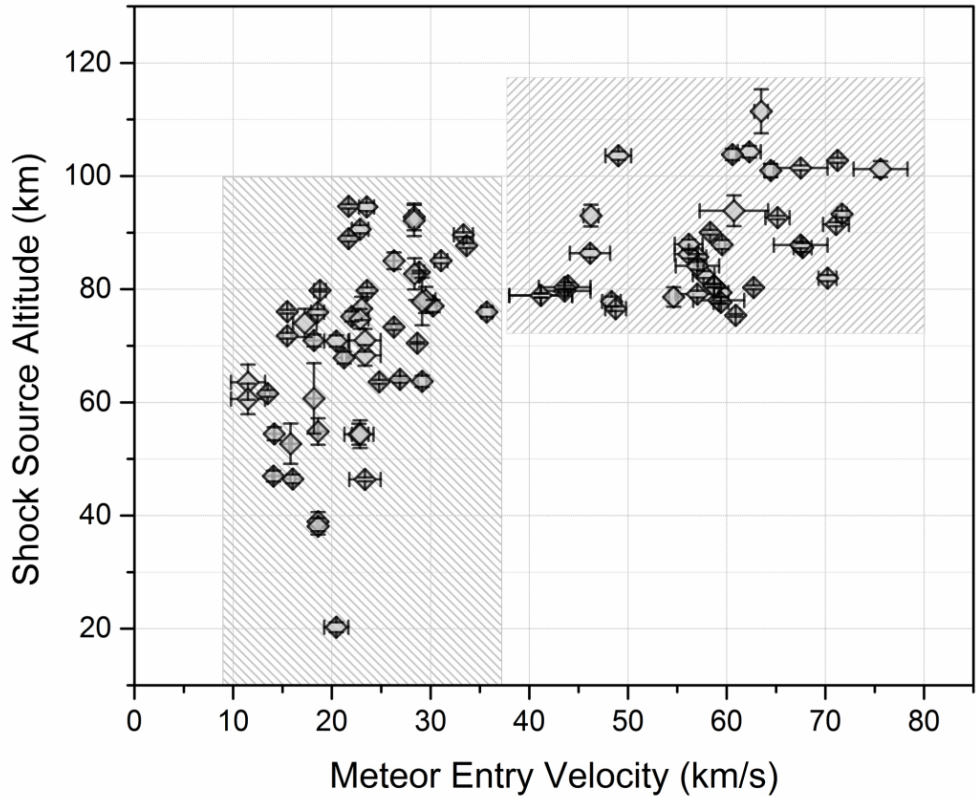

**Figure 5:** Pie charts showing the distribution of spherical (S) shocks for events with ray deviations β ≤117°, β>117°, and cylindrical (C) shocks, where identification of a probable spherical shock source is defined as association with an optical flare and vice versa. Only one arrival in our entire data set is of ambiguous nature (A). Nearly half of the multi arrivals are found to be associated with probable spherical shocks, while a much greater proportion of the single arrivals are generated by a cylindrical shock. The histograms represent the distribution of the $S_h$ parameter, separated by the shock source type.

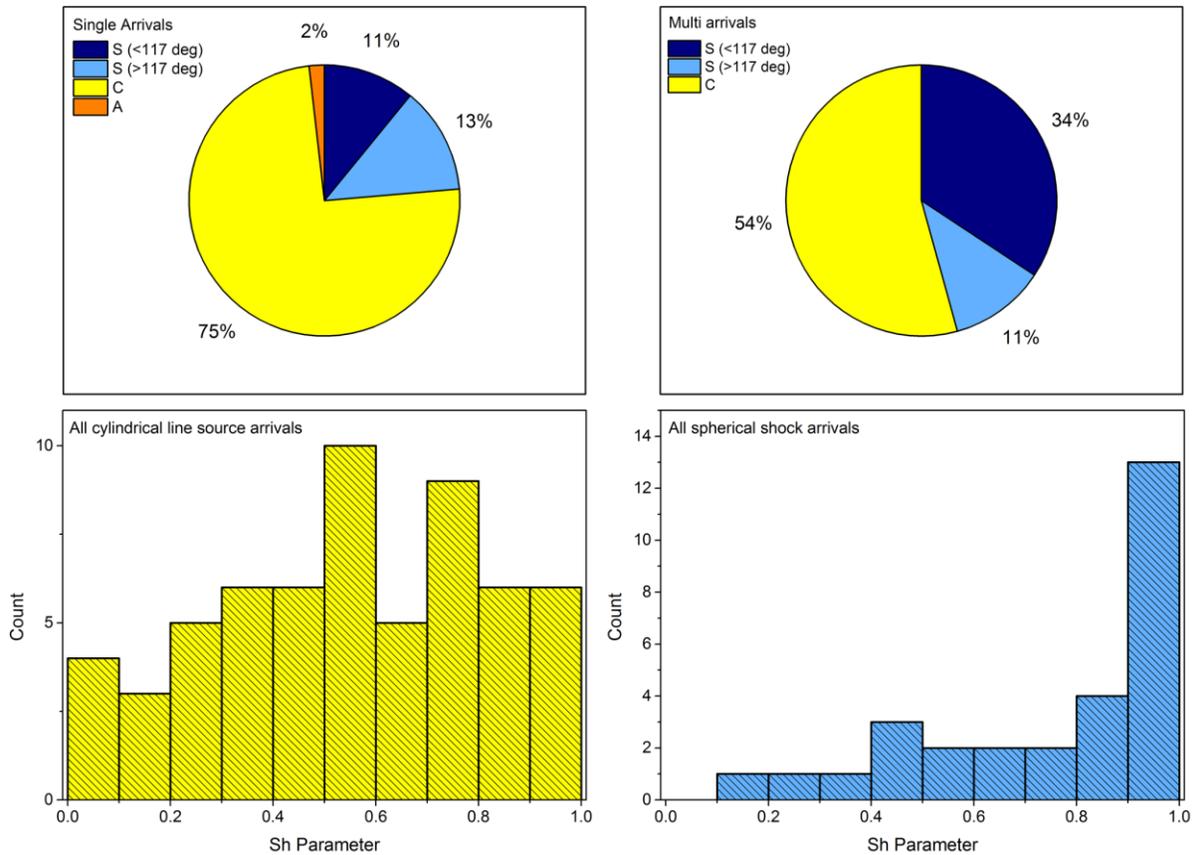

# Appendix A: Supplemental Material

*Silber, E. A., Brown, P. G. (2014) Optical Observations of Meteors Generating Infrasound – I: Acoustic Signal Identification and Phenomenology, JASTP, XX*

# Table of Contents

# S1. Supplemental Figures

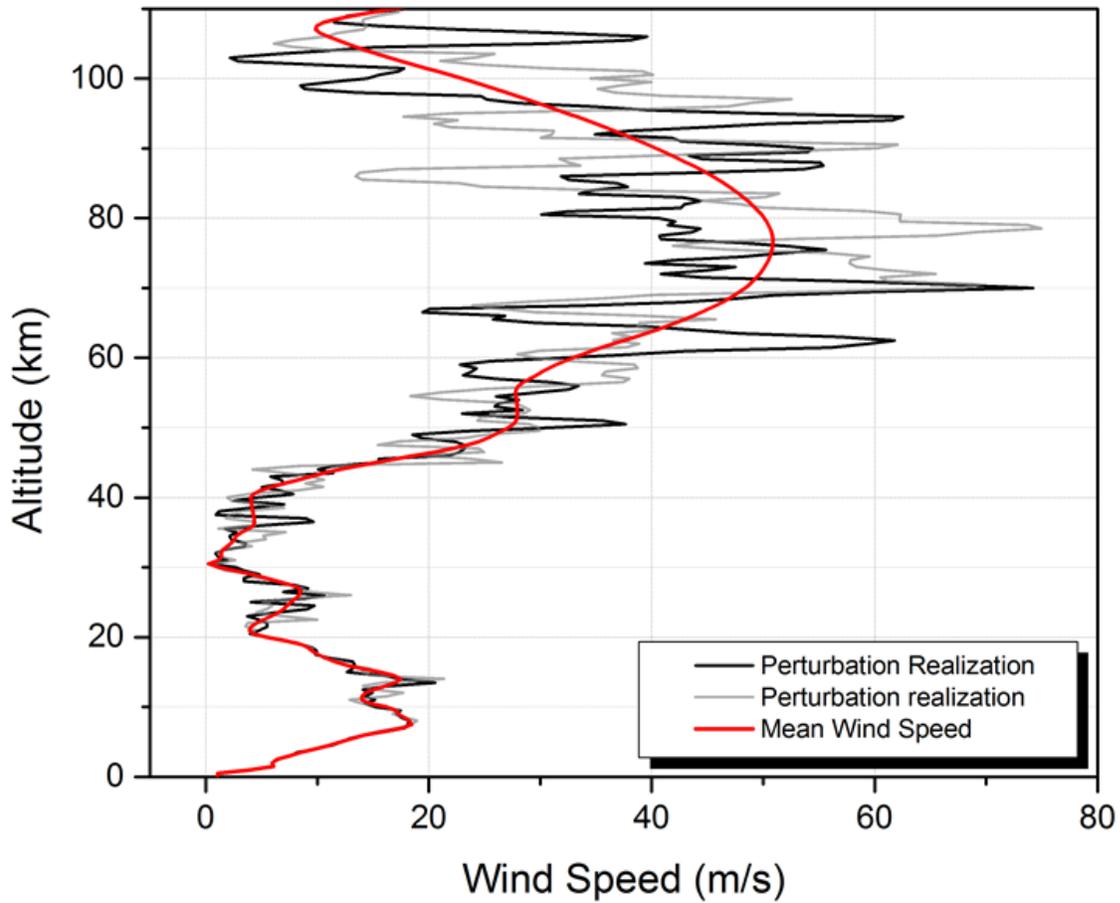

**Figure S1:** An example of a gravity wave perturbed wind speed as a function of altitude. The black and gray lines represent a sample of two separate realizations applied to the to the UKMO-HWM07 mean wind model (red line). For ray-trace modelling, 1400 realizations were applied to the mean atmospheric profile for each event and the resulting spread in travel times and backazimuths recorded.

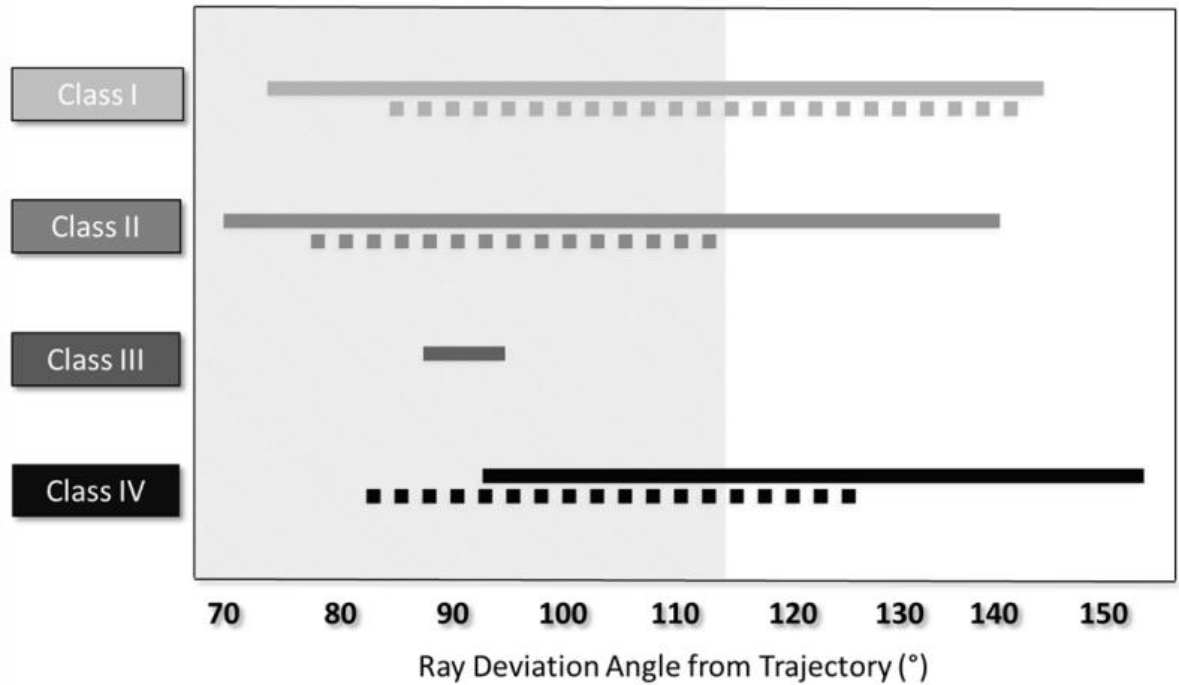

**Figure S2:** Ray launch deviation angles as a function of the signal Class type. Solid line: single arrival event category; Dotted line: multi arrival event category (all arrivals). The shaded region represents the ballistic region (90°±25°) as originally defined on theoretical ground by Brown et al (2007). Although the multi infrasound arrival sample is smaller (16 individual events showing a total of 35 arrivals) than the single infrasound arrival group, it is notable that there are no multi infrasound arrivals in Class III. This, however, may simply be due to small number statistics.

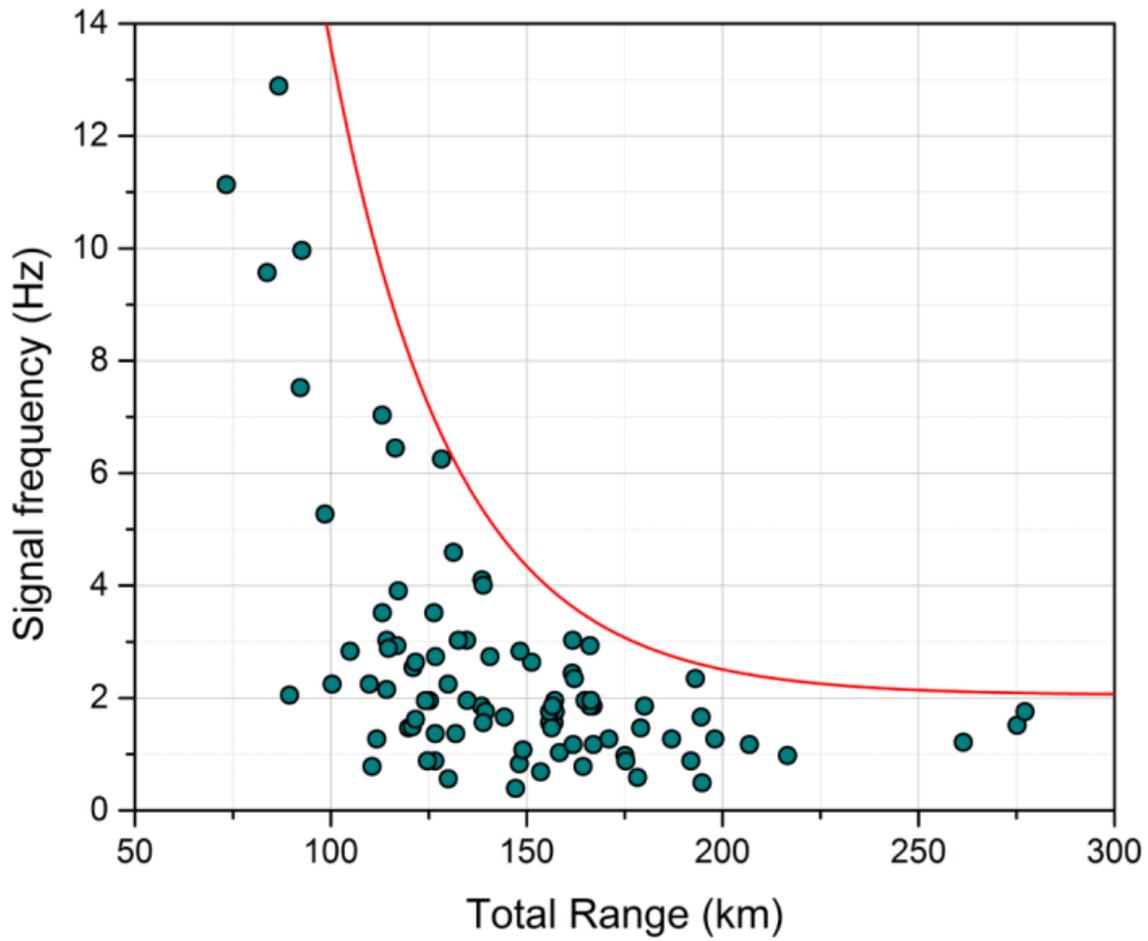

**Figure S3:** The observed fall-off in the dominant signal frequency of bright regional fireballs as a function of range.
The exponential decay curve is given by equation: *Frequency* = 2.05 + 290 exp(-R/31).

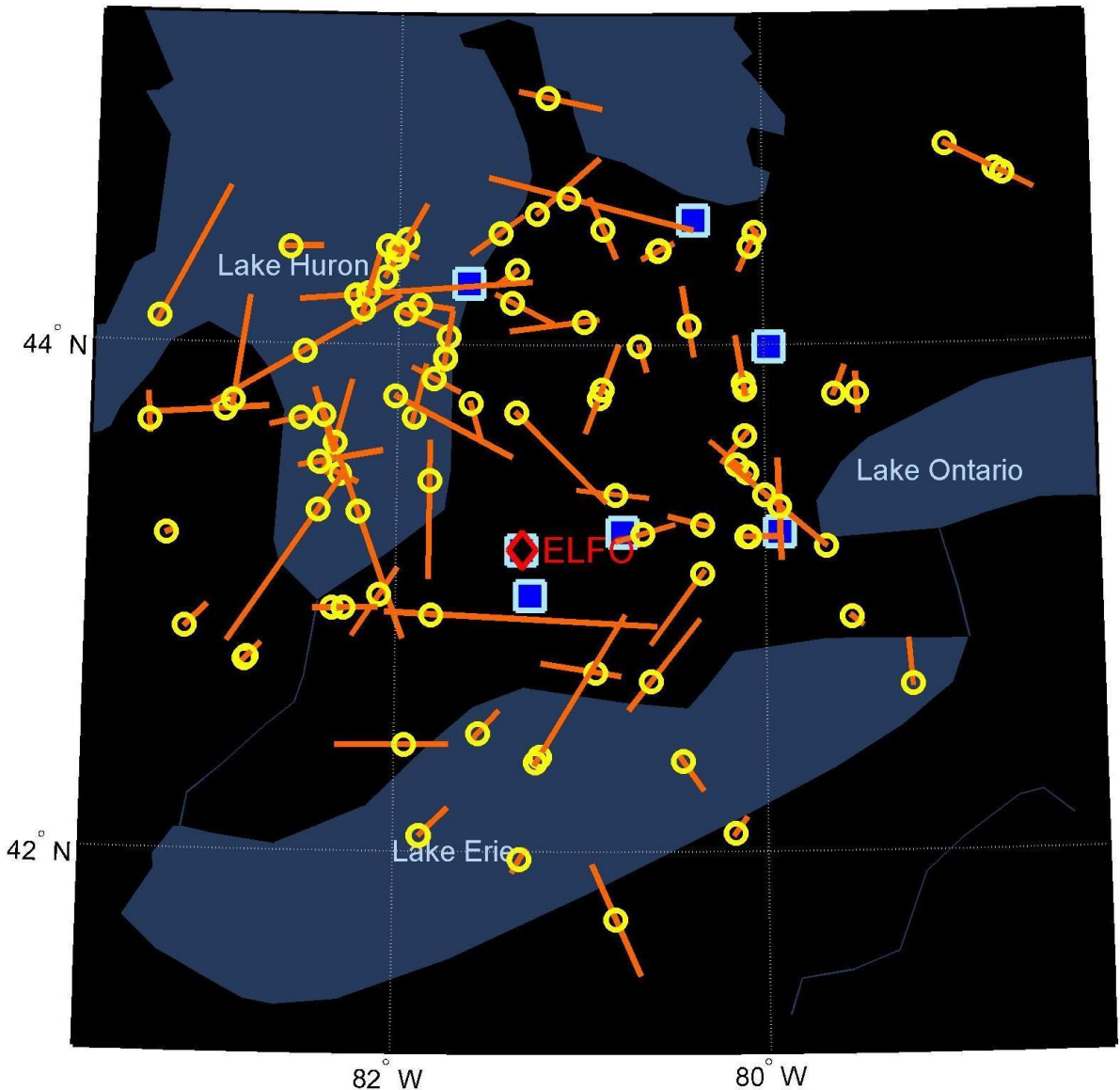

**Figure S4:** A map of Southwestern Ontario and surrounding lakes showing the spatial orientation of ground projected meteor trajectories for all events (orange lines), including the point along the trail where infrasound was produced and subsequently detected at ELFO (yellow circles). The location of the array is shown with the red/black diamond. The cameras in the All-Sky camera network, in operation during the study period (2006-2011), are shown with blue squares. Note that one of the cameras is installed at the same location as the infrasound array.

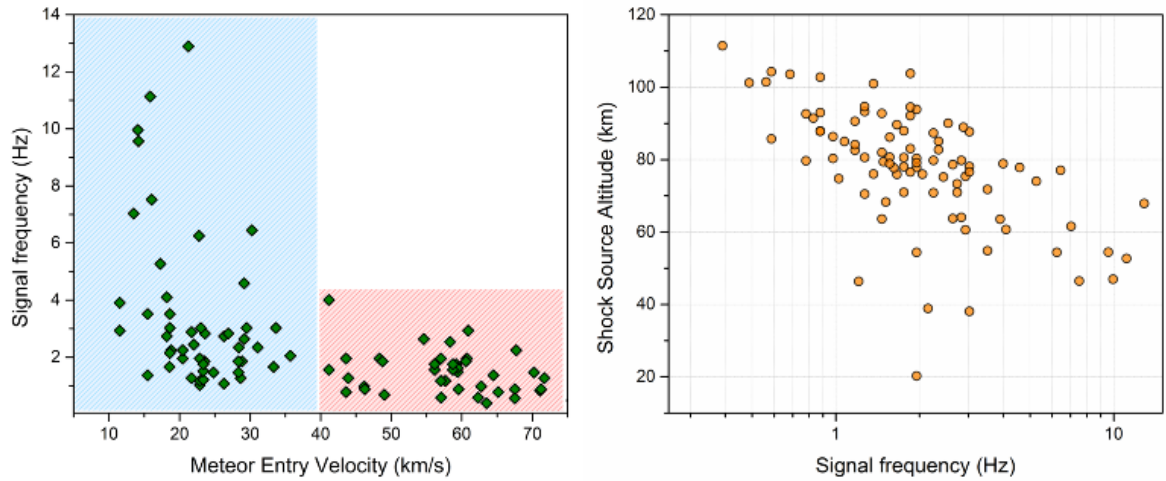

**Figure S5:** Left: The dominant signal frequency as a function of meteoroid entry velocity for meteor infrasound events in this study. The blue region denotes the slow velocity population (< 40 km/s), while the red region is the high velocity meteoroid population (> 40 km) which correspond roughly to asteroidal and cometary meteoroids, respectively. Right: Shock source altitude vs. dominant signal frequency.

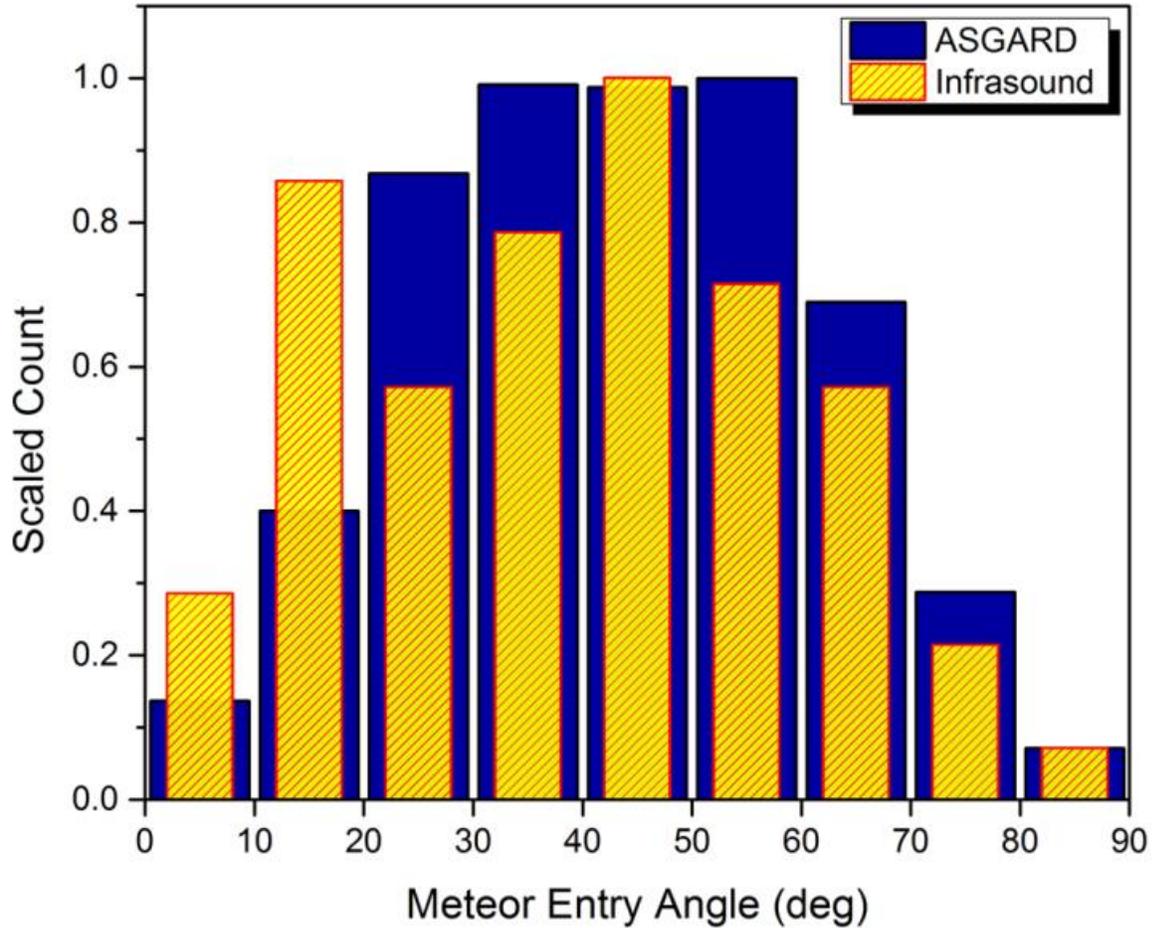

**Figure S6:** The scaled distributions of meteoroid entry angles (from the vertical) for all optically detected events (ASGARD data) and simultaneously detected events in this study. In ASGARD data, the maximum is between 30° – 60°, while the simultaneously detected meteors in this study show a peak at 45° and another one at 15°. The meteoroids with small entry angles (as measured from the vertical) are more likely to produce infrasound via spherical shock than cylindrical line source. In our dataset, the limiting meteoroid radiant zenith angle for spherical shocks is 64° (average 38°) and for cylindrical line source it is 88° (average 49°).

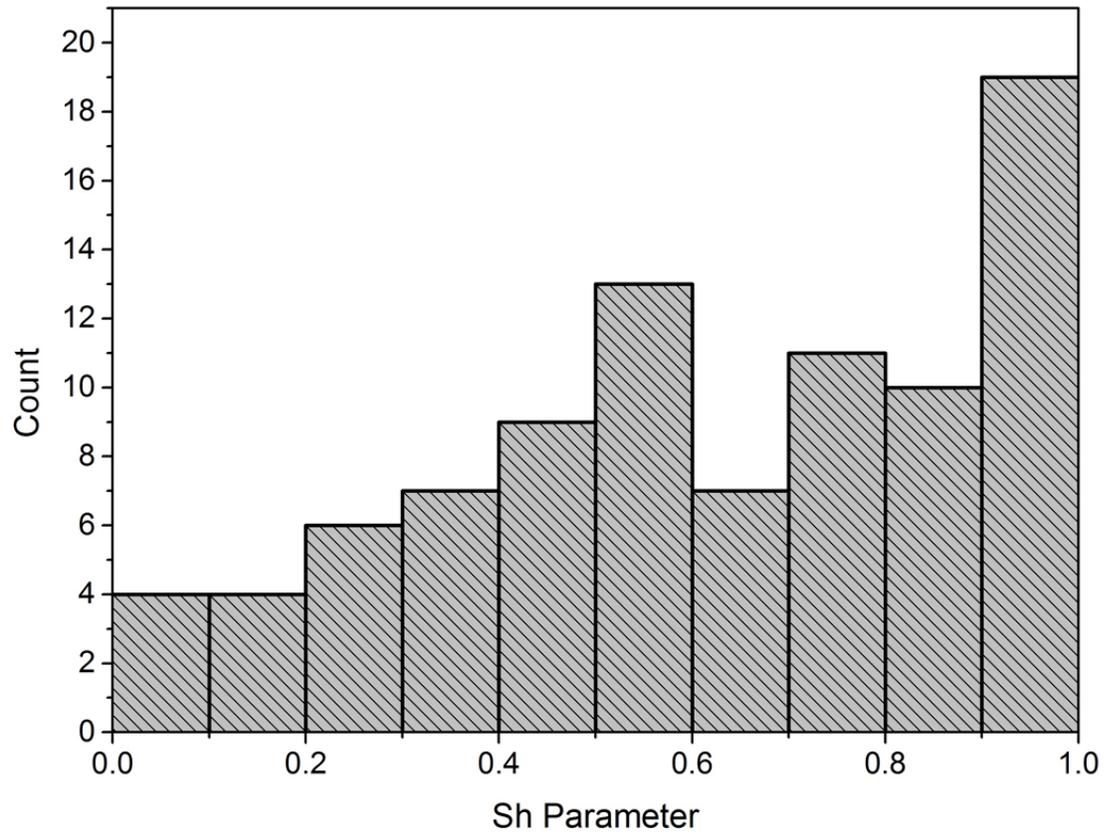

**Figure S7:** The distribution of $S_h$ parameter shown for all arrivals.

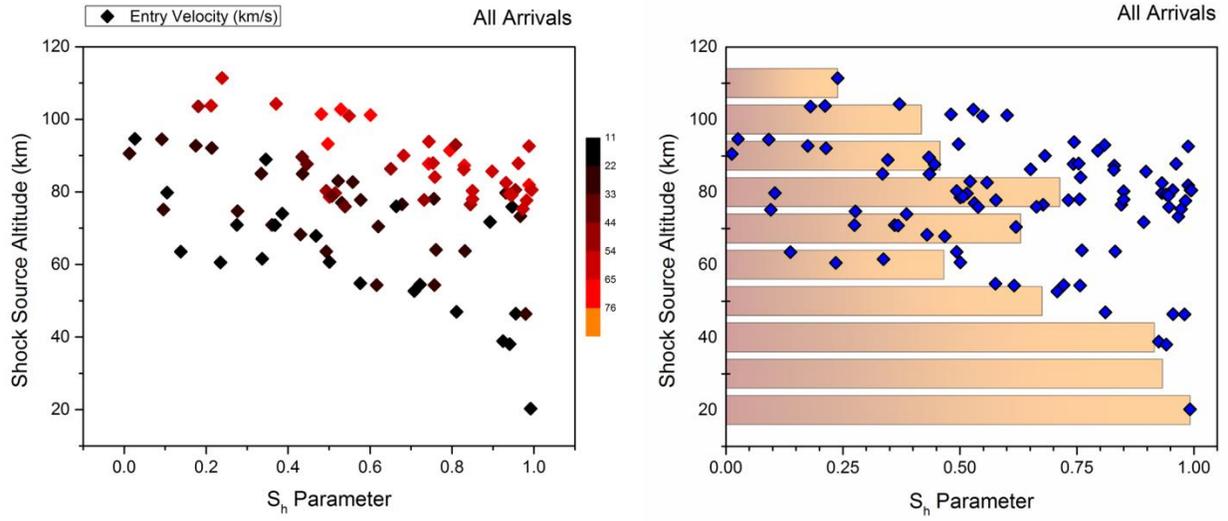

**Figure S8:** Left: Altitude vs. $S_h$ parameter, where the colour coded diamonds represent meteor entry velocity (high intensity red denotes highest velocity events, gradually shifting towards dark shades and to black, denoting slowest velocities). Right: The mean value of $S_h$ parameter as a function of height (orange bars) in 10 km increments (bins) overlaid with $S_h$ parameter as a function of height (blue diamonds).

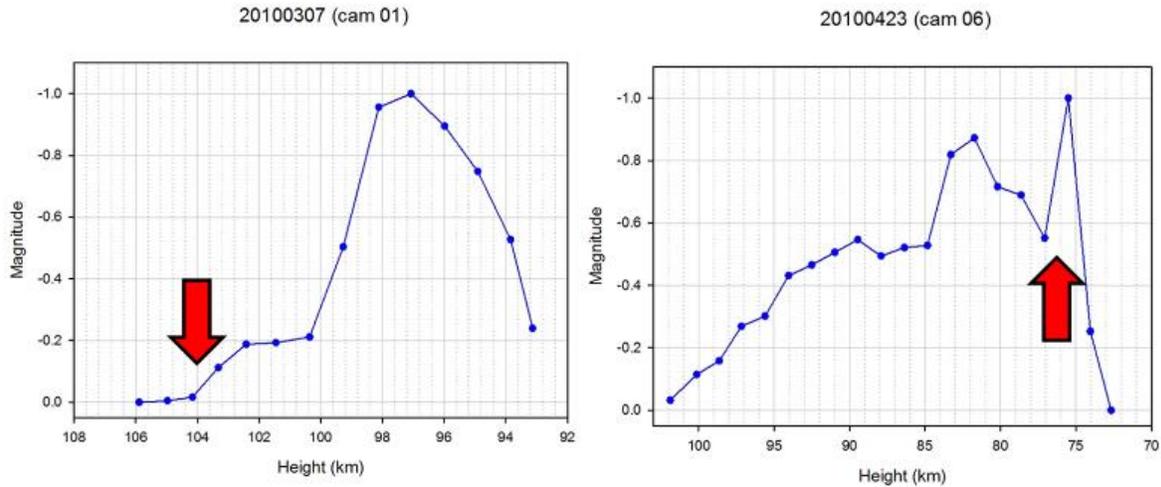

**Figure S9:** Examples of relative photometric lightcurves. The photometric lightcurve for each event was produced following standard photometric procedures for the best camera site (longest optical trail, most optimal sky conditions) and corrected for the lens roll off, air mass, range and saturation. Since the visual magnitude was not calibrated to an absolute visual magnitude at this stage, it was normalized to -1 for each event and the differential lightcurve used to identify flares. Left: The shock source height was determined to have occurred at 104 km. Thus, this event most likely produced infrasound via hypersonic shock (cylindrical line source) as the shock source height does not correlate with any local brightening (flare) suggestive of a fragmentation episode. Right: The onset of the flare (fragmentation) coincides with the best estimate of the source height, suggesting that the signal was most likely produced by a spherical shock. On both panels the red arrow shows the most probable shock source height. For the first event, the ray deviation angle is 100° and for the event on the right, it is 136° degrees.

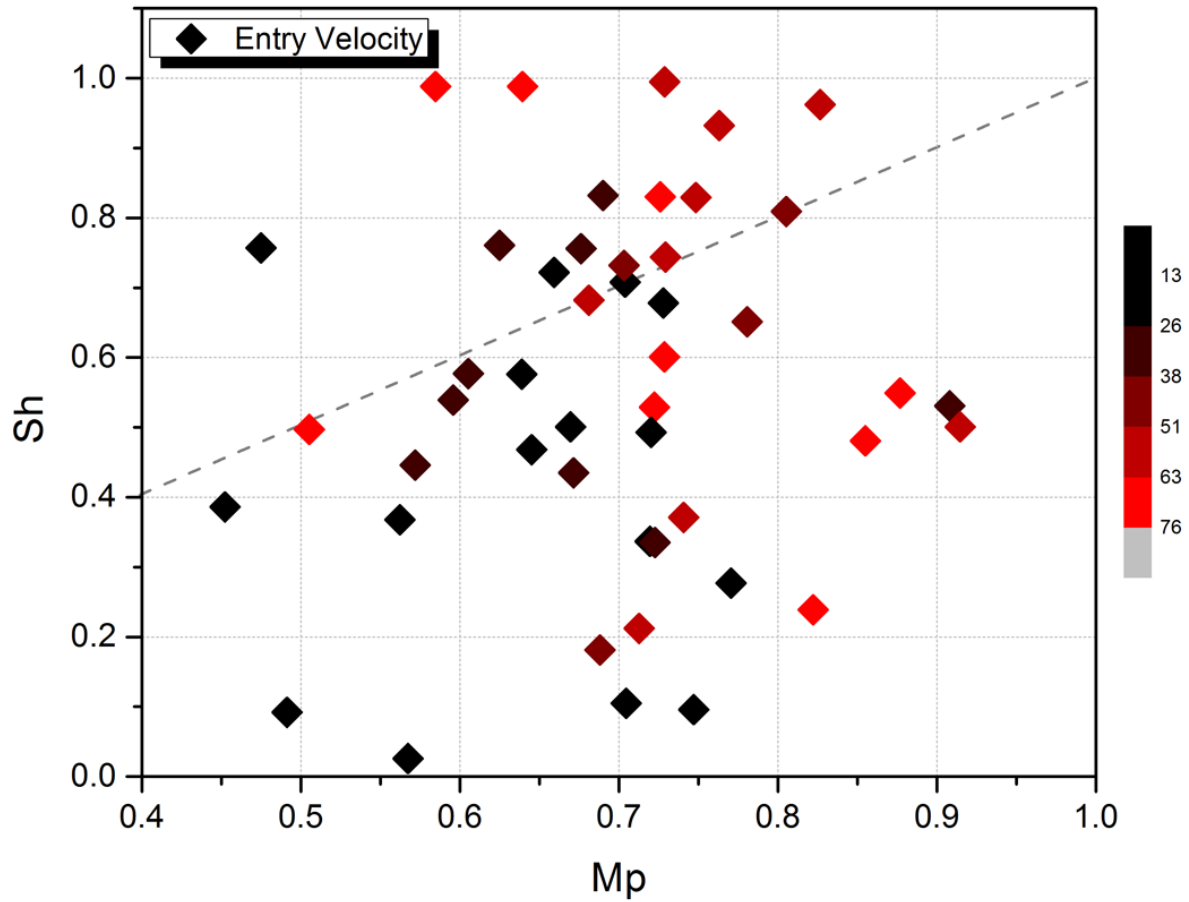

**Figure S10:** The $S_h$ parameter vs $M_p$ for 50 arrivals (single arrival and multi arrival populations combined) generated by a cylindrical line source (i.e. not showing a shock source height which can be obviously correlated with a local maximum in the lightcurve). Due to complex light curve features (e.g. flares), it was not possible to determine the height of the peak brightness with certainty for the remaining 10 arrivals. Therefore, those arrivals are excluded from the plot. The dotted line is the 1:1 line.

# S2. Supplemental Tables

**Table S1:** Infrasound signal measurements for meteors optically observed and which produced a single infrasonic arrival. Event time denotes the onset of luminous trail as seen by the All-Sky cameras. The remaining columns represent the signal parameters observed at the infrasound array.

| Date | Event Time hh | mm | ss | Observed Travel Time s | Observed Back Azimuth ° | ± ° | Trace Velocity km/s | Dominant Freq Hz | Dominant Period s | ± s | Max Amp Pa | ± Pa | P2P Amp Pa | ± Pa | Amp SNR | Integrated Energy SNR |
|---|---|---|---|---|---|---|---|---|---|---|---|---|---|---|---|---|
| 20060419 | 4 | 21 | 28 | 345 | 319.9 | 2.8 | 0.488 | 2.2 | 0.53 | 0.02 | 0.05 | 0.01 | 0.09 | 0.03 | 29.5 | 3.5 |
| 20060419 | 7 | 5 | 56 | 280 | 144.5 | 0 | 0.42 | 9.6 | 0.12 | 0.07 | 0.10 | 0.04 | 0.13 | 0.07 | 29.7 | 4.3 |
| 20060813 | 9 | 42 | 26 | 554 | 147.0 | 0.3 | 0.358 | 1.0 | 0.98 | 0.06 | 0.03 | 0.01 | 0.05 | 0.02 | 18.3 | 8.1 |
| 20061101 | 6 | 46 | 12 | 527 | 275.3 | 1.4 | 0.421 | 0.6 | 1.55 | 0.44 | 0.05 | 0.01 | 0.08 | 0.02 | 13.4 | 4.9 |
| 20061104 | 3 | 29 | 29 | 354 | 293.1 | 0.1 | 0.439 | 6.4 | 0.16 | 0.00 | 0.06 | 0.02 | 0.09 | 0.03 | 29.7 | 5.2 |
| 20061121 | 10 | 45 | 46 | 526 | 133.5 | 1.8 | 0.395 | 0.8 | 1.18 | 0.11 | 0.03 | 0.01 | 0.06 | 0.01 | 16.7 | 4.3 |
| 20070125 | 10 | 2 | 5 | 398 | 289.4 | 4.1 | 0.601 | 0.9 | 1.35 | 0.03 | 0.06 | 0.01 | 0.08 | 0.02 | 7.6 | 5.6 |
| 20070421 | 9 | 21 | 1 | 636 | 36.7 | 0.6 | 0.379 | 1.7 | 0.64 | 0.04 | 0.02 | 0.00 | 0.03 | 0.01 | 18.8 | 6.8 |
| 20070511 | 7 | 41 | 14 | 439 | 326.1 | 1.1 | 0.442 | 1.4 | 0.80 | 0.04 | 0.01 | 0.00 | 0.02 | 0.01 | 14.9 | 6.9 |
| 20070725 | 4 | 42 | 44 | 392 | 25.6 | 2.4 | 0.425 | 2.7 | 0.38 | 0.06 | 0.03 | 0.02 | 0.05 | 0.04 | 31.7 | 6.0 |
| 20070727 | 4 | 51 | 58 | 467 | 29.9 | 0.4 | 0.41 | 1.1 | 0.94 | 0.08 | 0.06 | 0.01 | 0.09 | 0.03 | 20.0 | 5.4 |
| 20070917 | 7 | 55 | 43 | 572 | 259.3 | 0.3 | 0.368 | 0.9 | 1.07 | 0.12 | 0.06 | 0.01 | 0.09 | 0.01 | 28.7 | 11.8 |
| 20071004 | 4 | 55 | 9 | 557 | 19.5 | 0.2 | 0.352 | 1.9 | 0.52 | 0.03 | 0.13 | 0.01 | 0.15 | 0.02 | 41.8 | 12.0 |
| 20071004 | 5 | 19 | 27 | 307 | 98.6 | 0.7 | 0.345 | 7.5 | 0.15 | 0.04 | 0.05 | 0.05 | 0.08 | 0.10 | 43.5 | 6.3 |
| 20071021 | 10 | 26 | 25 | 700 | 38.9 | 3.1 | 0.347 | 0.5 | 2.21 | 0.03 | 0.11 | 0.00 | 0.20 | 0.01 | 51.0 | 72.4 |
| 20071130 | 10 | 28 | 49 | 645 | 64.0 | 0.4 | 0.332 | 1.3 | 0.81 | 0.18 | 0.10 | 0.01 | 0.12 | 0.02 | 12.5 | 6.2 |
| 20071215 | 11 | 18 | 13 | 318 | 66.3 | 4.4 | 0.397 | 2.1 | 0.42 | 0.06 | 0.11 | 0.05 | 0.17 | 0.10 | 39.0 | 10.6 |
| 20080325 | 0 | 42 | 3 | 341 | 305.0 | 4 | 0.379 | 7.0 | 0.14 | 0.01 | 0.10 | 0.06 | 0.14 | 0.12 | 47.2 | 9.3 |
| 20080520 | 3 | 29 | 47 | 288 | 80.9 | 0.5 | 0.439 | 10.0 | 0.11 | 0.00 | 0.05 | 0.05 | 0.09 | 0.10 | 47.2 | 7.9 |
| 20080602 | 5 | 44 | 44 | 505 | 221.6 | 1.6 | 0.378 | 0.4 | 2.56 | 0.36 | 0.07 | 0.01 | 0.11 | 0.01 | 24.2 | 15.2 |
| 20080801 | 4 | 22 | 20 | 348 | 329.9 | 6.4 | 0.494 | 2.8 | 0.33 | 0.01 | 0.07 | 0.02 | 0.10 | 0.03 | 33.7 | 13.2 |
| 20080801 | 8 | 19 | 30 | 375 | 350.0 | 2.4 | 0.628 | 0.8 | 1.40 | 0.09 | 0.09 | 0.01 | 0.14 | 0.03 | 19.4 | 12.3 |
| 20080804 | 4 | 17 | 14 | 393 | 148.4 | 0.3 | 0.439 | 2.5 | 0.44 | 0.02 | 0.01 | 0.01 | 0.03 | 0.01 | 10.7 | 3.8 |
| 20080812 | 6 | 0 | 34 | 424 | 197.4 | 0.5 | 0.451 | 2.0 | 0.53 | 0.05 | 0.02 | 0.00 | 0.03 | 0.01 | 18.5 | 10.1 |
| 20080908 | 4 | 3 | 12 | 592 | 292.0 | 1.5 | 0.352 | 0.6 | 1.63 | 0.10 | 0.04 | 0.01 | 0.07 | 0.01 | 30.5 | 16.2 |
| 20081005 | 1 | 50 | 12 | 406 | 337.8 | 0.3 | 0.372 | 6.3 | 0.19 | 0.01 | 0.06 | 0.03 | 0.10 | 0.06 | 42.1 | 7.2 |
| 20081018 | 6 | 53 | 39 | 292 | 83.7 | 0.2 | 0.367 | 12.9 | 0.07 | 0.00 | 0.07 | 0.13 | 0.12 | 0.26 | 50.4 | 16.4 |
| 20081028 | 3 | 17 | 35 | 240 | 306.5 | 0.7 | 0.379 | 11.1 | 0.10 | 0.01 | 0.06 | 0.06 | 0.10 | 0.12 | 33.1 | 7.1 |
| 20081102 | 6 | 13 | 26 | 576 | 292.2 | 9.5 | 0.372 | 2.3 | 0.44 | 0.03 | 0.08 | 0.04 | 0.12 | 0.08 | 32.6 | 8.1 |
| 20081107 | 7 | 34 | 16 | 378 | 330.4 | 0.6 | 0.494 | 1.5 | 0.59 | 0.05 | 0.06 | 0.01 | 0.09 | 0.01 | 23.9 | 8.5 |
| 20090126 | 7 | 16 | 24 | 320 | 231.4 | 0 | 0.831 | 2.2 | 0.34 | 0.09 | 0.13 | 0.04 | 0.21 | 0.08 | 32.2 | 16.7 |
| 20090523 | 7 | 7 | 25 | 428 | 62.2 | 0.1 | 0.41 | 3.0 | 0.35 | 0.02 | 0.20 | 0.04 | 0.32 | 0.07 | 64.3 | 18.1 |
| 20090530 | 6 | 35 | 20 | 312 | 254.3 | 0.6 | 0.451 | 5.3 | 0.19 | 0.09 | 0.07 | 0.04 | 0.11 | 0.07 | 24.4 | 3.3 |
| 20090709 | 5 | 24 | 23 | 460 | 357.1 | 3.3 | 0.349 | 1.7 | 0.58 | 0.08 | 0.10 | 0.01 | 0.15 | 0.02 | 15.0 | 4.5 |
| 20090813 | 6 | 47 | 53 | 540 | 356.9 | 0.2 | 0.366 | 2.9 | 0.31 | 0.03 | 0.02 | 0.00 | 0.03 | 0.00 | 15.0 | 1.8 |
| 20090906 | 1 | 27 | 55 | 460 | 269.0 | 1.8 | 0.409 | 1.9 | 0.51 | 0.10 | 0.02 | 0.01 | 0.03 | 0.01 | 6.8 | 2.0 |
| 20090917 | 1 | 20 | 38 | 429 | 358.0 | 2.1 | 0.401 | 3.0 | 0.37 | 0.12 | 0.08 | 0.02 | 0.13 | 0.04 | 22.4 | 13.5 |
| 20091025 | 11 | 5 | 58 | 579 | 15.0 | 1.3 | 0.349 | 1.3 | 0.91 | 0.11 | 0.03 | 0.01 | 0.05 | 0.02 | 14.1 | 8.0 |
| 20100111 | 5 | 41 | 5 | 450 | 313.8 | 0.4 | 0.383 | 2.6 | 0.38 | 0.13 | 0.15 | 0.08 | 0.21 | 0.16 | 4.5 | 0.5 |
| 20100307 | 6 | 3 | 26 | 480 | 300.1 | 1.2 | 0.417 | 0.7 | 1.54 | 0.46 | 0.08 | 0.01 | 0.11 | 0.01 | 8.2 | 2.6 |
| 20100309 | 7 | 40 | 44 | 376 | 295.0 | 0.9 | 0.416 | 2.0 | 0.54 | 0.04 | 0.15 | 0.07 | 0.25 | 0.13 | 37.1 | 3.5 |
| 20100316 | 5 | 7 | 44 | 492 | 297.3 | 0.1 | 0.362 | 2.4 | 0.42 | 0.04 | 0.10 | 0.02 | 0.13 | 0.04 | 13.8 | 1.8 |
| 20100421 | 4 | 49 | 43 | 709 | 5.7 | 0.5 | 0.36 | 1.0 | 1.21 | 0.27 | 0.05 | 0.01 | 0.07 | 0.01 | 20.9 | 8.9 |
| 20100423 | 8 | 32 | 35 | 567 | 100.5 | 0.6 | 0.345 | 1.9 | 0.51 | 0.04 | 0.03 | 0.00 | 0.05 | 0.01 | 11.1 | 3.1 |
| 20100429 | 5 | 21 | 35 | 617 | 320.4 | 0.5 | 0.363 | 0.9 | 0.99 | 0.35 | 0.07 | 0.01 | 0.11 | 0.02 | 24.7 | 8.0 |
| 20100814 | 3 | 0 | 42 | 686 | 298.2 | 1.2 | 0.332 | 1.2 | 0.76 | 0.09 | 0.04 | 0.02 | 0.07 | 0.05 | 5.4 | 1.1 |
| 20100914 | 5 | 42 | 26 | 667 | 107.2 | 0.3 | 0.357 | 1.3 | 0.73 | 0.17 | 0.04 | 0.01 | 0.05 | 0.02 | 16.5 | 4.5 |
| 20101129 | 9 | 12 | 44 | 425 | 22.4 | 1.5 | 0.349 | 2.6 | 0.38 | 0.03 | 0.09 | 0.02 | 0.10 | 0.05 | 13.6 | 5.9 |
| 20110208 | 8 | 59 | 33 | 437 | 340.3 | 0.7 | 0.381 | 1.8 | 0.53 | 0.04 | 0.13 | 0.03 | 0.24 | 0.05 | 24.1 | 4.7 |
| 20110402 | 8 | 46 | 52 | 529 | 0.0 | 0.1 | 0.374 | 2.3 | 0.39 | 0.04 | 0.05 | 0.00 | 0.08 | 0.01 | 24.2 | 4.1 |
| 20110520 | 6 | 2 | 9 | 565 | 62.5 | 0.3 | 0.385 | 1.9 | 0.65 | 0.02 | 0.03 | 0.01 | 0.06 | 0.01 | 17.3 | 5.3 |
| 20110630 | 3 | 39 | 38 | 535 | 186.2 | 0.3 | 0.348 | 3.0 | 0.40 | 0.00 | 0.03 | 0.01 | 0.04 | 0.02 | 8.6 | 3.4 |
| 20110808 | 5 | 22 | 6 | 565 | 170.2 | 0.4 | 0.368 | 1.5 | 0.50 | 0.03 | 0.03 | 0.01 | 0.04 | 0.02 | 12.0 | 4.4 |
| 20111005 | 5 | 8 | 53 | 407 | 306.6 | 0.8 | 0.379 | 4.6 | 0.18 | 0.01 | 0.13 | 0.06 | 0.20 | 0.12 | 68.9 | 15.5 |
| 20111202 | 0 | 31 | 4 | 449 | 339.1 | 0.8 | 0.381 | 2.8 | 0.38 | 0.11 | 0.15 | 0.06 | 0.21 | 0.12 | 24.3 | 7.0 |

**Table S2:** Infrasound signal measurements for optical meteors which produced more than one distinct infrasound arrival.

| Date | Event Time hh | mm | ss | Observed Travel Time s | Observed Back Azimuth ° | ± ° | Trace Velocity km/s | Dominant Freq Hz | Dominant Period s | ± s | Max Amp Pa | ± Pa | P2P Amp Pa | ± Pa | Amp SNR | Integrated Energy SNR |
|---|---|---|---|---|---|---|---|---|---|---|---|---|---|---|---|---|
| 20060305 | 5 | 15 | 36 | 349 | 342.6 | 0.9 | 0.381 | 3.5 | 0.33 | 0.02 | 0.11 | 0.04 | 0.18 | 0.08 | 50.8 | 3.6 |
| 20060305 | 5 | 15 | 36 | 358 | 343.7 | 1.2 | 0.357 | 2.1 | 0.44 | 0.09 | 0.05 | 0.01 | 0.07 | 0.01 | 17.0 | 3.8 |
| 20060305 | 5 | 15 | 36 | 359 | 342.8 | 1.2 | 0.357 | 3.0 | 0.29 | 0.04 | 0.04 | 0.01 | 0.06 | 0.02 | 16.4 | 4.3 |
| 20060405 | 3 | 3 | 27 | 394 | 86.4 | 0.4 | 0.335 | 3.9 | 0.34 | 0.07 | 0.10 | 0.07 | 0.15 | 0.14 | 33.0 | 7.1 |
| 20060405 | 3 | 3 | 27 | 397 | 86.6 | 0.3 | 0.339 | 2.9 | 0.31 | 0.05 | 0.08 | 0.03 | 0.10 | 0.05 | 25.2 | 1.1 |
| 20060805 | 8 | 38 | 50 | 427 | 255.5 | 1.3 | 0.372 | 0.9 | 1.19 | 0.10 | 0.49 | 0.10 | 0.68 | 0.20 | 162.9 | 30.8 |
| 20060805 | 8 | 38 | 50 | 450 | 257 | 0.2 | 0.41 | 0.6 | 2.01 | 0.17 | 0.10 | 0.02 | 0.18 | 0.05 | 44.5 | 0.4 |
| 20061223 | 6 | 27 | 26 | 479 | 341.4 | 1.6 | 0.381 | 2.0 | 0.53 | 0.07 | 0.05 | 0.03 | 0.09 | 0.05 | 28.9 | 9.3 |
| 20061223 | 6 | 27 | 26 | 489 | 346.3 | 1 | 0.351 | 1.0 | 0.96 | 0.05 | 0.09 | 0.04 | 0.15 | 0.07 | 21.2 | 8.7 |
| 20061223 | 6 | 27 | 26 | 516 | 346.4 | 0.9 | 0.386 | 1.2 | 0.74 | 0.06 | 0.18 | 0.07 | 0.31 | 0.14 | 53.3 | 48.8 |
| 20070102 | 10 | 42 | 3 | 579 | 33.1 | 0.3 | 0.358 | 0.8 | 1.30 | 0.17 | 0.04 | 0.01 | 0.07 | 0.02 | 11.1 | 2.9 |
| 20070102 | 10 | 42 | 3 | 581 | 32.8 | 0.3 | 0.372 | 2.0 | 0.51 | 0.07 | 0.04 | 0.02 | 0.05 | 0.05 | 9.8 | 3.4 |
| 20080511 | 4 | 22 | 17 | 371 | 24.8 | 0.4 | 0.516 | 1.3 | 0.77 | 0.03 | 0.03 | 0.01 | 0.04 | 0.01 | 13.8 | 2.8 |
| 20080511 | 4 | 22 | 17 | 381 | 24.8 | 0.4 | 0.425 | 2.9 | 0.36 | 0.00 | 0.01 | 0.00 | 0.02 | 0.01 | 13.2 | 2.7 |
| 20080612 | 5 | 58 | 29 | 388 | 71.2 | 0.2 | 0.408 | 3.5 | 0.25 | 0.00 | 0.04 | 0.01 | 0.06 | 0.02 | 13.8 | 0.3 |
| 20080612 | 5 | 58 | 29 | 391 | 64.7 | 0.2 | 0.425 | 1.4 | 0.72 | 0.09 | 0.10 | 0.04 | 0.16 | 0.07 | 42.4 | 20.3 |
| 20080812 | 3 | 27 | 25 | 400 | 170.3 | 0.3 | 0.446 | 1.5 | 0.62 | 0.13 | 0.20 | 0.06 | 0.35 | 0.12 | 110.1 | 20.2 |
| 20080812 | 3 | 27 | 25 | 406 | 176.2 | 0.6 | 0.431 | 1.6 | 0.66 | 0.17 | 0.11 | 0.01 | 0.15 | 0.02 | 3.9 | 0.3 |
| 20080812 | 8 | 19 | 29 | 554 | 249.4 | 0.1 | 0.349 | 1.6 | 0.59 | 0.05 | 0.02 | 0.00 | 0.03 | 0.01 | 16.5 | 6.5 |
| 20080812 | 8 | 19 | 29 | 557 | 249.8 | 0.5 | 0.349 | 1.8 | 0.58 | 0.01 | 0.02 | 0.00 | 0.03 | 0.01 | 14.0 | 2.8 |
| 20090428 | 4 | 43 | 37 | 456 | 53.6 | 1.7 | 0.345 | 4.1 | 0.21 | 0.03 | 0.16 | 0.03 | 0.22 | 0.06 | 45.1 | 4.6 |
| 20090428 | 4 | 43 | 37 | 460 | 55.3 | 1.5 | 0.345 | 2.7 | 0.32 | 0.01 | 0.06 | 0.02 | 0.11 | 0.04 | 26.3 | 6.4 |
| 20090812 | 7 | 55 | 58 | 522 | 204.5 | 1.2 | 0.372 | 1.6 | 0.59 | 0.05 | 0.09 | 0.02 | 0.16 | 0.03 | 26.6 | 1.0 |
| 20090812 | 7 | 55 | 58 | 525 | 204.3 | 1.1 | 0.372 | 1.8 | 0.58 | 0.01 | 0.07 | 0.01 | 0.09 | 0.02 | 10.9 | 1.0 |
| 20090825 | 1 | 14 | 35 | 811 | 46.2 | 0.1 | 0.349 | 1.2 | 0.76 | 0.10 | 0.07 | 0.03 | 0.12 | 0.05 | 18.0 | 1.6 |
| 20090825 | 1 | 14 | 35 | 835 | 47.2 | 0 | 0.348 | 1.5 | 0.54 | 0.01 | 0.03 | 0.01 | 0.05 | 0.02 | 5.6 | 1.4 |
| 20090825 | 1 | 14 | 35 | 842 | 47.4 | 0.4 | 0.348 | 1.8 | 0.58 | 0.03 | 0.03 | 0.01 | 0.05 | 0.02 | 6.0 | 1.3 |
| 20090926 | 1 | 2 | 58 | 424 | 82.8 | 0.8 | 0.347 | 2.0 | 0.59 | 0.01 | 0.28 | 0.02 | 0.49 | 0.04 | 112.2 | 50.0 |
| 20090926 | 1 | 2 | 58 | 439 | 79.7 | 0.1 | 0.363 | 2.2 | 0.49 | 0.03 | 0.05 | 0.01 | 0.08 | 0.02 | 3.2 | 0.4 |
| 20100530 | 7 | 0 | 31 | 530 | 324.4 | 0.4 | 0.358 | 1.5 | 0.64 | 0.01 | 0.03 | 0.01 | 0.06 | 0.02 | 16.1 | 1.7 |
| 20100530 | 7 | 0 | 31 | 535 | 325.2 | 0.1 | 0.358 | 1.9 | 0.84 | 0.21 | 0.03 | 0.01 | 0.05 | 0.01 | 10.9 | 1.7 |
| 20100802 | 7 | 18 | 25 | 406 | 74.5 | 0.3 | 0.417 | 4.0 | 0.22 | 0.00 | 0.03 | 0.01 | 0.06 | 0.02 | 16.1 | 1.7 |
| 20100802 | 7 | 18 | 25 | 410 | 73.7 | 0.1 | 0.417 | 1.6 | 0.30 | 0.01 | 0.03 | 0.01 | 0.05 | 0.01 | 10.9 | 1.7 |
| 20110815 | 5 | 50 | 16 | 550 | 328.3 | 0.4 | 0.347 | 2.0 | 0.66 | 0.02 | 0.04 | 0.02 | 0.06 | 0.04 | 14.5 | 2.1 |
| 20110815 | 5 | 50 | 16 | 556 | 328.9 | 0.4 | 0.372 | 1.2 | 0.87 | 0.09 | 0.03 | 0.01 | 0.05 | 0.02 | 9.0 | 1.1 |

**Table S3:** All astrometric measurements. N is the number of cameras which recorded the event, meteor azimuth and zenith angles refer to the location of the meteor radiant (direction in the sky), and meteor trajectory length is the ground projected length.

| Date | Event Time | | | N | Entry Velocity | ± | Meteor Azimuth | Meteor Zenith | Begin Height | End Height | Meteor Trajectory Length | Meteor Flight Time | Camera Error Time |
|---|---|---|---|---|---|---|---|---|---|---|---|---|---|
| | hh | mm | ss | | km/s | km/s | ° | ° | ° | ° | km | s | s |
| 20060419 | 4 | 21 | 28 | 2 | 18.8 | 0.6 | 13.1 | 64.1 | 81.3 | 68.1 | 26.7 | 1.57 | 0.22 |
| 20060419 | 7 | 5 | 56 | 3 | 14.2 | 0.1 | 98.9 | 56.1 | 72.0 | 47.7 | 35.6 | 3.30 | 0.53 |
| 20060813 | 9 | 42 | 26 | 3 | 62.7 | 0.3 | 216.1 | 19.2 | 105.4 | 75.8 | 10.1 | 0.50 | 0.04 |
| 20061101 | 6 | 46 | 12 | 4 | 57.1 | 0.9 | 245.2 | 14.7 | 106.0 | 83.4 | 5.8 | 0.40 | 0.05 |
| 20061104 | 3 | 29 | 29 | 2 | 30.3 | 0.4 | 293.8 | 35.7 | 89.9 | 65.8 | 17.1 | 1.03 | 0.01 |
| 20061121 | 10 | 45 | 46 | 5 | 71.1 | 1.3 | 325.2 | 25.8 | 124.3 | 83.0 | 19.5 | 0.63 | 0.32 |
| 20070125 | 10 | 2 | 5 | 2 | 71.2 | 0.4 | 340.0 | 75.0 | 119.2 | 88.5 | 109.5 | 1.64 | 0.30 |
| 20070421 | 9 | 21 | 1 | 3 | 33.3 | 1.0 | 144.5 | 14.6 | 104.4 | 70.2 | 8.7 | 1.00 | 0.13 |
| 20070511 | 7 | 41 | 14 | 3 | 64.5 | 0.6 | 297.4 | 52.9 | 111.1 | 92.7 | 23.8 | 0.47 | 0.21 |
| 20070725 | 4 | 42 | 44 | 3 | 26.3 | 0.2 | 344.6 | 34.2 | 91.5 | 72.8 | 12.5 | 0.87 | 0.09 |
| 20070727 | 4 | 51 | 58 | 3 | 26.3 | 0.0 | 350.8 | 51.2 | 96.2 | 70.6 | 31.3 | 1.57 | 0.79 |
| 20070917 | 7 | 55 | 43 | 2 | 59.5 | 0.3 | 224.4 | 40.5 | 103.7 | 87.1 | 13.9 | 0.33 | 0.09 |
| 20071004 | 4 | 55 | 9 | 3 | 28.9 | 0.5 | 156.5 | 47.0 | 97.3 | 69.6 | 29.1 | 1.37 | 0.13 |
| 20071004 | 5 | 19 | 27 | 3 | 16.1 | 0.2 | 36.2 | 53.1 | 75.9 | 45.1 | 40.5 | 3.50 | 0.35 |
| 20071021 | 10 | 26 | 25 | 6 | 75.6 | 2.7 | 24.5 | 22.4 | 130.8 | 81.7 | 19.8 | 0.70 | 0.06 |
| 20071130 | 10 | 28 | 49 | 3 | 71.7 | 0.5 | 357.3 | 25.0 | 112.9 | 73.5 | 18.1 | 0.60 | 0.02 |
| 20071215 | 11 | 18 | 13 | 5 | 35.7 | 0.5 | 96.3 | 44.6 | 93.6 | 60.7 | 31.8 | 1.30 | 0.41 |
| 20080325 | 0 | 42 | 3 | 6 | 13.5 | 0.3 | 15.3 | 45.3 | 76.2 | 32.8 | 43.3 | 5.21 | 0.47 |
| 20080520 | 3 | 29 | 47 | 4 | 14.1 | 0.1 | 104.2 | 29.5 | 75.2 | 40.4 | 19.4 | 3.24 | 16.26 |
| 20080602 | 5 | 44 | 44 | 6 | 63.5 | 0.7 | 269.1 | 61.2 | 118.1 | 90.0 | 49.6 | 0.90 | 2.92 |
| 20080801 | 4 | 22 | 20 | 4 | 23.6 | 0.4 | 344.0 | 50.6 | 93.3 | 78.9 | 17.3 | 0.93 | 0.24 |
| 20080801 | 8 | 19 | 30 | 7 | 65.2 | 1.2 | 314.9 | 68.3 | 115.5 | 92.4 | 56.5 | 0.97 | 0.98 |
| 20080804 | 4 | 17 | 14 | 5 | 58.3 | 0.3 | 218.2 | 59.8 | 110.7 | 80.3 | 50.8 | 1.00 | 0.49 |
| 20080812 | 6 | 0 | 34 | 5 | 60.7 | 3.5 | 222.6 | 48.5 | 105.9 | 89.6 | 18.2 | 0.43 | 0.47 |
| 20080908 | 4 | 3 | 12 | 5 | 62.3 | 1.2 | 265.5 | 56.1 | 117.6 | 81.7 | 52.0 | 1.04 | 0.10 |
| 20081005 | 1 | 50 | 12 | 3 | 22.8 | 1.5 | 278.0 | 24.0 | 88.4 | 43.4 | 18.4 | 2.10 | 0.37 |
| 20081018 | 6 | 53 | 39 | 3 | 21.3 | 0.3 | 74.1 | 38.9 | 84.1 | 49.8 | 27.2 | 2.24 | 0.57 |
| 20081028 | 3 | 17 | 35 | 3 | 15.8 | 0.1 | 0.4 | 57.1 | 81.2 | 41.1 | 60.9 | 5.21 | 0.19 |
| 20081102 | 6 | 13 | 26 | 6 | 31.1 | 0.7 | 356.7 | 28.2 | 96.5 | 62.6 | 17.9 | 1.33 | 0.08 |
| 20081107 | 7 | 34 | 16 | 6 | 70.2 | 0.9 | 297.2 | 62.0 | 113.5 | 81.5 | 58.5 | 0.97 | 0.34 |
| 20090126 | 7 | 16 | 24 | 5 | 67.7 | 0.9 | 272.4 | 75.0 | 115.6 | 81.7 | 119.6 | 1.87 | 0.49 |
| 20090523 | 7 | 7 | 25 | 4 | 29.5 | 1.0 | 40.2 | 42.4 | 95.9 | 72.4 | 21.1 | 1.14 | 0.12 |
| 20090530 | 6 | 35 | 20 | 4 | 17.3 | 1.2 | 213.4 | 59.6 | 81.7 | 60.3 | 35.9 | 2.47 | 0.80 |
| 20090709 | 5 | 24 | 23 | 5 | 18.6 | 0.7 | 53.6 | 39.9 | 90.2 | 75.2 | 12.4 | 1.03 | 0.15 |
| 20090813 | 6 | 47 | 53 | 6 | 60.9 | 0.4 | 228.4 | 42.1 | 116.2 | 74.2 | 37.0 | 0.94 | 0.01 |
| 20090906 | 1 | 27 | 55 | 4 | 60.6 | 0.4 | 213.8 | 69.9 | 110.9 | 77.4 | 88.4 | 2.50 | 0.61 |
| 20090917 | 1 | 20 | 38 | 6 | 23.0 | 0.4 | 296.9 | 65.4 | 85.7 | 72.4 | 28.5 | 1.60 | 0.90 |
| 20091025 | 11 | 5 | 58 | 4 | 28.7 | 0.3 | 284.1 | 88.4 | 72.6 | 69.2 | 92.2 | 3.40 | 0.46 |
| 20100111 | 5 | 41 | 5 | 3 | 54.6 | 0.4 | 239.0 | 73.0 | 93.9 | 63.4 | 95.5 | 1.47 | 0.88 |
| 20100307 | 6 | 3 | 26 | 4 | 49.0 | 1.3 | 257.0 | 53.3 | 105.8 | 93.0 | 16.9 | 0.43 | 0.16 |
| 20100309 | 7 | 40 | 44 | 6 | 48.3 | 0.9 | 259.2 | 44.1 | 106.9 | 67.2 | 37.6 | 1.17 | 0.10 |
| 20100316 | 5 | 7 | 44 | 6 | 22.0 | 0.3 | 8.3 | 60.5 | 78.0 | 48.6 | 51.2 | 2.08 | 0.30 |
| 20100421 | 4 | 49 | 43 | 6 | 46.2 | 2.0 | 281.5 | 48.1 | 108.5 | 74.6 | 37.0 | 1.07 | 0.09 |
| 20100423 | 8 | 32 | 35 | 6 | 48.8 | 1.1 | 308.6 | 13.5 | 103.4 | 71.6 | 7.5 | 0.63 | 0.13 |
| 20100429 | 5 | 21 | 35 | 4 | 46.3 | 0.8 | 268.4 | 49.3 | 105.7 | 89.9 | 18.0 | 0.50 | 0.05 |
| 20100814 | 3 | 0 | 42 | 4 | 57.7 | 1.1 | 207.4 | 65.6 | 112.9 | 80.3 | 69.8 | 1.34 | 0.09 |
| 20100914 | 5 | 42 | 26 | 4 | 43.9 | 0.1 | 175.7 | 40.5 | 104.4 | 79.4 | 20.9 | 0.73 | 0.09 |
| 20101129 | 9 | 12 | 44 | 3 | 29.2 | 0.4 | 82.7 | 42.7 | 100.1 | 56.5 | 39.4 | 2.10 | 0.03 |
| 20110208 | 8 | 59 | 33 | 3 | 59.2 | 2.6 | 290.2 | 28.2 | 112.8 | 71.9 | 21.5 | 0.80 | 0.04 |
| 20110402 | 8 | 46 | 52 | 4 | 28.4 | 0.4 | 53.9 | 53.4 | 94.8 | 73.3 | 28.4 | 1.30 | 0.42 |
| 20110520 | 6 | 2 | 9 | 3 | 23.5 | 0.8 | 21.6 | 51.5 | 95.7 | 84.1 | 14.3 | 0.80 | 0.22 |
| 20110630 | 3 | 39 | 38 | 6 | 33.7 | 0.3 | 209.2 | 23.6 | 100.5 | 71.7 | 12.4 | 0.87 | 0.16 |
| 20110808 | 5 | 22 | 6 | 5 | 24.8 | 0.4 | 156.1 | 49.3 | 86.6 | 39.9 | 53.3 | 2.84 | 0.10 |
| 20111005 | 5 | 8 | 53 | 3 | 29.2 | 0.2 | 342.2 | 43.4 | 96.2 | 64.5 | 29.5 | 1.50 | 0.12 |
| 20111202 | 0 | 31 | 4 | 4 | 26.9 | 0.1 | 265.1 | 67.8 | 97.0 | 53.8 | 101.8 | 4.40 | 0.42 |
| 20060305 | 5 | 15 | 36 | 3 | 18.6 | 0.1 | 8.8 | 30.9 | 81.2 | 35.5 | 27.0 | 3.17 | 0.30 |
| 20060405 | 3 | 3 | 27 | 2 | 11.5 | 1.7 | 89.7 | 26.9 | 67.8 | 36.4 | 15.7 | 3.54 | 0.76 |
| 20060805 | 8 | 38 | 50 | 4 | 67.5 | 2.7 | 268.5 | 29.3 | 126.4 | 74.5 | 28.5 | 0.80 | 0.19 |
| 20061223 | 6 | 27 | 26 | 4 | 22.9 | 0.8 | 29.7 | 32.1 | 91.5 | 31.1 | 37.2 | 3.00 | 0.93 |
| 20070102 | 10 | 42 | 3 | 2 | 43.6 | 2.6 | 242.8 | 30.0 | 94.0 | 66.2 | 15.8 | 0.77 | 0.04 |
| 20080511 | 4 | 22 | 17 | 4 | 21.7 | 0.3 | 20.6 | 67.1 | 95.2 | 77.3 | 41.5 | 1.97 | 0.72 |
| 20080612 | 5 | 58 | 29 | 3 | 15.5 | 0.4 | 130.6 | 52.2 | 88.3 | 69.7 | 23.7 | 1.97 | 0.14 |
| 20080812 | 3 | 27 | 25 | 2 | 59.4 | 0.8 | 211.0 | 64.0 | 115.8 | 77.0 | 77.2 | 1.47 | 0.75 |
| 20080812 | 8 | 19 | 29 | 3 | 56.2 | 1.4 | 224.9 | 28.1 | 105.7 | 82.0 | 12.4 | 0.47 | 0.34 |
| 20090428 | 4 | 43 | 37 | 4 | 18.2 | 0.0 | 351.0 | 32.8 | 83.5 | 38.0 | 29.0 | 2.97 | 0.29 |
| 20090812 | 7 | 55 | 58 | 3 | 58.7 | 0.2 | 225.7 | 32.6 | 108.5 | 80.4 | 17.7 | 0.57 | 0.80 |
| 20090825 | 1 | 14 | 35 | 4 | 23.4 | 1.6 | 297.0 | 48.5 | 85.4 | 45.7 | 44.1 | 2.50 | 0.40 |
| 20090926 | 1 | 2 | 58 | 7 | 20.5 | 1.2 | 129.4 | 36.3 | 100.5 | 19.6 | 56.3 | 6.07 | 0.03 |
| 20100530 | 7 | 0 | 31 | 4 | 28.4 | 0.5 | 16.3 | 61.9 | 96.0 | 78.3 | 32.5 | 1.33 | 0.06 |
| 20100802 | 7 | 18 | 25 | 3 | 41.2 | 3.2 | 358.3 | 57.8 | 93.6 | 64.8 | 44.9 | 1.27 | 0.18 |
| 20110815 | 5 | 50 | 16 | 2 | 57.1 | 2.2 | 292.1 | 30.5 | 104.3 | 77.7 | 15.4 | 0.57 | 0.21 |

**Table S4:** Summary of infrasound signal characteristics by taxonomic class

| | Source Altitude (km) | Signal celerity (km/s) | Total range (km) | Horiz range (km) | Ray deviation (°) | Sh parameter | Meteor Entry Velocity (km/s) | Begin Altitude (km) | End Altitude (km) | Flight Time (s) | Trace Velocity (km/s) | Dominant Freq (Hz) | Dominant Period (s) | Max Amplitude (Pa) | P2P Amplitude (Pa) | Bolide Integrated Energy SNR |
|---|---|---|---|---|---|---|---|---|---|---|---|---|---|---|---|---|
| **Class I** *(51 arrivals)* | | | | | | | | | | | | | | | | |
| Avg | 77.5 | 0.306 | 140.6 | 115.3 | 102.0 | 0.6 | 38.5 | 97.6 | 65.6 | 1.7 | 0.394 | 2.64 | 0.70 | 0.09 | 0.14 | 8.50 |
| Std. dev. | 15.6 | 0.015 | 32.8 | 36.3 | 16.7 | 0.3 | 20.1 | 15.5 | 17.2 | 1.2 | 0.080 | 2.56 | 0.50 | 0.07 | 0.10 | 11.16 |
| Min | 38.1 | 0.274 | 73.3 | 47.4 | 75.2 | 0.1 | 11.5 | 67.8 | 31.1 | 0.4 | 0.332 | 0.39 | 0.10 | 0.02 | 0.03 | 0.28 |
| Max | 111.4 | 0.334 | 216.6 | 198.6 | 146.7 | 1.0 | 75.6 | 130.8 | 93.0 | 5.2 | 0.831 | 12.89 | 2.60 | 0.49 | 0.68 | 72.40 |
| **Class II** *(20 arrivals)* | | | | | | | | | | | | | | | | |
| Avg | 78.2 | 0.313 | 151.8 | 127.8 | 99.4 | 0.6 | 37.3 | 97.6 | 63.2 | 2.2 | 0.393 | 2.29 | 0.60 | 0.07 | 0.12 | 9.15 |
| Std. dev. | 17.0 | 0.015 | 28.0 | 33.4 | 13.7 | 0.3 | 17.7 | 12.0 | 22.0 | 1.8 | 0.057 | 1.70 | 0.30 | 0.06 | 0.11 | 14.13 |
| Min | 20.2 | 0.288 | 113.1 | 72.5 | 70.3 | 0.0 | 13.5 | 72.6 | 19.6 | 0.3 | 0.332 | 0.88 | 0.10 | 0.01 | 0.02 | 0.31 |
| Max | 102.7 | 0.336 | 206.9 | 189.7 | 122.6 | 1.0 | 71.2 | 119.2 | 88.5 | 6.1 | 0.601 | 7.03 | 1.30 | 0.28 | 0.49 | 50.03 |
| **Class III** *(6 arrivals)* | | | | | | | | | | | | | | | | |
| Avg | 77.0 | 0.298 | 118.1 | 87.9 | 92.2 | 0.8 | 35.8 | 95.2 | 71.7 | 1.5 | 0.429 | 3.63 | 0.40 | 0.05 | 0.08 | 9.59 |
| Std. dev. | 19.7 | 0.008 | 16.0 | 15.1 | 3.2 | 0.1 | 21.3 | 13.3 | 20.1 | 1.1 | 0.049 | 3.16 | 0.20 | 0.03 | 0.04 | 3.24 |
| Min | 46.9 | 0.287 | 92.6 | 68.2 | 87.0 | 0.5 | 14.1 | 75.2 | 40.4 | 0.4 | 0.349 | 1.37 | 0.10 | 0.01 | 0.02 | 5.92 |
| Max | 101.0 | 0.307 | 132.6 | 108.2 | 95.9 | 0.9 | 64.5 | 111.1 | 92.7 | 3.2 | 0.494 | 9.96 | 0.80 | 0.09 | 0.13 | 13.55 |
| **Class IV** *(13 arrivals)* | | | | | | | | | | | | | | | | |
| Avg | 78.6 | 0.309 | 174.8 | 150.6 | 111.7 | 0.6 | 40.8 | 99.6 | 65.3 | 1.6 | 0.386 | 2.01 | 0.70 | 0.04 | 0.06 | 3.94 |
| Std. dev. | 17.9 | 0.019 | 57.9 | 69.9 | 17.8 | 0.3 | 19.0 | 12.2 | 19.5 | 1.0 | 0.050 | 1.43 | 0.40 | 0.02 | 0.03 | 2.46 |
| Min | 46.4 | 0.282 | 111.7 | 59.3 | 83.9 | 0.0 | 21.7 | 85.4 | 31.1 | 0.4 | 0.348 | 0.59 | 0.20 | 0.01 | 0.03 | 1.34 |
| Max | 103.8 | 0.338 | 277.1 | 267.9 | 151.5 | 1.0 | 71.1 | 124.3 | 83.4 | 3.0 | 0.52 | 6.25 | 1.60 | 0.07 | 0.12 | 9.26 |

**Table S5:** The raytracing results and signal classification for single arrival events. All raytracing parameters represent the modelled quantities at shock source height, which are then compared to the observed quantities to determine which source heights best match the observations. Here Solution type code is: NS - no source height solution; D - degenerate source height solution and S - unique source height solution.

| Date | Event Time hh | mm | ss | Source H_mean km | ± km | Total Range km | Hor. Range km | Ray Dev ° | Back Azimuth ° | Travel Time s | Sol Type | Sh Param | Signal Class |
|---|---|---|---|---|---|---|---|---|---|---|---|---|---|
| 20060419 | 4 | 21 | 28 | 79.8 | 0.2 | 109.8 | 75.4 | 88.2 | 327.4 | 349 | NS | 0.11 | I |
| 20060419 | 7 | 5 | 56 | 54.4 | 1.1 | 83.7 | 63.5 | 80.7 | 143.9 | 279 | D | 0.72 | I |
| 20060813 | 9 | 42 | 26 | 80.3 | 0.4 | 175.1 | 155.6 | 122.6 | 146.8 | 554 | S | 0.85 | II |
| 20061101 | 6 | 46 | 12 | 85.7 | 0.5 | 178.3 | 156.4 | 121.8 | 273.6 | 527 | S | 0.90 | IV |
| 20061104 | 3 | 29 | 29 | 77.0 | 1.1 | 116.4 | 87.2 | 104.3 | 294.1 | 354 | S | 0.53 | II |
| 20061121 | 10 | 45 | 46 | 91.4 | 0.4 | 148.1 | 116.7 | 151.5 | 136.0 | 526 | S | 0.80 | IV |
| 20070125 | 10 | 2 | 5 | 102.7 | 0.5 | 126.6 | 73.7 | 89.3 | 287.5 | 399 | S | 0.53 | II |
| 20070421 | 9 | 21 | 1 | 89.6 | 0.4 | 194.6 | 172.7 | 130.8 | 36.5 | 636 | S | 0.43 | I |
| 20070511 | 7 | 41 | 14 | 101.0 | 1.2 | 131.9 | 84.8 | 95.9 | 331.7 | 439 | S | 0.55 | III |
| 20070725 | 4 | 42 | 44 | 73.3 | 0.5 | 126.7 | 103.3 | 105.5 | 28.6 | 403 | NS | 0.97 | I |
| 20070727 | 4 | 51 | 58 | 85.0 | 1.5 | 149.0 | 122.5 | 94.6 | 30.2 | 469 | S | 0.44 | I |
| 20070917 | 7 | 55 | 43 | 87.9 | 0.8 | 175.3 | 151.7 | 95.6 | 257.8 | 589 | NS | 0.96 | II |
| 20071004 | 4 | 55 | 9 | 83.0 | 0.2 | 166.9 | 144.9 | 146.7 | 18.1 | 557 | S | 0.52 | I |
| 20071004 | 5 | 19 | 27 | 46.4 | 0.8 | 92.2 | 79.7 | 82.1 | 96.6 | 311 | NS | 0.96 | I |
| 20071021 | 10 | 26 | 25 | 101.2 | 1.4 | 194.8 | 166.3 | 105.2 | 39.8 | 700 | S | 0.60 | I |
| 20071130 | 10 | 28 | 49 | 93.2 | 0.6 | 187.0 | 162.1 | 114.2 | 68.2 | 645 | S | 0.50 | I |
| 20071215 | 11 | 18 | 13 | 75.9 | 0.9 | 89.5 | 47.4 | 94.0 | 65.3 | 328 | NS | 0.54 | I |
| 20080325 | 0 | 42 | 3 | 61.6 | 0.6 | 113.1 | 94.9 | 107.4 | 304.9 | 343 | S | 0.34 | II |
| 20080520 | 3 | 29 | 47 | 46.9 | 1.0 | 92.6 | 79.8 | 87.0 | 82.2 | 305 | NS | 0.81 | III |
| 20080602 | 5 | 44 | 44 | 111.4 | 3.9 | 147.2 | 96.2 | 89.0 | 212.9 | 508 | S | 0.24 | I |
| 20080801 | 4 | 22 | 20 | 79.8 | 0.5 | 104.9 | 68.2 | 93.2 | 332.5 | 348 | NS | 0.93 | III |
| 20080801 | 8 | 19 | 30 | 92.6 | 0.3 | 110.5 | 60.3 | 86.3 | 346.4 | 373 | S | 0.99 | I |
| 20080804 | 4 | 17 | 14 | 90.0 | 0.4 | 121.0 | 80.8 | 108.6 | 140.7 | 392 | S | 0.68 | IV |
| 20080812 | 6 | 0 | 34 | 93.8 | 2.7 | 125.2 | 82.8 | 94.4 | 200.8 | 430 | NS | 0.74 | III |
| 20080908 | 4 | 3 | 12 | 104.3 | 1.0 | 178.3 | 144.6 | 75.2 | 294.2 | 614 | NS | 0.37 | I |
| 20081005 | 1 | 50 | 12 | 54.3 | 1.8 | 128.3 | 116.3 | 102.4 | 341.1 | 405 | S | 0.76 | IV |
| 20081018 | 6 | 53 | 39 | 67.9 | 1.0 | 86.7 | 53.3 | 101.4 | 84.6 | 303 | NS | 0.47 | I |
| 20081028 | 3 | 17 | 35 | 52.7 | 3.6 | 73.3 | 50.9 | 92.2 | 307.2 | 239 | S | 0.71 | I |
| 20081102 | 6 | 13 | 26 | 85.0 | 0.5 | 193.1 | 173.4 | 117.3 | 292.4 | 576 | S | 0.34 | II |
| 20081107 | 7 | 34 | 16 | 81.9 | 0.6 | 119.9 | 87.6 | 85.3 | 332.0 | 377 | D | 0.99 | I |
| 20090126 | 7 | 16 | 24 | 87.3 | 0.8 | 100.3 | 49.1 | 88.4 | 219.9 | 329 | NS | 0.83 | I |
| 20090523 | 7 | 7 | 25 | 78.1 | 2.3 | 134.7 | 109.8 | 96.0 | 60.8 | 429 | S | 0.76 | I |
| 20090530 | 6 | 35 | 20 | 74.0 | 2.5 | 98.5 | 65.7 | 88.1 | 256.1 | 325 | NS | 0.39 | I |
| 20090709 | 5 | 24 | 23 | 75.9 | 0.5 | 144.3 | 122.7 | 107.3 | 352.3 | 468 | NS | 0.95 | I |
| 20090813 | 6 | 47 | 53 | 75.4 | 0.2 | 166.1 | 148.0 | 144.4 | 359.1 | 541 | S | 0.97 | I |
| 20090906 | 1 | 27 | 55 | 103.8 | 1.0 | 138.5 | 91.5 | 92.1 | 282.6 | 469 | NS | 0.21 | IV |
| 20090917 | 1 | 20 | 38 | 76.6 | 2.1 | 132.6 | 108.2 | 90.3 | 358.4 | 437 | NS | 0.68 | III |
| 20091025 | 11 | 5 | 58 | 70.5 | 0.2 | 171.0 | 155.8 | 91.4 | 14.6 | 595 | NS | 0.62 | II |
| 20100111 | 5 | 41 | 5 | 78.6 | 1.7 | 151.3 | 129.2 | 100.3 | 326.1 | 458 | S | 0.50 | I |
| 20100307 | 6 | 3 | 26 | 103.6 | 0.8 | 153.5 | 113.4 | 99.9 | 304.3 | 480 | S | 0.18 | I |
| 20100309 | 7 | 40 | 44 | 77.8 | 0.9 | 124.2 | 96.6 | 102.4 | 297.3 | 375 | S | 0.73 | I |
| 20100316 | 5 | 7 | 44 | 75.2 | 0.9 | 161.6 | 143.1 | 94.3 | 301.3 | 493 | S | 0.10 | I |
| 20100421 | 4 | 49 | 43 | 86.3 | 0.8 | 216.6 | 198.6 | 109.4 | 5.7 | 709 | S | 0.65 | I |
| 20100423 | 8 | 32 | 35 | 76.5 | 0.4 | 166.2 | 147.5 | 136.5 | 100.8 | 566 | S | 0.85 | I |
| 20100429 | 5 | 21 | 35 | 93.0 | 1.9 | 191.9 | 167.9 | 94.6 | 322.3 | 622 | NS | 0.81 | I |
| 20100814 | 3 | 0 | 42 | 82.5 | 0.6 | 206.9 | 189.7 | 109.2 | 300.7 | 711 | NS | 0.93 | II |
| 20100914 | 5 | 42 | 26 | 80.6 | 0.6 | 198.1 | 180.9 | 105.8 | 108.1 | 667 | S | 0.96 | I |
| 20101129 | 9 | 12 | 44 | 63.7 | 1.0 | 121.6 | 103.5 | 92.2 | 25.3 | 426 | S | 0.83 | III |
| 20110208 | 8 | 59 | 33 | 78.0 | 0.5 | 139.5 | 115.6 | 109.8 | 340.6 | 437 | S | 0.85 | II |
| 20110402 | 8 | 46 | 52 | 82.7 | 2.8 | 162.1 | 139.4 | 89.5 | 359.6 | 529 | S | 0.56 | II |
| 20110520 | 6 | 2 | 9 | 94.5 | 0.7 | 180.5 | 153.2 | 88.0 | 62.0 | 570 | NS | 0.09 | II |
| 20110630 | 3 | 39 | 38 | 87.7 | 0.5 | 161.7 | 135.8 | 113.6 | 186.0 | 535 | S | 0.45 | IV |
| 20110808 | 5 | 22 | 6 | 63.6 | 0.3 | 179.1 | 167.4 | 70.3 | 169.5 | 565 | S | 0.49 | II |
| 20111005 | 5 | 8 | 53 | 77.8 | 4.2 | 131.3 | 105.7 | 97.1 | 307.2 | 407 | S | 0.58 | I |
| 20111202 | 0 | 31 | 4 | 64.0 | 0.6 | 148.3 | 133.7 | 91.3 | 339.4 | 448 | D | 0.76 | I |

**Table S6:** The raytracing results and signal classification for multi arrival events. For the purpose of quantifying the shock source parameters, each infrasound arrival was treated as a separate event. All raytracing parameters represent the modelled quantities at shock source height, which are then compared to the observed quantities to determine the solution viability. Here Solution type code is: NS - no source height solution; D - degenerate source height solution and S - unique source height solution.

| Date | Event Time | | | Source $H_{mean}$ | ± | Total Range | Hor. Range | Ray Dev | Back Azimuth | Travel Time | Sol Type | Sh Param | Signal Class |
|---|---|---|---|---|---|---|---|---|---|---|---|---|---|
| | hh | mm | ss | km | km | km | km | ° | ° | s | | | |
| 20060305 | 5 | 15 | 36 | 54.8 | 2.3 | 113.1 | 99.0 | 91.3 | 346.6 | 349 | S | 0.58 | I |
| 20060305 | 5 | 15 | 36 | 38.9 | 1.7 | 114.2 | 107.4 | 93.7 | 346.0 | 358 | D | 0.93 | I |
| 20060305 | 5 | 15 | 36 | 38.1 | 1.4 | 114.3 | 107.7 | 93.3 | 346.1 | 358 | S | 0.94 | I |
| 20060405 | 3 | 3 | 27 | 63.5 | 3.1 | 117.2 | 98.6 | 90.7 | 86.2 | 400 | D | 0.14 | I |
| 20060405 | 3 | 3 | 27 | 60.6 | 2.7 | 116.9 | 100.3 | 87.9 | 86.0 | 398 | D | 0.24 | I |
| 20060805 | 8 | 38 | 50 | 87.8 | 0.6 | 124.7 | 88.3 | 110.5 | 254.8 | 427 | S | 0.74 | I |
| 20060805 | 8 | 38 | 50 | 101.4 | 0.4 | 130.0 | 81.2 | 112.8 | 254.1 | 450 | S | 0.48 | I |
| 20061223 | 6 | 27 | 26 | 54.3 | 2.4 | 157.2 | 147.6 | 98.1 | 341.9 | 480 | S | 0.62 | IV |
| 20061223 | 6 | 27 | 26 | 74.7 | 1.7 | 158.4 | 139.7 | 103.9 | 341.4 | 489 | S | 0.28 | I |
| 20061223 | 6 | 27 | 26 | 90.6 | 0.7 | 161.9 | 133.9 | 112.9 | 338.3 | 508 | NS | 0.01 | II |
| 20070102 | 10 | 42 | 3 | 79.7 | 0.2 | 164.4 | 143.7 | 143.2 | 31.8 | 579 | S | 0.52 | I |
| 20070102 | 10 | 42 | 3 | 80.3 | 0.3 | 164.9 | 144.0 | 144.6 | 32.6 | 581 | S | 0.49 | I |
| 20080511 | 4 | 22 | 17 | 94.6 | 0.4 | 111.7 | 59.3 | 83.9 | 25.5 | 372 | S | 0.03 | IV |
| 20080511 | 4 | 22 | 17 | 88.9 | 0.5 | 114.7 | 72.5 | 78.6 | 25.1 | 382 | S | 0.35 | II |
| 20080612 | 5 | 58 | 29 | 71.8 | 0.5 | 126.3 | 104.0 | 101.2 | 70.2 | 388 | S | 0.89 | I |
| 20080612 | 5 | 58 | 29 | 76.0 | 0.3 | 126.7 | 101.3 | 104.6 | 67.3 | 391 | S | 0.66 | I |
| 20080812 | 3 | 27 | 25 | 79.4 | 0.7 | 120.7 | 91.0 | 100.5 | 181.4 | 400 | D | 0.94 | I |
| 20080812 | 3 | 27 | 25 | 77.7 | 0.2 | 121.6 | 93.6 | 101.9 | 182.3 | 405 | S | 0.98 | II |
| 20080812 | 8 | 19 | 29 | 86.2 | 0.8 | 157.0 | 131.3 | 103.8 | 249.6 | 555 | S | 0.83 | IV |
| 20080812 | 8 | 19 | 29 | 87.9 | 0.8 | 157.3 | 130.4 | 104.0 | 250.1 | 558 | S | 0.75 | IV |
| 20090428 | 4 | 43 | 37 | 60.7 | 6.2 | 138.6 | 124.8 | 95.6 | 52.4 | 454 | S | 0.50 | I |
| 20090428 | 4 | 43 | 37 | 70.9 | 1.1 | 140.6 | 121.5 | 105.1 | 55.2 | 459 | S | 0.28 | I |
| 20090812 | 7 | 55 | 58 | 80.6 | 0.3 | 155.9 | 133.4 | 101.1 | 205.5 | 526 | NS | 0.99 | I |
| 20090812 | 7 | 55 | 58 | 80.5 | 0.3 | 155.9 | 133.4 | 101.1 | 205.5 | 526 | S | 1.00 | I |
| 20090825 | 1 | 14 | 35 | 46.4 | 0.3 | 261.4 | 257.2 | 116.0 | 46.0 | 811 | S | 0.98 | IV |
| 20090825 | 1 | 14 | 35 | 68.3 | 1.8 | 275.2 | 266.5 | 127.4 | 47.9 | 846 | NS | 0.43 | IV |
| 20090825 | 1 | 14 | 35 | 70.9 | 2.0 | 277.1 | 267.9 | 129.0 | 48.4 | 852 | S | 0.36 | IV |
| 20090926 | 1 | 2 | 58 | 20.2 | 0.9 | 134.8 | 133.2 | 88.3 | 89.0 | 423 | S | 0.99 | II |
| 20090926 | 1 | 2 | 58 | 70.8 | 1.0 | 129.9 | 109.1 | 103.9 | 77.0 | 440 | D | 0.37 | II |
| 20100530 | 7 | 0 | 31 | 92.7 | 2.4 | 156.4 | 125.9 | 87.2 | 322.7 | 535 | NS | 0.18 | I |
| 20100530 | 7 | 0 | 31 | 92.1 | 2.8 | 156.6 | 126.7 | 86.9 | 323.2 | 536 | S | 0.21 | II |
| 20100802 | 7 | 18 | 25 | 78.9 | 0.3 | 138.9 | 114.3 | 110.8 | 81.3 | 422 | NS | 0.51 | II |
| 20100802 | 7 | 18 | 25 | 78.9 | 0.3 | 138.9 | 114.3 | 110.8 | 81.3 | 422 | NS | 0.51 | II |
| 20110815 | 5 | 50 | 16 | 79.1 | 0.6 | 166.4 | 146.4 | 102.6 | 332.5 | 550 | S | 0.95 | I |
| 20110815 | 5 | 50 | 16 | 84.1 | 0.7 | 167.0 | 144.3 | 105.2 | 333.4 | 556 | S | 0.76 | I |

**Table S7:** Optical meteor events producing single infrasonic arrivals divided into the slow (< 40 km/s) and the fast (> 40 km/s) meteor entry velocity population. Multi arrival events are excluded from this tabulation.

|  | Slow Population (30 events) | | | Fast Population (25 events) | | |
| --- | --- | --- | --- | --- | --- | --- |
|  | Mean ± error | Min | Max | Mean ± error | Min | Max |
| Entry Velocity (km/s) | 24.4 ± 6.3 | 13.5 | 35.7 | 59.9 ± 8.8 | 43.9 | 75.6 |
| Begin Point Altitude (km) | 88.7 ± 9.2 | 72.0 | 104.4 | 111.2 ± 7.6 | 93.9 | 130.8 |
| End Point Altitude (km) | 60.7 ± 13.7 | 32.8 | 84.1 | 81.0 ± 8.0 | 63.4 | 93 |
| Meteor Flight Time (s) | 2.1 ± 1.3 | 0.8 | 5.2 | 0.9 ± 0.5 | 0.3 | 2.5 |
| Meteor Radiant Zenith Angle (deg) | 47.0 ± 15.5 | 14.6 | 88.4 | 47.4 ± 19.3 | 13.5 | 75 |
| Shock Source Altitude (km) | 72.3 ± 12.7 | 46.4 | 94.5 | 90.0 ± 10.3 | 75.4 | 111.4 |
| Total Range (km) | 132.7 ± 34.9 | 73.3 | 194.6 | 156.6 ± 32.0 | 100.3 | 216.6 |
| Ray Deviation (deg) | 97.8 ± 15.2 | 70.3 | 146.7 | 104.9 ± 18.4 | 75.2 | 151.5 |
| Signal Trace Velocity (km/s) | 0.392 ± 0.040 | 0.345 | 0.494 | 0.422 ± 0.111 | 0.332 | 0.831 |
| Signal Celerity (km/s) | 0.310 ± 0.015 | 0.274 | 0.336 | 0.305 ± 0.016 | 0.278 | 0.338 |
| Dominant Signal Frequency (Hz) | 4.24 ± 3.25 | 1.07 | 12.89 | 1.35 ± 0.71 | 0.39 | 2.93 |
| Dominant Signal Period (s) | 0.38 ± 0.23 | 0.07 | 0.94 | 0.98 ± 0.58 | 0.31 | 2.56 |
| Signal Duration (s) | 6.3 ± 1.5 | 3.0 | 10.0 | 7.9 ± 2.8 | 5.0 | 19.7 |

**Table S8:** Average statistics for single arrival events and multi arrival events. The 16 multi arrival events include statistics for all 35 arrivals, except for meteor entry velocity, meteor flight time and meteor zenith angle, which are global across all arrivals in each event.

| | Single Arrival Events | Min | Max | Multi Arrival Events | Min | Max | All Events |
|---|---|---|---|---|---|---|---|
| Shock Source Height (km) | 80.7 ± 14.6 | 46.4 | 111.4 | 73.2 ± 17.8 | 20.2 | 101.4 | 77.8 ± 16.2 |
| Begin Altitude (km) | 99.3 ± 14.1 | 72.0 | 130.8 | 95.24 ± 13.7 | 67.8 | 126.4 | 97.7 ± 14.0 |
| End Altitude (km) | 70.3 ± 15.2 | 32.8 | 93.0 | 57.7 ± 21.1 | 19.6 | 82.0 | 65.4 ± 18.6 |
| Signal Celerity (km/s) | 0.307 ± 0.015 | 0.274 | 0.338 | 0.307 ± 0.016 | 0.282 | 0.329 | 0.307 ± 0.016 |
| Total Range (km) | 144.0 ± 35.4 | 73.3 | 216.6 | 150.5 ± 41.8 | 111.7 | 277.1 | 146.5 ± 37.9 |
| Horizontal Range (km) | 117.7 ± 41.3 | 47.4 | 198.6 | 128.4 ± 47.5 | 59.3 | 267.9 | 121.3 ± 43.3 |
| Horizontal Trajectory Length (km) | 37.0 ± 27.4 | 5.8 | 119.6 | 32.7 ± 16.6 | 12.4 | 77.3 | 35.3 ± 23.8 |
| Ray Deviation (deg) | 101.2 ± 17.0 | 70.3 | 151.5 | 103.8 ± 15.0 | 78.6 | 144.6 | 102.2 ± 16.2 |
| Meteor Entry Velocity (km/s) | 41.2 ± 19.4 | 13.5 | 75.6 | 34.1 ± 18.3 | 11.5 | 67.5 | 38.4 ± 19.2 |
| Meteor Fight Time (s) | 1.55 ± 1.15 | 0.33 | 5.21 | 2.10 ± 1.42 | 0.47 | 6.07 | 1.8 ± 1.3 |
| Meteor Radiant Zenith Angle (deg) | 47.2 ± 17.2 | 13.5 | 88.3 | 42.2 ± 14.6 | 26.9 | 67.1 | 46.1 ± 16.7 |
| Signal Dominant Period (s) | 0.7 ± 0.5 | 0.1 | 2.6 | 0.6 ± 0.4 | 0.2 | 2.0 | 0.6 ± 0.5 |
| Signal Dominant Frequency (Hz) | 2.9 ± 2.8 | 0.4 | 12.9 | 2.0 ± 0.9 | 0.6 | 4.1 | 2.5 ± 2.3 |
| Maximum Amplitude (Pa) | 0.07 ±0.04 | 0.01 | 0.20 | 0.09 ± 0.08 | 0.01 | 0.49 | 0.07 ± 0.06 |
| Peak-to-peak Amplitude (Pa) | 0.10 ± 0.06 | 0.02 | 0.32 | 0.13 ± 0.10 | 0.02 | 0.68 | 0.12 ± 0.10 |
| Observed Travel Time (s) | 466 ± 119 | 240 | 709 | 489 ± 127 | 349 | 842 | 475 ± 122 |

# S3 Supplement to Section 2 (Methodology)

## S3.1 Astrometry (supplement to Section 2.1)

To make precise measurements of the position in the sky and movement of a meteor detected by a camera, calibrations of the plates use known stars to establish plates which map x,y pixel coordinates to local coordinates (elevation and azimuth). Since cameras can move slightly and lose calibration over time, for the highest degree of accuracy in astrometry it is helpful to ensure that the camera plates are made from stellar observations as close to the time of the meteor event as possible. Each camera produces a number of calibration images throughout the night (typically every 20 – 30 minutes) which can later be used for making astrometric plates. For the automated system, new plates are normally generated every 30 – 45 days using METeor AnaLysis (*METAL*), in-house software (Weryk et al, 2007; Weryk and Brown, 2012), which uses the RedSky routine to define the plate (Borovička et al, 1995). The sensitivity of the cameras allows use of stars to magnitude +3.5 for calibration in 30 second image stacks, where the magnitude refers to a stellar magnitude in the R band found in the SKY2000v4 catalogue (Myers et al, 2002).

In order to do astrometric measurements for each optical/infrasonic meteor in this study, it was first necessary to make new plates for each camera and for each night having an event. To make useable plates for any given camera, the all-sky calibration images have to satisfy the condition that there have to at least nine identifiable stars throughout the entire image, but less than ~50, after which plate residuals slowly increase due to random errors. However, it was not always possible to achieve this due to weather conditions. To overcome this shortcoming, the plates were made using multiple images spanning several hours. For those nights which remain cloudy throughout, the plates are made on the closest clear night (ideally a day or two before or after the actual event date). There are instances of two or more meteor events analysed in this study occurring within a time frame of order a week; however, the new plates were still produced for each night to ensure astrometric solution accuracy. Since astrometric stellar fits undergo significant degradation at low elevations (high zenith angle), it is preferable to choose calibration stars at elevations 20º or more above the horizon, where a plate fit solution has smaller stellar residuals (< 0.2 degrees). However, since many optically detected meteors which produce infrasound tend to be low to the horizon, it is necessary to select a good balance of stars throughout the entire image and at all elevations. In this process we use plates where the mean residuals (difference between the fit position and the actual position for stars used in the fit) do not exceed 0.2 degrees. *METAL* displays the star residuals on the screen, thus allowing for outliers to be removed interactively. The plate can be fit and re-fit at any point until a desired average or maximum stellar residual is achieved. The interested reader is referred to Weryk et al. (2007), Brown et al. (2010), and Weryk and Brown (2012) for further details about *METAL*.

While it was possible to generate astrometric trajectory solutions using automated picks, these solutions generally had high residuals (>0.2 km) and often were affected by unusual effects, such as hot pixels, weather conditions (i.e. overcast), blemishes or reflections on the dome, meteor fragmentation, very bright flares and occasional insects that adversely affect the quality of the automated picks (positions of the meteor as a function of time in the plane of the sky). Therefore, manual reductions were performed for the final set of complete trajectory solutions for those meteor events having a probable infrasound signal association based on the initial automated solution.

Once the plates were made, in-house programs/functions written in *IDL* were used to generate astrometric solutions for each event. The procedure includes: (i) selecting images containing the meteor; (ii) using either automated meteor position picks or a rough version of the manual picks as an approximate guideline in selecting new meteor picks manually, (iii) apply the plate; (iv) generate a trajectory solution using the software *MILIG* (Borovička, 1990); (v) verify if the solution is acceptable (i.e. residuals from each camera are less than 0.2 km from the trajectory straight line solution and good (<10%) average speed agreement among all cameras); (vi) repeat as necessary until the solution meets the residual and interstation speed consistency . Each frame had the meteor position measured using a manual centroid, since an automatic centroid may suffer from undesirable shifting under certain conditions, such as pixels being close to a star, in very dim and/or overly bright regions (e.g. blooming and oversaturation).

The other criteria we use to define a good astrometric solution also include: the intersecting planes of any two cameras have to be at an angle (Q) of more than 20º (in most instances the trajectory solution is unstable and unreliable otherwise), the entire meteor trail should be clearly visible, the meteor should be at elevation of >20º above the horizon from cameras used in the solution and the event lasts at least 10 frames. Complications in reduction occur due to poor sky conditions, flares on the meteor trail and spurious reflections on the camera dome. These complications are dealt with manually on an event by event basis. For example, many optically detected meteors which produce infrasound tend to be low to the horizon, making the astrometric reductions less accurate as the pixel scale is larger at low elevations. In these cases, the trajectory solution is produced using local plate fits, obtained by concentrating on the sky region around the meteor and choosing nearby stars rather than stars throughout the image. Due to poor sky conditions and/or poor camera angle view geometry, astrometric solutions were judged to be of low quality for four optical events having simultaneous infrasound signals and rejected from the final simultaneous infrasound - optical meteor data set.

We computed the event time, accounting for both any interstation camera time discrepancies and time of the first detected frame for each camera. Establishing absolute timing is important in raytracing analysis, when the ray travel time (time between the airwave arrival time and the event start time measured by the camera) has to be known accurately. We estimate the absolute time from any one camera based upon the first frame used for astrometry by subtracting from the trigger time any additional frames from a particular camera where manual examination shows the meteor to be visible. All event start times from all cameras included in the astrometric solution were averaged to give one global event start time. The standard deviations of the averaged camera times were generally less than one second, except in two cases, which were manually corrected when raytracing was performed. This is the estimated maximum uncertainty in our travel time due to uncertainty in the absolute camera event time. The time error was included in our overall travel time uncertainty for each event start time using the standard deviation between cameras calculated in the previous step and applied to the total uncertainty/error in the observed signal travel time.

## S3.2 Meteor Infrasound Signal Identification and Measurements (supplement to Section 2.1)

*MatSeis* (Figure S3.1) implements the standard form of cross-correlation of the output between each element of the array and performs beamforming of the signals across the array (Evers and Haak, 2001). *PMCC* (Figure S3.2) is sensitive to signals with very low signal-to-noise ratio (SNR) and uses element pair-wise correlation techniques to declare detections on the basis of signal coherency and back azimuth identifying return 'families' in time and frequency space (Cansi and Klinger, 1997; Cansi and Le Pichon, 2009).

Even though the near field events, especially if they are discrete point source explosions (expected to be produced from fragmenting meteoroids), tend to produce diverging spherical waves, the plane wave geometry approximation in both *MatSeis* and *PMCC* is still fairly good when calculating the back azimuth from sources many times further away than the array size; however, it should be noted that the wave distortion effects and atmospheric variability, winds in particular, may produce additional uncertainty. For example, the observed back azimuth deviations for far field infrasonic events due to the variability of atmospheric winds can be as large as ±15° (Garcés, 2013).

To search for possible signals from typical, small regional meteor events using MatSeis, we used the following detection parameter ranges: window size 7 – 10s, window overlap 50 – 70%, Butterworth bandpass $2^{nd}$ order filter cutoffs between 0.2-1 Hz on the lower end and 2 Hz up to 45 Hz on the upper end. A series of separate independent runs employing different filter and window settings within these ranges for each possible event are used for every meteor to isolate a possible associated infrasound signal recorded by ELFO. Additionally, the correlation and Fisher F-statistics (Melton and Bailey, 1957) have to be above the background values (F-statistics >3) for a positive detection to be declared. Before any coherent event is declared a possible correlated detection, it has to satisfy the additional conditions that the back azimuth and trace velocity have small standard deviations (<2º and < 0.010 km/s, respectively), without any abrupt variations or spread within the correlation window. Other nearby moving signal sources (e.g. storms, helicopters) generally give infrasonic signals with a spread over several or even tens of degrees in the back azimuth, last for many tens of seconds and in some cases show significant variations in trace velocity; characteristics not typically expected for most meteors. Therefore, such coherent signals would not qualify as possible correlated meteor detections. Based on our experience examining meteor infrasound signals in the past, we also rejected signals which had the following characteristics: signals containing only high frequency (>15 Hz) content, repetitive signals (more than 4 impulsive instances in 15 or less seconds), signals with trace velocity < 0.30 km/s, and long duration signal clusters (>30s).

Typical detection parameter settings for PMCC were as follows: 5 – 8s window length with 5s time step with default 'family' settings (e.g. Brachet et al., 2010; Cansi and LePichon, 2009). The Chebyshev filter parameters were: $2^{nd}$ order, 15 bands with ripple size of 0.005. As with Matseis, a series of independent runs are performed to search for 'families'. Once a 'family' is detected, additional runs are performed to narrow down the frequency range, arrival time, duration and other signal characteristics.

The results from both *MatSeis* and *PMCC* are then compared to look for inconsistencies with the detected signal and determine whether a given positive detection is associated within the window range of our criteria for each meteor. All signal properties found by *PMCC* were recorded and then used as a secondary means of event and signal measurement verification by comparing to *MatSeis* results.

The uncertainty in the signal onset (arrival time) was set to 1s, in order to account for potential windowing and intrinsic biases, including the probable discrepancies between *MatSeis* and *PMCC*, though we expect in most cases we have localized the start time with better precision. We found empirically that the signal arrival times as calculated in both *MatSeis* and *PMCC* were found to generally agree within ~1s for the majority of our meteor infrasound events.

The dominant signal period was calculated using two separate approaches; first, by measuring the zero crossings of the waveform at the maximum Hilbert envelope (max peak-to-peak amplitude) and second, by finding the inverse of the frequency at the maximum signal power spectral density (PSD) after subtracting the noise. In contrast to the Ens et al. (2012) procedure, where the signal is stacked using a best beam across the array and where long duration, often high SNR infrasonic bolide signals on large arrays were examined, here the signal measurements on each separate channel were performed. This included calculating the maximum amplitude, peak-to-peak amplitude and the period at maximum amplitude to check for any intra-array discrepancies. As there were periods when one of the elements (Centre Element) experienced digitizing issues and thus the amplitude was either systematically higher or lower by some factor (~2x or 0.5) than the rest of the elements, this approach ensured our amplitudes, in particular, did not suffer biases due to equipment problems. In cases where one element was rejected from amplitude measurements, while all four channels were used in cross-correlating, beamforming the period and isolating the signal, the remaining three channels were used for the maximum and peak-to-peak amplitude measurements.

## S3.3 Analysis of Raytracing Results (supplement to Section 2.3)

Even for events showing unique source height solutions, the two quantities used to define source height independently (travel time and back azimuth) would frequently have height solutions which differed from each other, in some instances by up to as much as 20 km much, larger than the uncertainty from the various realizations). To reconcile these differences we developed a set of heuristic rules to try and best objectively quantify the height uncertainty based on our experience with past solutions. For events showing multiple signal arrivals, each arrival was treated as a distinct infrasonic 'signal'. The algorithmic logic for deciding on a best estimate for source height and error is summarized in Figure S3.3 For each modelled quantity, the absolute value of the difference between each model run and observation is first computed, as shown in equation (S3.1) - we refer to these as the model residuals. Here, $res$ is the residual, $Q_{obs}$ is the observed and experimentally measured quantity, while $Q_{model}$ is the modelled quantity:

$$res = |Q_{obs} - Q_{model}| \qquad (S3.1)$$

All residuals lying within a window containing the observed value and its uncertainty were found. A model residual of zero indicates quantitative agreement between the model value and the measurement (Figure S3.4). This produces a 2-dimensional grid with all the possible heights along one dimension, and the model residuals along the other dimension where all residuals are weighted equally. Even though the arrival ray elevation angle and launch ray deviation angle from the trajectory were not used to establish the source height, the residuals were still calculated in order to get an overall sense of the goodness of the solution in qualitative terms. To make a best estimate for source height, we use equally weighted measurement residuals of source height from the travel time and backazimuths produced by the model runs (equation (S3.1)), first using only those residuals within the measurement error (red points in Figure S3.4).

If no residuals are found within the measurement uncertainty window the range of allowable residuals was increased beyond the one sigma in measurement uncertainty. Here we used expected uncertainties in the atmospheric models and back azimuth uncertainties to gauge the size of the initial allowable increases starting at 1% for travel time (which translates to 1% variation in signal celerity), 1.5° for back azimuth, 5° for elevation angle and 10° for the ray deviation angle. These values are chosen based on the expectation that the signal celerity may vary by a few percent and back azimuth is often not exactly a plane wave in the near field (Pierce and Posey, 1970; Picone et al., 1997; Brown et al, 2003). The choices for the elevation angle and ray deviation were arbitrarily set to the above mentioned values, as these quantities were not used to determine the shock source height but only as a secondary check. The final shock source height selection was made primarily based on the arrival times, with the back azimuth as a secondary measurement, except for degenerate solutions. The implementation of the atmospheric variability in both the direction along and transverse to the propagation path from source to receiver delineates our maximum expected deviations in travel time and back azimuth.

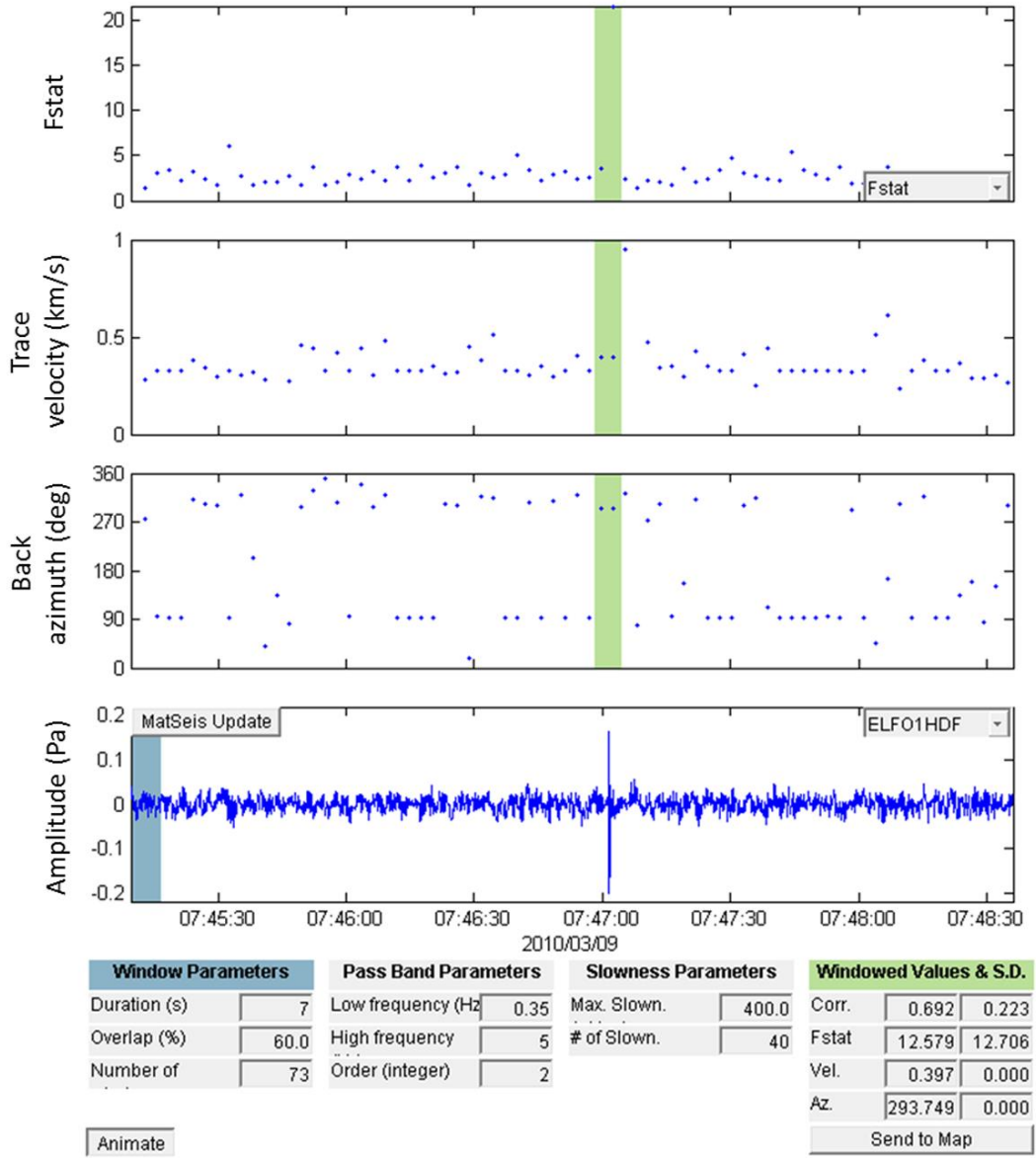

**Figure S3.1:** An example of a meteor infrasound signal displayed in *InfraTool* (*MatSeis 1.7*). The top window is the F-statistic, a measure of the coherency of the signal across the array elements, the second window is the apparent trace velocity of the infrasound signal across the array in the direction of the peak F-stat, and the third window shows the best estimate for the signal back-azimuth. The fourth window shows the bandpassed raw pressure signal for the Centre Element of ELFO.

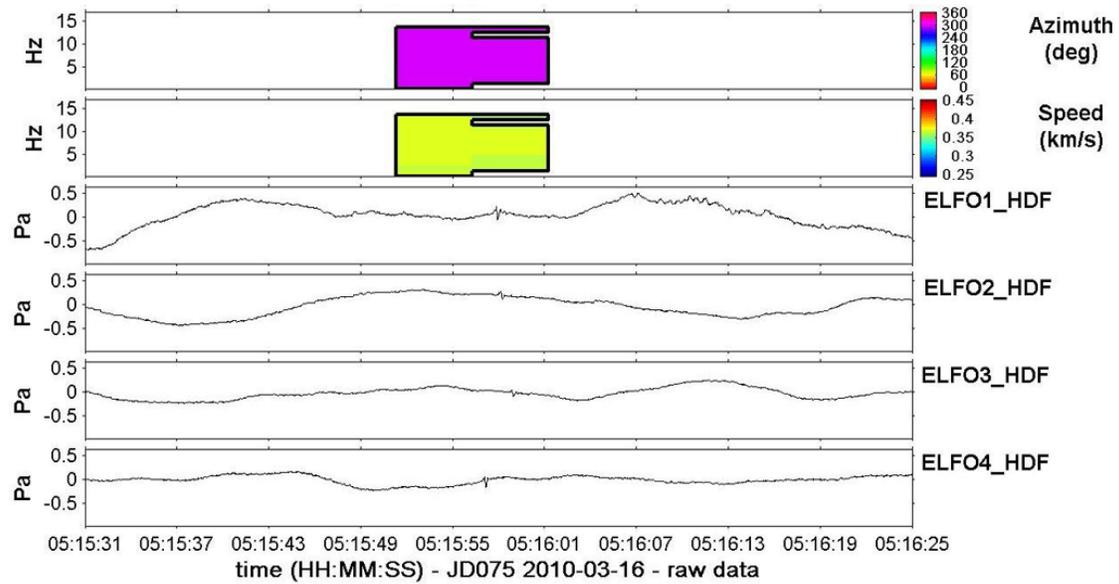

**Figure S3.2:** Results from array processing using the *PMCC* algorithm. The top window gives the observed azimuth, the middle window represents the trace velocity of the signal, while the bottom window shows the bandpassed raw pressure signal for all four array elements.

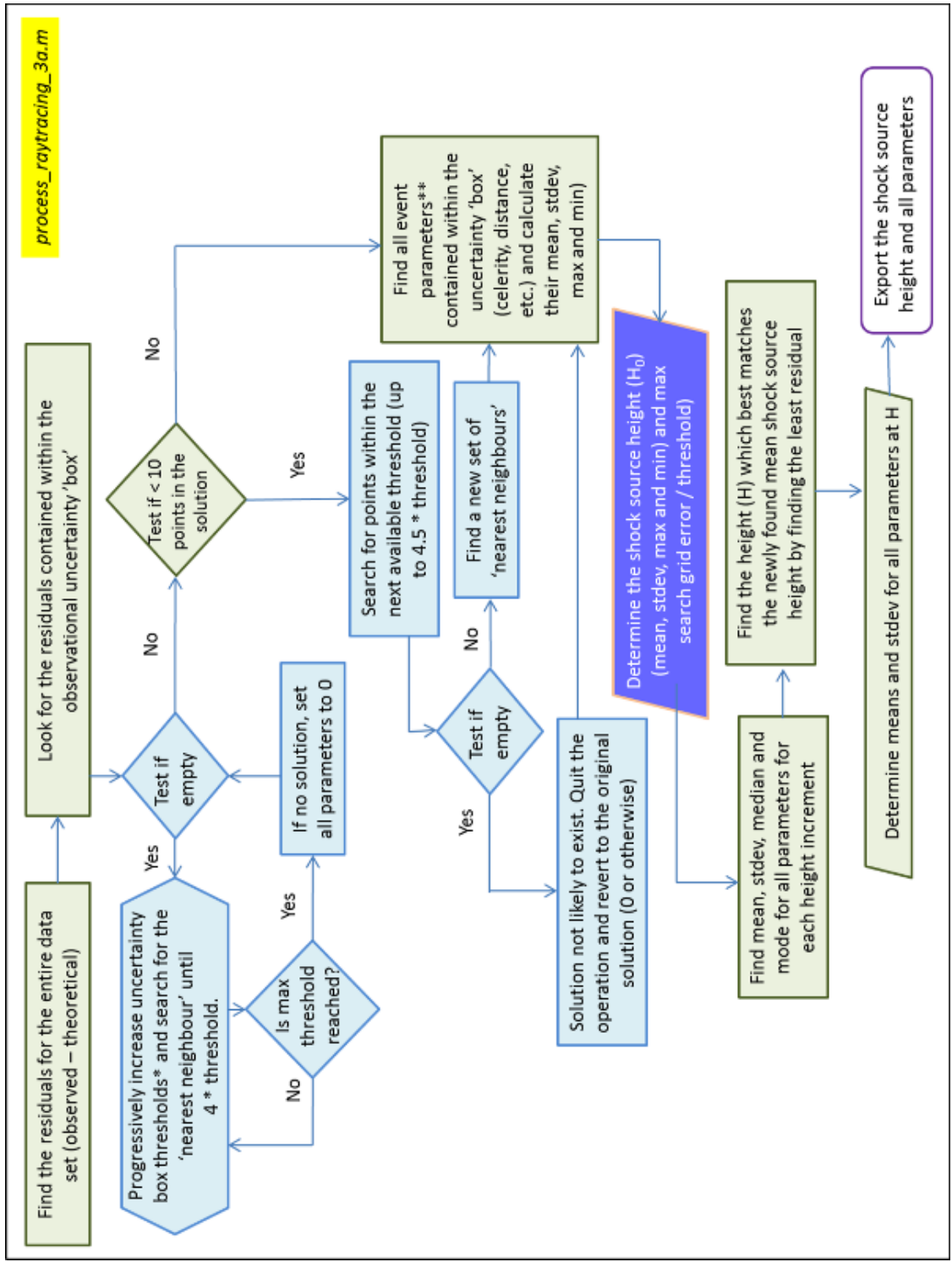

**Figure S3.3:** The flow chart showing the logic used for calculating the meteor shock source heights and associated parameters.

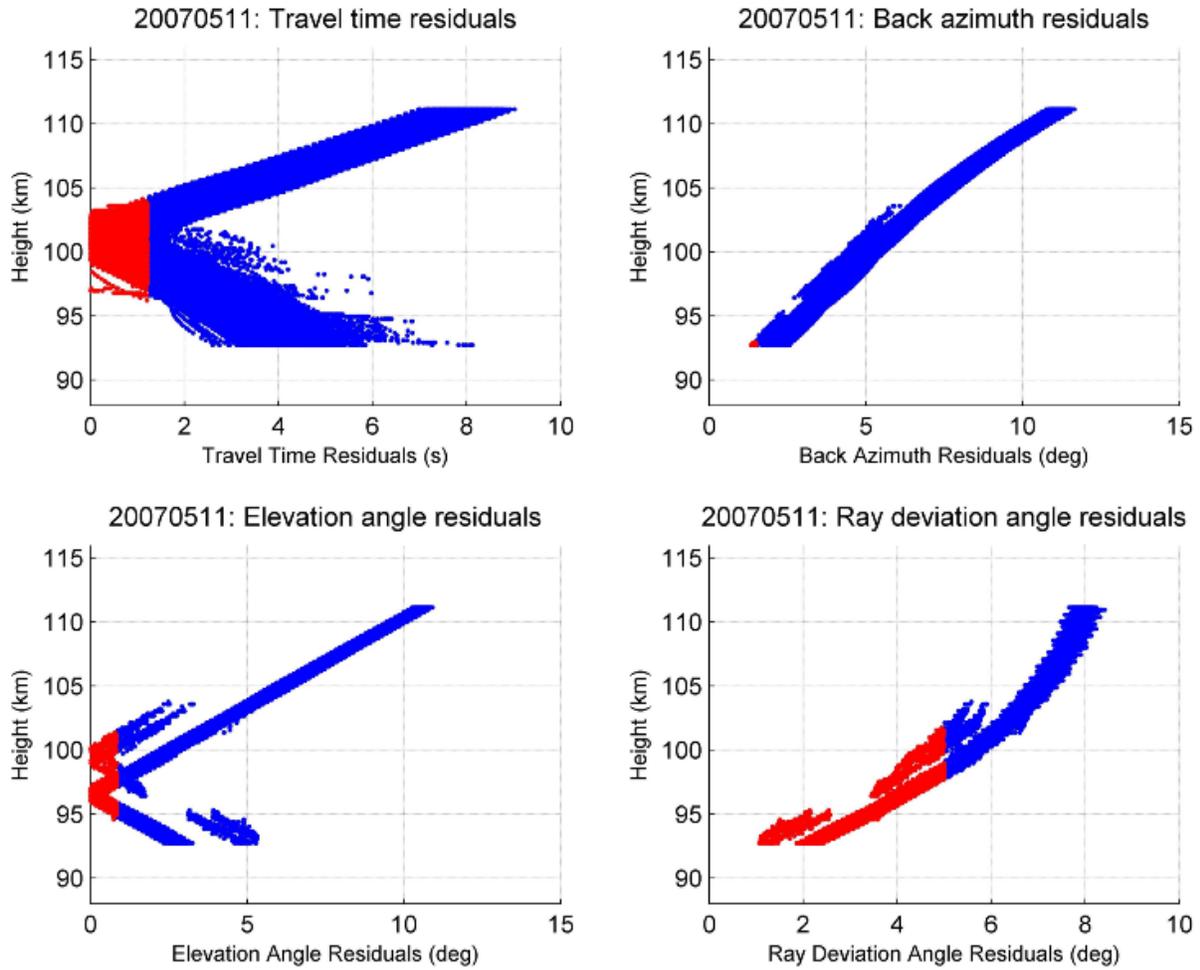

**Figure S3.4:** A residuals composite plot showing the difference between the model values for each realization and the observed value. Red points show the residuals contained within the height-residual grid where model points agree with observations within the observational uncertainty. Here the ray deviation angle residuals are relative to a ballistic arrival (i.e. residual of zero = 90 degree ray deviation angle).

# S4 Supplement to Section 3 (Results)

## S4.1 Detection Efficiency

The lowermost limit on the detection rate of the array during the 6 year period covered in this study is approximately 1% - out of the total sample of 6989 optically detected meteors, ~80 were also detected by infrasound. This translates into an average of one meteor optical detection on our network producing noticeable infrasound at ELFO per month. However, not all these automated solutions represent physically reasonable solutions as evidenced by unreasonable begin height, end height and entry velocity. Furthermore, the true total infrasonic detection rate is 3-4 times higher accounting for daylight, weather conditions and moonlight which restrict the optical detections to a 20%-30% duty cycle throughout the entire year. This implies that for a site with the noise characteristics of ELFO (Appendix A) approximately one regional meteor per week is expected to be detected infrasonically as a direct arrival. Though rarer, bolides (energies above 1 kiloton TNT) producing ducted infrasonic arrivals on a nearly global scale are expected a few times per year (Brown et al., 2013). On this basis we expect direct arrival meteor infrasound from regional sources to be the dominant type of meteor infrasound detectable at any one station by close to an order of magnitude. We emphasize that identification of an infrasound signal with a meteor or bolide requires separate cuing (in our case optical detection) as the meteor infrasound signal is indistinguishable from most other infrasound sources. In this sense, most meteor infrasound signals can be expected to go unrecognized in the absence of other information. The distribution of begin and end heights for all optically detected meteors, irrespective of infrasound production, as well as the infrasound detected fraction as a function of end height for our study, are shown in Figure S4.1. It is clear that deep penetrating fireballs are much more likely to produce infrasound detectable at the ground, compared to those ending at higher altitudes; however, only a very small number of all optical meteors detected by the SOMN network penetrate below 50 km. Note that the end height for most of our events is not the same as the source height. For example, only about 13% of all SOMN detected meteors have their end height below 70 km, 4% reach below 60 km, while <1% of our optically detected gram-sized and larger meteors have their end heights between 20-40 km. We note, however, that the sole event with an end height below 30 km is the meteorite-dropping Grimsby fireball (Brown et al., 2011); the rest of our data have end heights above 30 km.

## S4.2 Final Source Height Solution Quality

To determine the quality of the final source height solution, the travel time and back azimuth residuals were compared. The quality of the solutions were classified by the amount the estimated source height primary parameter deviates (percent in travel time and degrees in back azimuth) from the observed quantity. If both travel time and back azimuth were within 1% and 1°, respectively, then the solution was deemed well constrained and of high quality (solution quality Type $S_A$). For the remainder of the solution quality categories ($S_B$ = 2% or 2°, $S_C$ = 3% or 3°, $S_D$ = 4-5% or 4-5°, $S_F$ = >5% or >5°), the event was assigned based on the highest deviation of either travel time or back azimuth. For example, if the travel time was within 2%, but back azimuth deviation was greater than 5°, the event was assigned to the $S_F$ quality category (Table S4.1).

We classify our raytracing solutions according to the types (unique solution (S), no solution (NS) and degenerate solution (D)) for all events: in the single arrival category, there are 60% S, 35% NS and 5% D solutions, while in the multi arrival group, there are 69% S, 17% NS and 14% D solutions. For events falling into the no solution category, to find an estimate using the travel time and thus determine the best estimated shock source altitude, the residuals threshold grid had to be increased. Degenerate solutions are found only in those events which occur at ranges of less than 150 km. We note that almost half (45%) of single arrival events lie within 150 km, while nearly all (94%) multi arrival events are found within that range.

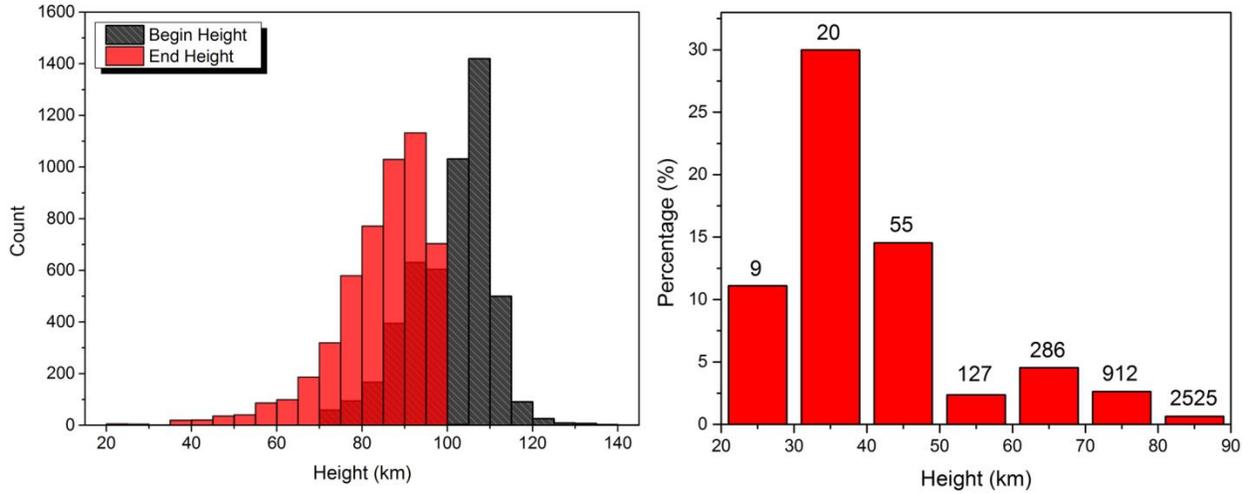

**Figure S4.1:** Left: Distribution of begin and end heights for ASGARD meteors from 2006-mid 2013. The transparent red histogram (end heights) overlays the gray histogram (begin heights). Outliers and events beyond the upper and lower cutoff limits are excluded. Outliers arise as a result of non-converging automated solutions or incorrect picks. Right: Detection rate of infrasonic meteor signals from simultaneously observed optical events for the entire network as a function of end height during the period between 2006 and 2011 in 10 km bins. The percentage represents the ratio between the number with end height of our final data set of all infrasound producing meteors (71 events) and all optically detected meteors in the given height bin during the 6 year period as a function of height. The count number (given at the top of the vertical bar in each bin) represents the total number of optically detected meteors in that height bin (3934 out of 6989 events were in the end height range between 20 – 90 km).

**Table S4.1:** All events, categorized based on their solution quality.

| Solution Viability | Travel time (%) | Back azimuth (°) | Single arrivals | Multi arrivals | All |
|---|---|---|---|---|---|
| $S_A$ | < 1 | < 1 | 19 | 11 | 30 |
| $S_B$ | $1 \leq n < 2$ | $1 \leq n < 2$ | 7 | 8 | 15 |
| $S_C$ | $2 \leq n < 3$ | $2 \leq n < 3$ | 9 | 5 | 14 |
| $S_D$ | $3 \leq n < 5$ | $3 \leq n < 5$ | 12 | 5 | 17 |
| $S_F$ | $\geq 5$ | $\geq 5$ | 8 | 6 | 14 |

# S5 Instrumentation

## S5.1 Instrumentation: Infrasound Array

The Elginfield Infrasound Array (ELFO), situated near the town of Elginfield (43º.1907N, 81º.3152W, 322 m), some 20 km north of London, Ontario, Canada, is a four sensor tripartite (microbarograph) array, positioned in a traditional triangular formation with an off-centre central element (Figure S5.1).

Since it is expected from theory (ReVelle, 1976) that infrasound from regional (< 300 km range) meteors will have a peak infrasound frequency in the range of ~1 Hz, the array element spacing was optimized for this frequency. Each microbarograph is placed in a vault designed to protect it from the elements and minimize temperature variations and all vaults are located in forest to reduce noise. The sensors are 12-port Chapparals, model 2.5 made by Chapparal Physics, with a flat response (3 dB points) from 0.1 to 200 Hz. Three elements use 15m long porous garden hoses laid out in a star pattern to minimize the local wind noise (Christie and Campus, 2010), while the fourth element (Northwest Element) features a wind shelter, built in August 2007. A snow fence is installed around all elements to further reduce local wind noise. Data from each element is digitized at 100 Hz and transmitted via a buried and steel armoured TECK cable to a centralized data system, where it is stored locally and streamed to Natural Resources Canada in Ottawa. A GPS antenna at each element enables timing to be embedded in the data stream.

Since beginning operation on January 25, 2006, the array has been continuously collecting infrasound data, capturing signals produced by a number of phenomena, such as machinery, lightning, storms, mining activities, local explosions, etc. During the time period of this study, the array has experienced occasional temporary equipment issues. For example, the sensor at the Centre Element experienced systematic gain problems, where the amplitude was either higher or lower by some factor (~0.5 – 2x). In the summer of 2009, a delay in the sensor replacement resulted in only three functional array channels for a period of several months. Regular preventative maintenance visits are conducted in order to inspect all equipment, perform repairs (e.g. re-install snow fence) and replacements if necessary (e.g. garden hoses can break down due to elements or animal interference).

The most prominent source of seasonally dependent persistent background noise is the Niagara Falls, located about 150 km NE from London. From late April to early October, it produces a constant coherent signal with the mean frequency of 2 Hz, which falls within the same frequency range as most meteors, and hence reduces detection efficiency during those months. Figure S5.2 shows the infrasound noise level and variation as a function of local time of day at ELFO during the summer of 2006. In contrast with earlier studies (e.g. Kraemer, 1977) which detected ~1 meteor/year, during our study we find on average about one optically measured meteor is infrasonically detected per month at ELFO. This meteor infrasound data base is particularly unique as meteor infrasound is correlated with meteor optical data obtained with a multi-station all-sky camera network. This has allowed detection of small, short-range meteor infrasound events and ensures robust confirmation of each meteor infrasound event based on timing and directionality determined from optical data as described later.

We note that not every meteor will produce infrasound detectable at the ground, and not every meteor that does produce infrasound is detected by all-sky cameras, the latter being limited to night-time operations under clear skies. Here we consider only those events which are simultaneously recorded by at least two stations of the all-sky camera network (thus permitting trajectory solutions) as well as have an associated infrasound signal. We provide average detection frequency estimates and lower energy bounds in the results section based on these considerations, updating the earlier work by Edwards et al. (2008).

## S5.2 Instrumentation: All Sky Camera System

The All-Sky and Guided Automatic and Realtime Detection (ASGARD) camera uses 8-bit HiCam HB-710E cameras with Sony Hole Accumulation Diode (HAD) CCDs and Rainbow L163VDC4 1.6-3.4 mm f/1.4 lenses producing all-sky views from each station (see Figure S5.3 for an of example all-sky view from a typical camera).

The cameras operate with a gamma setting of 0.45. All cameras are enclosed in a waterproof 30cm in diameter acrylic dome, and set up to observe the entire sky (Weryk et al., 2007) during the night. Each site records video at 29.97 frames per second with 640 x 480 pixel resolution, corresponding to a pixel scale of 0.2 degrees resulting in typical trajectory solutions on order of ~ 250 m precision. The system hardware and software are described in Weryk et al. (2007) and Brown et al (2010).

The all-sky camera system uses an automated detection algorithm in real time (Weryk et al., 2007), triggered by bright visual meteors (brighter than -2 magnitude). Meteors of this brightness correspond roughly to masses ranging from 100g (at 15 km/s) to 0.1g at 70 km/s (Jacchia et al., 1967). When a meteor is detected by two or more cameras, an automated astrometric solution (Table S5.1 and Figure S5.4) is also produced and saved together with raw video comprising 30 frames prior to and 30 frames post event. All stations have Network Time Protocol (NTP) calibrated time using GPS signals.

For this study, automated astrometric solutions were used only for initial optical meteor association to locate infrasound detections; all final astrometric solutions were obtained using manual processing in IDL.

The meteor astrometric measurements for this study are used to establish the begin and end points of the luminous meteor trail (latitude, longitude and height), radiant (apparent direction in the sky a meteor emanates) as well as the meteor speed at any point of the trajectory. These quantities are used to associate potential infrasound signals with meteors and identify source heights of the infrasound signal.

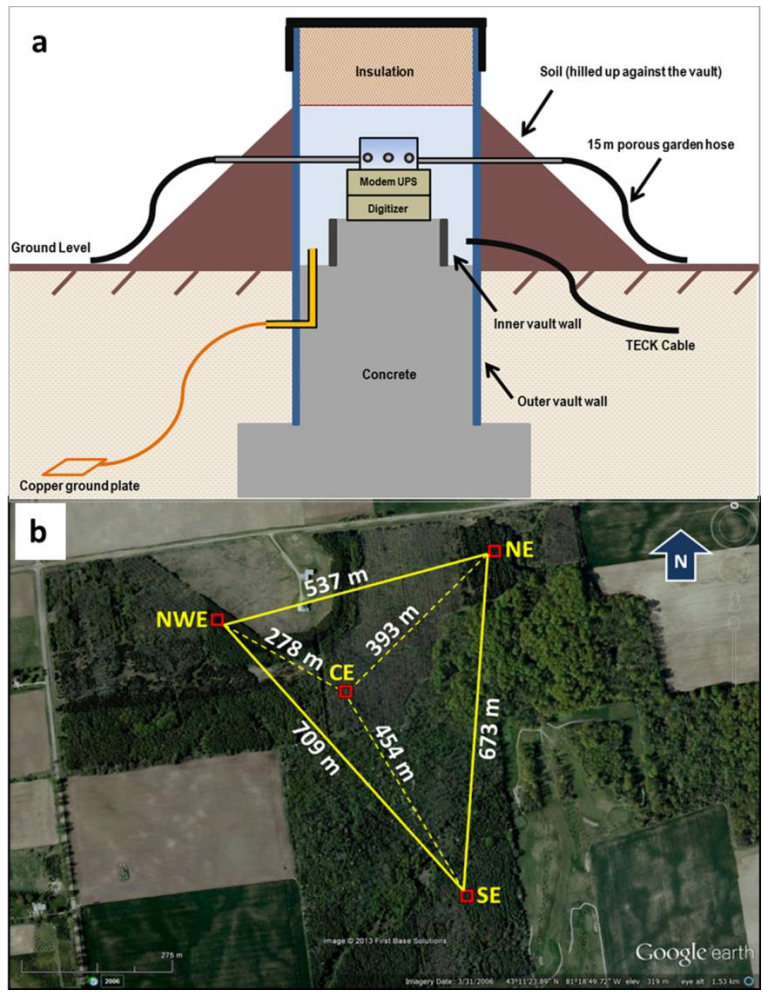

**Figure S5.1:** (a) ELFO vault diagram; (b) plane view of the array configuration of ELFO.

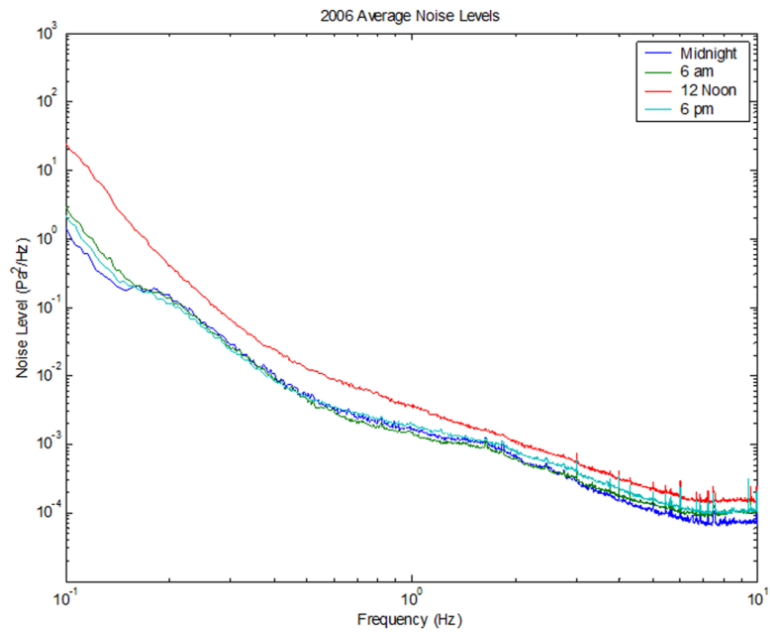

**Figure S5.2:** A power spectral density (PSD) plot for the entire array for the summer 2006 showing the noise levels as a function of day/night hour. The average noise levels at ELFO at 10 Hz, 1 Hz and 0.1 Hz are ~$10^{-4}$ Pa$^2$/Hz, ~$10^{-3}$ Pa$^2$/Hz and ~$10^{-1}$ Pa$^2$/Hz, respectively.

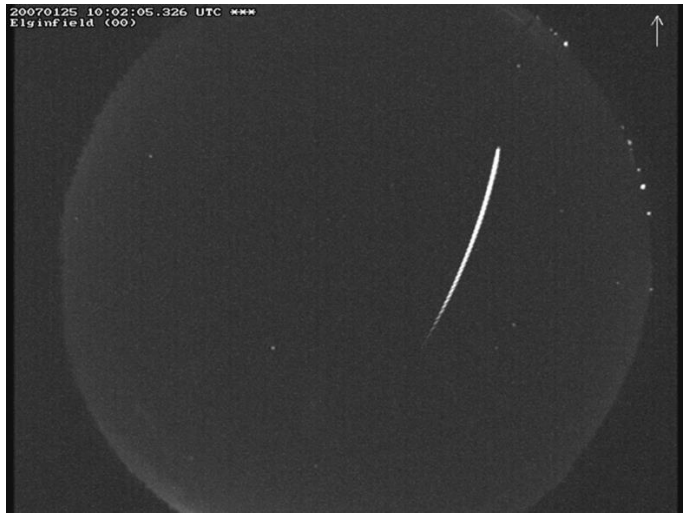

**Figure S5.3:** An example of a stacked (100 frames) video image showing a meteor captured by one of the all sky cameras. North is shown by the arrow. This particular event has a long trail. Most of the events have much shorter trails and are often low in the horizon, where the atmospheric collecting area is largest.

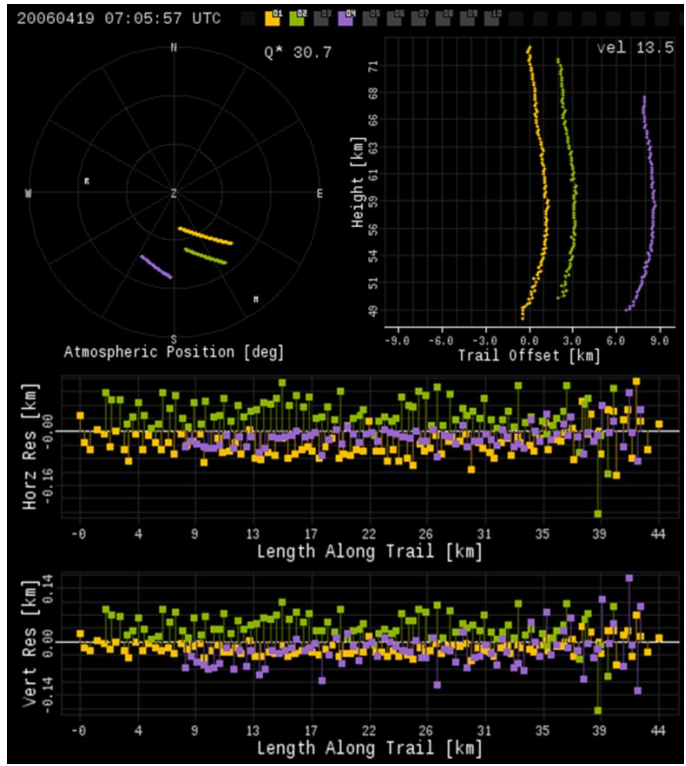

**Figure S5.4:** An automated trajectory solution for a meteor event recorded by three cameras of the ASGARD system. The top left panel shows the apparent path of the meteor as seen from the three different camera sites where the event was detected. The upper right panel shows the apparent height vs. model height of the meteor where the latter uses an average constant speed of 13.5 km/s - the curved lines demonstrate that the meteor noticeably decelerated. The bottom plots show the individual meteor picks on each frame projected to the meteor trail - deviations are shown from the horizontal and vertical relative to the best fit straight line solution in the atmosphere.

**Table S5.1:** A sample output from the automated optical meteor astrometric solutions. N is the number of cameras detecting the event and used in the trajectory solution, Q* is the maximum angle between cameras local observation planes of the meteor, shw indicates if the event is associated with a known meteor shower using the three letter codes from the International Astronomical Union (http://www.ta3.sk/IAUC22DB/MDC2007/), vel and err are the entry speed and error (in km/s), respectively, H_beg and H_end are the begin and end heights in km, respectively.

| date | time | N | Q* | shw | vel | err | H_beg | H_end |
|---:|---:|---:|---:|:---:|---:|---:|---:|---:|
| 20131125 | 10:44:03 | 2 | 75.8 | ... | 57.4 | 3.9 | 105.5 | 94.1 |
| 20131125 | 10:33:26 | 3 | 48.2 | ... | 60.6 | 2.4 | 106 | 96.5 |
| 20131125 | 08:44:44 | 4 | 89.5 | ... | 68.6 | 0.7 | 113 | 89.8 |
| 20131125 | 08:36:05 | 2 | 48.6 | ... | 67.7 | 3.2 | 107.1 | 98.4 |
| 20131125 | 08:14:50 | 2 | 79.5 | LEO | 67.8 | 1.4 | 112.4 | 102.5 |
| 20131125 | 07:41:19 | 2 | 41.7 | ... | 30.4 | 0 | 87.7 | 72.6 |
| 20131125 | 07:28:17 | 2 | 24.5 | NOO | 44.8 | 1.1 | 95.2 | 81.9 |
| 20131125 | 06:35:39 | 2 | 57.9 | NOO | 40.1 | 1.3 | 89.2 | 76.4 |
| 20131125 | 03:54:53 | 3 | 74.5 | ... | 57.8 | 5 | 95.9 | 90.7 |
| 20131125 | 03:53:16 | 3 | 76.2 | ... | 60.6 | 0.4 | 108.2 | 86.6 |
| 20131125 | 03:42:27 | 2 | 13.1 | ... | 29 | 1.6 | 88.2 | 80 |
| 20131124 | 23:54:28 | 4 | 88.6 | ... | 28.4 | 1 | 99.1 | 65 |
| 20131124 | 11:24:58 | 2 | 40.6 | ... | 53.4 | 3.2 | 94.2 | 84.3 |
| 20131124 | 10:44:56 | 2 | 55.5 | LEO | 67.4 | 6.5 | 106.7 | 90.4 |